%% file: main.tex
\documentclass[12pt,draftcls, onecolumn]{IEEEtran}
\usepackage[utf8]{inputenc}
\usepackage{amsmath,amsfonts,amssymb}
\usepackage{multicol}
\usepackage{tikz}
\tikzset{>=stealth}
\usepackage{braket}
\usepackage{pgfplots}
\pgfplotsset{compat=1.18}
\usepgfplotslibrary{fillbetween}
\usetikzlibrary{spy}
\usepackage{physics}
\usepackage{cases}
\usepackage{bbm}
\usetikzlibrary{decorations.pathreplacing,calligraphy}
\usepackage[linesnumbered,ruled,vlined]{algorithm2e}
\usepackage[short]{optidef}
\SetKwRepeat{Do}{do}{while}
\renewcommand{\vec}[1]{\ensuremath{\boldsymbol{#1}}}
\usepackage[thinc]{esdiff}
\allowdisplaybreaks
\title{Multi-Player Resource-Sharing Games with Fair Reward Allocation}
\author{Mevan Wijewardena, Michael J. Neely
\thanks{The authors are with the Electrical Engineering department at the University of Southern California.}
\thanks{This work was supported in part by one or more of: NSF CCF-1718477, NSF SpecEES 1824418.}
}
\date{June 2023}
\usepackage{pgfplots}
\newtheorem{theorem}{Theorem}
\newtheorem{lemma}{Lemma}

\begin{document}
\maketitle
%%%%%Congestion Games
\begin{abstract}
This paper considers an online multi-player resource-sharing game with bandit feedback. Multiple players choose from a finite collection of resources in a time slotted system. In each time slot, each resource brings a random reward that is equally divided among the players who choose it. The reward vector is independent and identically distributed over the time slots. The statistics of the reward vector are unknown to the players. During each time slot, for each resource chosen by the first player, they receive as feedback the reward of the resource and the number of players who chose it, after the choice is made. We develop a novel Upper Confidence Bound (UCB) algorithm that learns the mean rewards using the feedback and maximizes the worst-case time-average expected reward of the first player. The algorithm gets within $\mathcal{O}(\log(T)/\sqrt{T})$ of optimality within $T$ time slots. The simulations depict fast convergence of the learnt policy in comparison to the worst-case optimal policy.
\end{abstract}
\section{Introduction}\label{sec1}
In this paper, we consider the following game with $m \geq 2$ players numbered $1,2,\dots,m$, and $n \geq 2$ resources numbered $1,2,\cdots,n$. The game evolves in slotted time $t \in \{1,2,\dots\}$. The vector $\vec{W}(t) \in \mathbb{R}^n$ denotes the random reward vector at time $t \in \{1,2,\dots\}$. In particular, for each $i\in \{1,2,\dots,n\}$ and each $t \in \{1,2,\dots\}$, $W_i(t) \geq 0$ denotes the reward offered by resource $i$ at time $t$. We assume that $\vec{W}(t)$ are i.i.d. with $\mathbb{E}\{\vec{W}(t)\} = \vec{E} = [E_1,E_2,\dots,E_n]$. The vector $\vec{E}$ is unknown to the players. During each time slot, each player selects $r$ resources without knowing the other player's selections (assume that $0 < r \leq n$), and without knowledge of $\vec{W}(t)$. During time slot $t$, for each $k \in [1:n]$, each player selecting resource $k$ receives a reward of $W_k(t)/S_k(t)$ from resource $k$, where $S_k(t)$ is the number of players choosing resource $k$ during time slot $t$. For each $i \in \{1,2,\dots,m\}$, let $\mathcal{A}_i(t)$ denote the set of resources chosen by player $i$ during time slot $t$. During time slot $t$, after the selection of resources, player $i$ receives $(W_k(t),S_k(t))$ for $k \in \mathcal{A}_i(t)$ as feedback. 

The total reward received by player $i$ during time slot $t$ is $\sum_{k \in \mathcal{A}_i(t)} W_k(t)/S_k(t)$. The time-average expected reward of player $i$ in a finite time horizon of $T$ time slots is
\begin{align}
    \frac{1}{T}\sum_{t=1}^T\mathbb{E}\left\{\sum_{k \in \mathcal{A}_i(t)} \frac{W_k(t)}{S_k(t)}\right\}.
\end{align}
The goal is to design policies to maximize the time-average expected reward of player 1. However, this is not possible since player 1 does not have control over the policies of the other players. Hence, we focus on maximizing the \textit{worst-case} time-average expected reward of player 1, which we define in the sections below. 

For $k \in \{1,2,\dots,n\}$ and $t \in \{1,2,\dots\}$, define $X_k(t) = \sum_{i=2}^m \mathbbm{1}_{[k \in \mathcal{A}_i(t)]}$. Hence, $X_k(t)$ is the number of players (other than player 1) choosing resource $k$ during time slot $t$. Also, we have $S_k(t) = \mathbbm{1}_{[k \in \mathcal{A}_1(t)]} + X_k(t)$. For each $t$, it can be shown that $\vec{X}(t) \in \mathcal{J}$, where
\begin{align}\label{eqn:J_def}
\mathcal{J} = \left\{\vec{x} \in \{0,1,\dots,m-1\}^n \Bigg{|} \sum_{j=1}^n x_j = (m-1)r \right\}.
\end{align}
Additionally, for a given $t \in \{1,2,\dots\}$, it can be easily shown that for any $\vec{x} \in \mathcal{J}$, there exists a way for players 2 to $m$ to choose resources such that $\vec{X}(t) = \vec{x}$.

\subsection{Time Average Expected Reward}
Define $\mathcal{H}(t) = \{(\mathcal{A}_1(\tau),\{W_k(\tau),S_k(\tau); 1 \leq k \leq n,k \in \mathcal{A}_1(\tau)\});1 \leq \tau < t\}$,  the history up to time $t$.
Given $\mathcal{H}(t)$, the action of player $1$ at time $t$ is conditionally independent of the other player's actions at time $t$. Define the random vector $\vec{p}(t)$ with components $p_k(t) = \mathbb{E}\{\mathbbm{1}_{[k \in \mathcal{A}_1(t)]}|\mathcal{H}(t)\}$.
Since $\sum_{k=1}^n\mathbbm{1}_{[k \in \mathcal{A}_1(t)]} = r$, it can be shown that $\vec{p}(t) \in \Delta_{n,r}$, where $\Delta_{n,r}$ is the $(n,r)$-hypersimplex given by
\begin{align}\label{def:hyp_simp_def}
    \Delta_{n,r} = \left\{\vec{p} \in \mathbb{R}^n_+: \sum_{i=1}^n p_i = r, p_i \in [0,1] \ \forall i\in \{1,2,\dots,n\}\right\}.
\end{align}
Also, notice that given $\vec{p} \in \Delta_{n,r}$, we can use the Madow's sampling technique (see, for example~\cite{pmlr-v151-mukhopadhyay22a}) to sample an action set $\mathcal{A} \subset \{1,2,\dots,n\}$ such that, $|\mathcal{A}| = r$, and $p_k = \mathbb{E}\{\mathbbm{1}_{[k \in \mathcal{A}]}\}$ for each $k \in \{1,2,\dots,n\}$.

Notice that we can write the time-average expected reward $R(T)$ of player 1 in a finite time horizon of $T$ time slots as
\begin{align}\label{eqn:regret}
    R(T) &= \frac{1}{T}\sum_{t=1}^T \sum_{k=1}^n \mathbb{E}\left\{\frac{W_k(t)\mathbbm{1}_{[k \in \mathcal{A}_1(t)]}}{1+X_k(t)}\right\}=_{(a)} \frac{1}{T}\sum_{t=1}^T \sum_{k=1}^n E_k\mathbb{E}\left\{\frac{\mathbbm{1}_{[k \in \mathcal{A}_1(t)]}}{1+X_k(t)}\right\} \nonumber\\& = \frac{1}{T}\sum_{t=1}^T \sum_{k=1}^n E_k \mathbb{E}\left\{\mathbb{E}\left\{\frac{\mathbbm{1}_{[k \in \mathcal{A}_1(t)]}}{1+X_k(t)} \Bigg{|}\mathcal{H}(t)\right\}\right\}\nonumber\\& =_{(b)} \frac{1}{T}\sum_{t=1}^T \sum_{k=1}^n E_k \mathbb{E}\left\{\mathbb{E}\left\{\frac{1}{1+X_k(t)} \Bigg{|}\mathcal{H}(t)\right\}\mathbb{E}\left\{ \mathbbm{1}_{[k \in \mathcal{A}_1(t)]}|\mathcal{H}(t)\right\}\right\}\nonumber\\& = \frac{1}{T}\sum_{t=1}^T \sum_{k=1}^n E_k \mathbb{E}\left\{\mathbb{E}\left\{\frac{1}{1+X_k(t)} \Bigg{|}\mathcal{H}(t)\right\}p_k(t)\right\}  \nonumber\\&=_{(c)} \frac{1}{T}\sum_{t=1}^T \sum_{k=1}^n E_k \mathbb{E}\left\{\frac{p_k(t)}{1+X_k(t)}\right\} =\frac{1}{T}\sum_{t=1}^T  \mathbb{E}\{f(\vec{p}(t),\vec{X}(t))\},
\end{align}
where (a) follows since $\vec{W}(t)$ is independent of the actions of players at time $t$, (b) follows since $\mathbbm{1}_{[k \in \mathcal{A}_1(t)]}$ is independent of $X_k(t)$ conditioned on $\mathcal{H}(t)$, (c) follows since $\vec{p}(t)$ is an $\mathcal{H}(t)$-measurable random variable, and the function $f: \mathbb{R}_+^n \times \mathbb{Z}_+^n \to \mathbb{R}$ is defined as
\begin{align}\label{eqn:f_def}
     f(\vec{p},\vec{x}) = \sum_{k=1}^{n}\frac{E_kp_k}{1+x_k}.
\end{align}
The time-average expected reward of player 1 is
\begin{align}\label{eqn:time_av_reward}
    R \coloneq \lim\inf_{T \to \infty} R(T).
\end{align}
\subsection{Worst-Case Time Average Expected Reward}
Notice that since player 1 does not have access to $\vec{X}(t)$ when taking action during time slot $t$, they cannot directly maximize $R$ defined in \eqref{eqn:time_av_reward}. But notice that for fixed $\vec{p} \in \Delta_{n,r}$, the worst-case value of $f(\vec{p},\vec{x})$ is $f^{\text{worst}}(\vec{p})$, where
\begin{align}\label{eqn:def_g}
    f^{\text{worst}}(\vec{p}) = \min_{\vec{x} \in \mathcal{J}} f(\vec{p} , \vec{x}).
\end{align}
Combining with \eqref{eqn:regret}, the worst-case time-average expected reward in a finite time horizon of $T$ time slots is given by
\begin{align}\label{eqn:r_worst}
    R^{\text{worst}}(T) = \frac{1}{T}\sum_{t=1}^T\mathbb{E}\{f^{\text{worst}}(\vec{p}(t))\}.
\end{align}
Hence, the worst-case time-average expected reward of player 1 is
\begin{align}\label{eqn:r_worst_t}
    R^{\text{worst}} =  \lim\inf_{T \to \infty} R^{\text{worst}}(T).
\end{align}
Instead of maximizing $R$, player 1 can take decisions to maximize $R^{\text{worst}}$ without knowledge of the decisions of other players.

From \eqref{eqn:r_worst} and \eqref{eqn:r_worst_t}, we have that the maximum possible value of $R^{\text{worst}}$ is $f^{\text{worst},*}$, where
\begin{align}\label{eqn:f_worst_star}
    f^{\text{worst},*} = \max_{\vec{p} \in \Delta_{n,r}}\min_{\vec{x} \in \mathcal{J}}f(\vec{p}, \vec{x}) = \max_{\vec{p} \in \Delta_{n,r}} f^{\text{worst}}(\vec{p}),
\end{align}
that is achieved by using the policy $\vec{p}(t) = \vec{p}^*$ in each time slot, where 
\begin{align}\label{eqn:p_worst_star}
    \vec{p}^* \in \arg\max_{\vec{p} \in \Delta_{n,r}} f^{\text{worst}} (\vec{p}).
\end{align}
The $f^{\text{worst}}$ function is unknown to player 1 because the function $f$ defined in \eqref{eqn:f_def} is in terms of the unknown $E_k$ values. Hence, we aim to design an algorithm that achieves a worst-case time-average expected reward close to $f^{\text{worst},*}$ using the bandit feedback. 

\noindent
\textbf{Note: }Notice that one can relax the constraint of each user choosing exactly $r$ resources by allowing each user to select at most $r$ resources. Since the rewards are assumed to be nonnegative, this will not affect $f^{\text{worst}}$. To formalize this, following the same analysis as before, it can be established that the worst-case expected utility in this case is
\begin{align}\label{eqn:tilde_f_worst_star}
    \tilde{f}^{\text{worst},*} = \max_{\vec{p} \in \tilde{\Delta}_{n,r}}\min_{\vec{x} \in \tilde{\mathcal{J}}}f(\vec{p}, \vec{x}),
\end{align}
where 
\begin{align}
    \tilde{\Delta}_{n,r} = \left\{\vec{p} \in [0,1]^n: \sum_{i=1}^n p_i \leq r\right\}
\end{align}
and
\begin{align}
    \tilde{\mathcal{J}} = \left\{\vec{x} \in \{0,1,\dots,m-1\}^n : \sum_{j=1}^n x_j \leq (m-1)r \right\}.
\end{align}
Now assume $(\vec{\tilde{p}}^*,\vec{\tilde{x}}^{*})$ achieves the maximin optimality for~\eqref{eqn:tilde_f_worst_star}. Now consider any $\vec{\hat{p}}^* \in \Delta_{n,r}$ such that $\vec{\hat{p}}^* \geq \vec{\tilde{p}}^*$, where the inequality is taken entry-wise and let $\vec{\hat{x}} \in \arg\min_{\vec{x} \in \tilde{\mathcal{J}}}f(\vec{\hat{p}}^*, \vec{x})$. Consider arbitrary $\vec{\hat{x}}^* \in \mathcal{J}$ such that $\vec{\hat{x}}^* \geq \vec{\hat{x}}$. First notice that $f(\vec{\hat{p}}^*,\vec{\hat{x}}^*) \leq f(\vec{\hat{p}}^*,\vec{\hat{x}})$, where the inequality follows combining definition of $f$ in \eqref{eqn:f_def} with $\vec{\hat{x}}^* \geq \vec{\hat{x}}$. Since, $\vec{\hat{x}} \in \arg\min_{\vec{x} \in \tilde{\mathcal{J}}}f(\vec{\hat{p}}^*, \vec{x})$, the above means $\vec{\hat{x}}^* \in \arg\min_{\vec{x} \in \tilde{\mathcal{J}}}f(\vec{\hat{p}}^*, \vec{x})$. Now, notice that $f(\vec{\hat{p}}^*, \vec{\hat{x}}^*) \geq f(\vec{\tilde{p}}^*, \vec{\hat{x}}^*) \geq f(\vec{\tilde{p}}^*, \vec{\tilde{x}}^*)$, where the first inequality follows  combining definition of $f$ in \eqref{eqn:f_def} with $\vec{\hat{p}}^* \geq \vec{\tilde{p}}^*$ and the second inequality follows since $\vec{\tilde{x}}^* \in \arg\min_{\vec{x} \in \tilde{\mathcal{J}}}f(\vec{\tilde{p}}^*, \vec{x})$. Hence, the pair $(\vec{\hat{p}}^*,\vec{\hat{x}}^*)$ is also a maximin optimal point for~\eqref{eqn:tilde_f_worst_star}. Since $\vec{\tilde{p}}^* \in \Delta_{n,r}$ and $\vec{\hat{x}}^* \in \mathcal{J}$ by definition, we have $\tilde{f}^{\text{worst},*} = f^{\text{worst},*}$.
\subsection{Related work}
The main challenge of applying online optimization techniques such as online gradient descent~\cite{orabona2023modernintroductiononlinelearning} to the above problem is due to the fact that we do not know the function $f$ since we do not know $\vec{E}$. The problem shares certain similarities with the problems of multi-armed bandit learning (MAB)~\cite{Lai1985,Auer2002FinitetimeAO}, adversarial bandit learning~\cite{lattimore_szepesvári_2020,odonoghue2021matrix}, online-convex optimization~\cite{Zinkevich2003}, online-convex optimization with multi-point bandit feedback~\cite{Agarwal2010OptimalAF}, and stochastic convex optimization~\cite{Hazan2014}.

Multi-armed bandit learning is extensively studied in the literature. The classical MAB problem consists of a fixed number of arms each with fixed mean reward. A player chooses an arm in each iteration of the game, without knowledge about the mean rewards, where after the choice is made the reward of the chosen arm is revealed to the player. The goal is to learn to choose the arm with the highest mean reward. An algorithm for the MAB problem has to explore all the arms in order to learn the best arm. But in doing so, the player also chooses arms with low mean reward, which affects the long term reward of the player. Upper confidence bound based algorithms, where the algorithm maintains an upper bound on the mean cost of each arm, is a popular in the MAB literature~\cite{Bubeck2012,lattimore_szepesvári_2020}. Our problem cannot be addressed using classical MAB approaches since the reward not only depends on the chosen resource, but also on the choices of other players. Another related problem is adversarial bandit learning. Unlike the worst-case approach, the adversarial bandit framework cannot be used to obtain utility guarantees for player 1 that are independent of the actions of the other players.

The framework of online optimization also shares similarities with our work since our goal is to design an online algorithm to minimize $f^{\text{worst}}(\vec{p})$. However, notice that $f^{\text{worst}}$ depends on the unknown vector $\vec{E}$. We also do not have access to an unbiased estimate or an unbiased gradient estimate of the function $f^{\text{worst}}$ due to its definition in \eqref{eqn:def_g}. Hence, the work on online-convex optimization where partial information on the underlying reward functions are revealed, such as online-convex optimization with multi-point bandit feedback, and the approaches based on stochastic gradient descent are also not applicable. Our problem is more similar to the work of~\cite{odonoghue2021matrix} on zero-sum matrix games with bandit feedback. However, the above work considers a two-player scenario where both players receive the actions and the rewards of themselves and the opponent as feedback.

Our game model has been studied for the offline non-stochastic case with full information on $\vec{E}$ under the more general framework of resource-sharing games~\cite{Rosenthal1973}, also known as congestion games. In these games, the \textit{per-player reward} of a resource is a general function of the number of players selecting the resource. Also, an action for a player is a subset of the resources, where the allowed subsets make up the player's action space. Resource-sharing games have also been extended to various stochastic settings~\cite{Nikolova2011,Angelidakis2013}. Problems similar to our work have been studied in the context of adversarial resource-sharing games. The work of~\cite{harks2022multi} considers an adversarial resource-sharing game where each player chooses a single resource from a collection of resources, after which an adversary chooses the resource chosen by the maximum number of players. Also, non-atomic congestion games with malicious players have been considered through the work of~\cite{BABAIOFF200922}. The above works assume that $\vec{E}$ is known to all the players. 

We have simplified the general resource-sharing game model described above in two ways. First, we assume a fair-reward allocation model, where we have assumed the existence of a reward for each resource, which is divided equally between the players selecting it. Second, we have assumed simple action spaces for players by allowing each player to select an arbitrary subset of $r$ resources. Resource-sharing games with special \textit{per-player reward} definitions have been considered in the literature. One such notable case is when the \textit{per-player} reward of a resource is nondecreasing in the number of players selecting the resource. These games are called cost-sharing games~\cite{Syrgkanis2010}. The particular case when the total cost of a resource is divided equally among the players choosing it is called \textit{fair cost-sharing games}. In such a model, a player would prefer to select resources selected by many players. In the fair reward allocation model considered in our work, players have the opposite incentive to select resources selected by a small number of players. 

One application of our model is multiple access control (MAC) in communication systems, where multiple users access communication channels, and the data rate of a channel is shared amongst the users who select it~\cite{Akkarajitsakul2011,Garg2002, Felegyhazi2006}. Here, a channel can be shared using Time Division Multiple Access (TDMA) or Frequency Division Multiple Access (FDMA), where in TDMA, the channel is time-shared among the users~\cite{Aryafar2013,Felegyhazi2007}, whereas in FDMA, the channel is frequency-shared among the users~\cite{Li2015}. In both cases, the total data rate supported by the channel can be considered the reward of the channel. Here, limiting the number of channels accessed by a single user in a given time slot is desirable. Additionally, the channel data rate should be shared among the users accessing the channel. 

The worst-case expected reward is an important objective different from Nash-equilibrium~\cite{Nash1950,Nash1951} and correlated equilibrium~\cite{Aumann1974,Aumann1987,Osborne1994}.
The problem of finding an approximate Nash equilibrium of a congestion game with bandit feedback has been considered \cite{Cui2022LearningIC}. However, implementing the algorithms by \cite{Cui2022LearningIC} requires cooperation among players. In contrast, the worst-case approach requires no cooperation among the players. Additionally, player 1 does not have to make assumptions about other players' strategies. Hence, understanding the worst-case expected reward is important even when the other players are not necessarily playing to hurt player 1. However, in practice, some players play just to hurt others. One particular example arises in military communications. Consider a multiple access communication system used in a military setting (for instance, consider the TDMA scheme considered in~\cite{Felegyhazi2007}, which has a similar structure to our model). Here, some users may transmit to disrupt the communication capabilities of other users. Our formulation is applicable even when the other users cooperate to reduce the data rate of a single player. Another motivation for the worst-case objective of this paper is to quantify the degree of punishment that can be inflicted on a particular user. This value is useful, for example, in repeated game algorithms that design punishment modes into the strategy space in order to discourage deviant behavior~\cite{SOLAN2002362, Osborne1994}.
\subsection{Background on Resource-Sharing Games}
The resource-sharing game was first studied by~\cite{Rosenthal1973}. These games, also called congestion games, fall under the general category of potential games~\cite{MONDERER1996124}. In potential games, the effect of any player changing policies is captured by the change of a global potential function. Various extensions to the classical resource sharing game introduced by~\cite{Rosenthal1973} have been studied in the literature~\cite{CHIEN2011315}. Some such extensions are stochastic resource-sharing games~\cite{Nikolova2011,Angelidakis2013}, weighted resource-sharing games \cite{Bhawalkar2010}, games with player-dependent reward allocation~\cite {MILCHTAICH1996111}, games with resources having preferences over players~\cite{Ackermann2008}, and singleton games, where each player is only allowed to choose a single resource~\cite{FOTAKIS20093305,Gairing2004}. 

Also similar to resource-sharing games are resource allocation games~\cite{Grundel2013ResourceAG,grundel2018}. In these games, a resource must be fairly divided among claimants claiming a certain portion. There is also work combining resource-sharing games with bandits and strategic experimentation. The work of~\cite{d2021strategic} considers a two-player game where players continually choose between their private risky arm and a shared safe arm. Only one player can activate the safe arm at any given time, which guarantees a payoff. This congestion effect on the safe arm gives rise to strategic consideration among the players. These works are based on the model of multi-agent, multi-armed bandit problems introduced by~\cite{Bolton1999}. Here, multiple players are faced with the same multi-armed bandit problem. In contrast to the classic single-agent setting, players can learn from other players' feedback, resulting in some players being able to free-ride on other players' experiments. This phenomena induces strategic experimentation.

Resource-sharing games have applications in multiple-access~\cite{Akkarajitsakul2011,Seo2018_1,Seo2018_2}, network selection~\cite{Malanchini2013}, network design~\cite{Anshelevich2004},  spectrum sharing~\cite{Ahmad2009}, resource sharing in wireless networks~\cite{Liu2009}, load balancing networks~\cite{Zhang2021}, radio access selection \cite{Ibrahim2010}, service chains~\cite{Le2020}, and congestion control~\cite{Zhang2019}

%Removes references.
% Migration of species quint1994model. 
% Load balancing networks Caragiannis2006
\subsection{Contributions}
We study the problem of maximizing the worst-case time average expected reward of online resource-sharing games with a fair-reward allocation model in the presence of bandit feedback on the mean rewards of the resources. We assume a model where in each time slot, each player is allowed to choose any $r$ element subset of the $n$ available resources, and the reward of a resource is shared among the users selecting it. We propose a novel algorithm combining the upper confidence bound technique with Madow's sampling technique and Euclidean projection onto the $(n,r)$-hypersimplex, to maximize the worst-case time average expected reward of player 1. In particular, in each time slot of the algorithm, we find $\vec{p}(t)$ in the $(n,r)$-hypersimplex, after which we sample the $r$ resources for player 1 using Madow's sampling technique. The algorithm gets within $\mathcal{O}(\log(T)/\sqrt{T})$ of optimality in a finite time-horizon of $T$ time slots. The parameters of the algorithm do not depend on $T$. Hence, the above guarantee can be achieved even if the time horizon $T$ is unknown. 
\subsection{Notation}
We use calligraphic letters to denote sets. Vectors and matrices are denoted in boldface characters. For integers $n$ and $m$, we denote by $[n:m]$ the set of integers between $n$ and $m$ inclusive. Also, we use $\mathbb{N} = \{1,2,3,\dots\}$ to denote the set of positive integers and $\mathbb{N}_0 = \{0,1,2,\dots\}$ to denote the set of non-negative integers.
\section{Bandit Algorithm }
Now, we move on to the algorithm and analysis. Before introducing the algorithm, we begin with a few definitions and some preliminary results that are useful.

Our algorithm, provided in Algorithm~\ref{algo:2} below, uses the first $n$ time slots as an initial exploration phase that obtains at least one sample of the reward of each of the $n$ resources. The main part of the algorithm starts in time slot $n+1$.

For all $t \in \{n+1,n+2,\dots\}$ and $k \in [1:n]$ define $n_k(t)$ as the number of times player $1$ chooses resource $k$ before time slot $t$. Formally, $n_k(t) = \sum_{\tau = 1}^{t-1} \mathbbm{1}_{[k \in \mathcal{A}_1(\tau)]}$,
where $\mathcal{A}_1(t)$ denotes the set of resources chosen by player  1 during time-slot $t$. Notice that due to initial exploration phase of Algorithm~\ref{algo:2}, we have that $n_k(t) \geq 1$ for all $k \in [1:n]$ and $t \in \{n+1,n+2,\dots\}$. For each $t \in \{n+1,n+2,\dots\}$ and $k \in [1:n]$, define
\begin{align}\label{eqn:bar_e}
    \bar{E}_k(t) = \frac{1}{n_k(t)}\sum_{\tau=1}^{t-1} \mathbbm{1}_{[k \in \mathcal{A}_1(\tau)]}W_k(\tau).
\end{align}
Fix $\delta_t \in (0,1)$ for each $t \in \{n+1,n+2,\dots\}$ such that $\delta_t \geq \delta_{t+1}$ for all $t \in \{n+1,n+2,\dots\}$. For  each $t \in \{n+1,n+2,\dots\}$ and $k \in [1:n]$, define
\begin{align}\label{eqn:til_e}
    \tilde{E}_k(t) = \bar{E}_k(t)+\sqrt{\frac{2\log\frac{n_k(t)(n_k(t)+1)}{\delta_t}}{n_k(t)}}.
\end{align}

Also, define the functions $f_t: \mathbb{R}^n \times \mathbb{N}_0^n\to \mathbb{R}$ for $t \in \{n+1,n+2,\dots\}$ as
\begin{align}\label{eqn:def_f_t}
    f_t(\vec{p},\vec{x}) = \sum_{k=1}^{n} \frac{\tilde{E}_k(t)p_k}{1+x_k}.
\end{align}

Before moving on to the main result, we introduce the following well-known lemma.
\begin{lemma}\label{lemma:chernoff}
Given a sequence $\{X_t\}_{t=1}^{\infty}$ of independent zero-mean $1$-sub Gaussian random variables, a positive integer-valued random variable $G$ (possibly dependent on the sequence $\{X_t\}_{t=1}^{\infty}$) and $\epsilon \in (0,1)$, we have
\begin{align}
  \mathbb{P}\left\{\frac{1}{G}\sum_{i=1}^GX_i \geq \sqrt{\frac{2\log\frac{G(G+1)}{\epsilon}}{G}}\right\} \leq \epsilon,  \mathbb{P}\left\{\frac{1}{G}\sum_{i=1}^GX_i \leq -\sqrt{\frac{2\log\frac{G(G+1)}{\epsilon}}{G}}\right\} \leq \epsilon.
\end{align}
\begin{proof}
    This result is given as an exercise in the book~\cite{lattimore_szepesvári_2020}. We include the proof for completeness. Define $\bar{X}(t) = \frac{1}{t}\sum_{\tau = 1}^t X_{\tau}$ for each $t \in \mathbb{N}$. Notice that,
        \begin{align}
            &P\left\{\bar{X}(G) \geq \sqrt{\frac{2\log{\frac{G(G+1)}{\epsilon}}}{G}}\right\} =  \sum_{g=1}^{\infty}P\left\{\bar{X}(g) \geq \sqrt{\frac{2\log{\frac{g(g+1)}{\epsilon}}}{g}},G= g\right\} \nonumber\\& \leq_{(a)}   \sum_{g=1}^{\infty}P\left\{\bar{X}(g) \geq \sqrt{\frac{2\log{\frac{g(g+1)}{\epsilon}}}{g}}\right\} \leq_{(b)} \sum_{g=1}^{\infty}e^{-\frac{\left(\sqrt{\frac{2\log{\frac{g(g+1)}{\epsilon}}}{g}}\right)^2}{2/g}}\nonumber\\& = \sum_{g=1}^{\infty}\frac{\epsilon}{g(g+1)} = \sum_{g=1}^{\infty}\left(\frac{\epsilon}{g} - \frac{\epsilon}{g+1}\right) = \epsilon
        \end{align}
         where (a) follows since for any two events $A,B$, $P(A,B) \leq P(A)$, and (b) follows  since $\bar{X}(g)$ is $1/\sqrt{g}$-sub-Gaussian. The other inequality follows from a similar argument.
\end{proof}
\end{lemma}
Fix $t \in \{n+1,n+2,\dots\}$ and $k \in [1:n]$. For each $s \in [1:n_k(t)]$, define $\tilde{W}_k(s)$ as the reward obtained when the resource $k$ is chosen for the $s$-th time by player  $1$. Hence, notice that $\bar{E}_k(t) = \frac{1}{n_k(t)} \sum_{s=1}^{n_k(t)} \tilde{W}_k(s)$.

Notice that from assumption \textbf{A2}, the collection $\{\tilde{W}_k(s)-E_k\}_{s=1}^{n_k(t)}$ is a collection of independent $1$-sub Gaussian random variables. Applying Lemma~\ref{lemma:chernoff}-(c) to the sequence $\{\tilde{W}_k(t)-E_k\}_{t=1}^{n_k(t)}$ with $G = n_k(t)$ and $\epsilon = \delta_t$, we have
\begin{align}\label{eqn:app_lemma}
    \mathbb{P} \left\{ \frac{1}{ n_k(t)} \sum_{s=1}^{n_k(t)} (\tilde{W}_k(s) -E_k) \geq \sqrt{\frac{2\log{\frac{n_k(t)(n_k(t)+1)}{\delta_t}}}{n_k(t)}}\right\} \leq \delta_t,
\end{align}
and 
\begin{align}\label{eqn:app_lemma_2}
    \mathbb{P} \left\{ \frac{1}{ n_k(t)} \sum_{s=1}^{n_k(t)} (\tilde{W}_k(s) -E_k) \leq -\sqrt{\frac{2\log{\frac{n_k(t)(n_k(t)+1)}{\delta_t}}}{n_k(t) }}\right\} \leq \delta_t,
\end{align}
The above two inequalities translate to,
\begin{align}\label{eqn:up_bnd}
    \mathbb{P}\left\{E_k \geq \tilde{E}_k(t)\right\} \leq \delta_t,
\end{align}
and 
\begin{align}\label{eqn:low_bnd}
    \mathbb{P}\left\{E_k \leq \tilde{E}_k(t) - 2\sqrt{\frac{2\log{\frac{n_k(t)(n_k(t)+1)}{\delta_t}}}{n_k(t)}} \right\} \leq \delta_t.
\end{align}
for all $t \in \{n+1,n+2,\dots\}$ and $k \in [1:n]$, where $\tilde{{E}}_k(t)$ is defined in \eqref{eqn:til_e}.

Now consider the collection $\{G_{n+1},G_{n+2},\dots\}$ of events that shall be called ``good" events: For $t \in \{n+1,n+2,\dots\}$ the ``good" event $G_t$ is defined by the inequalities
\begin{align}\label{eqn:good_ev_1}
    E_k < \tilde{E}_k(t),
\end{align}
and 
\begin{align}\label{eqn:good_ev_2}
    E_k > \tilde{E}_k(t) - 2\sqrt{\frac{2\log{\frac{n_k(t)(n_k(t)+1)}{\delta_t}}}{n_k(t)}}
\end{align}
for $k \in [1:n]$. Specifically, $G_t$ is defined as the event that \eqref{eqn:good_ev_1} and \eqref{eqn:good_ev_2} hold for all $k \in [1:n]$. Combining \eqref{eqn:low_bnd} and \eqref{eqn:up_bnd} with the union bound, we have that 
\begin{align}\label{eqn:complment_ineq}
    \mathbb{P}\{G_t^c\} \leq  2n\delta_t.
\end{align}
Recall that
 \begin{align}\label{eqn:def_star_p_1}
     \vec{p}^* \in  \arg\max_{\vec{p} \in \Delta_{n,r}}f^{\text{worst}}(\vec{p}),
 \end{align}
 where the function $f^{\text{worst}}$ is defined in \eqref{eqn:def_g}. Let
 \begin{align}\label{eqn:def_star_x}
     \vec{x}^* \in \arg\min_{\vec{x} \in \mathcal{J}} f(\vec{p}^*,\vec{x}),
\end{align}
where the function $f$ is defined in \eqref{eqn:f_def}. Hence, we have that
\begin{align}\label{eqn:f_worst_star_rem}
    f^{\text{worst},*} = f(\vec{p}^*,\vec{x}^*),
\end{align}
where $f^{\text{worst},*}$ is defined in \eqref{eqn:f_worst_star}.
Before moving on to the Algorithm and the main theorem, we first prove the following lemma.
\begin{lemma}\label{lemma:prel_lemma}
    Fix $t \in \{n+1,n+2,\dots\}$. Assume that the ``good" event $G_t$ is true. Then we have that
    \begin{enumerate}
    \item[(a)] $ f_t(\vec{p}^*,\vec{x})  \geq f(\vec{p}^*,\vec{x}^*)$
    for every $\vec{x} \in \mathcal{J}$, where $f_t$ is defined in \eqref{eqn:def_f_t}, $\vec{p}^*$ is defined in \eqref{eqn:def_star_p_1} and $\vec{x}^*$ is defined in \eqref{eqn:def_star_x}.
    \item[(b)] Define
    \begin{align}\label{eqn:d_def}
         D_t = C + 2\sqrt{2\log{\frac{t(t+1)}{\delta_t}}},
    \end{align}
    where 
    \begin{align}\label{eqn:def_C}
        C = \max_{ k \in [1:n]} E_k.
    \end{align}
     We have that $\lVert \nabla_{\vec{p}}f_t(\vec{p},\vec{x})\lVert^2 \leq nD_t^2$ for every $\vec{p} \in \Delta_{n,r}$ and $\vec{x} \in \mathcal{J}$.
    \end{enumerate}
    \begin{proof} 
    We prove the two parts separately.
        \begin{enumerate}
            \item[(a)]We have that
            \begin{align}\label{eqn:sub_1}
                f_t(\vec{p}^*,\vec{x}) = \sum_{k=1}^n \frac{\tilde{E}_k(t)p^*_k}{1+x_k} \geq_{(a)} \sum_{k=1}^n \frac{E_kp^*_k}{1+x_k} = f(\vec{p}^*,\vec{x}) \geq f(\vec{p}^*,\vec{x}^*),
            \end{align}
            where (a) follows since we are in the ``good" event $G_t$ (so~\eqref{eqn:good_ev_1} holds) and the last inequality follows from the definition of $\vec{x}^*$ in \eqref{eqn:def_star_x}.
            \item[(b)] First, notice that when we are in the event $G_t$, we have from \eqref{eqn:good_ev_2} that,
            \begin{align}\label{eqn:grad_bound}
                 \tilde{E}_k(t) < E_k + 2\sqrt{\frac{2\log{\frac{n_k(t)(n_k(t)+1)}{\delta_t}}}{n_k(t)}} \leq_{(a)} C + 2\sqrt{2\log{\frac{t(t+1)}{\delta_t}}} = D_t,
            \end{align}
             for all $k \in [1:n]$, where (a) follows since $E_k \leq C$ by definition of $C$ in \eqref{eqn:def_C}, and $1 \leq n_k(t) \leq t$ for $t \in \{n+1,n+2,\dots\}$ by definition of $n_k(t)$. Hence,
            \begin{align}
                 \lVert \nabla_{\vec{p}}f_t(\vec{p},\vec{x})\lVert^2 = \sum_{k=1}^n \frac{\tilde{E}^2_k(t)}{(1+x_k)^2} \leq \sum_{k=1}^n \tilde{E}^2_k(t)  \leq nD_t^2.
            \end{align}
           
        \end{enumerate}
    \end{proof}
\end{lemma}
We summarize our approach in Algorithm~\ref{algo:2}. The algorithm relies on three key steps. 

First, we assume that given $\vec{p} \in \Delta_{n,r}$, we can sample a set $\mathcal{A} \subset [1:n]$ such that $|\mathcal{A}| = r$, and $\mathbb{E}\{\mathbbm{1}_{k \in \mathcal{A}}\} = p_k$ for all $k \in [1:n]$. This can be solved using the  Madow's sampling technique (\cite{pmlr-v151-mukhopadhyay22a}). In Appendix~\ref{app:madows}, we provide the algorithm for completeness. The correctness of the algorithm is established in~\cite{pmlr-v151-mukhopadhyay22a}.

Second, we assume we have an oracle that can compute a solution $\vec{y} \in \arg\min_{\vec{x} \in \mathcal{J}} \sum_{k=1}^n \frac{F_k p_k}{1+x_k}$,
where $F_i \geq 0$ for all $i \in [1:n]$, and $\vec{p} = [p_1,p_2,\dots,p_n] \in \Delta_{n,r}$. This problem is nonconvex due to the fact that $\mathcal{J}$ is a discrete set. Nevertheless, the problems of above type can be solved explicitly (we describe a simple method in Appendix~\ref{app:solving_selection}).

Finally, we assume that given $\vec{x} \in \mathbb{R}^{n}_+$, we can find the projection $ \Pi_{\Delta_{n,r}}(\vec{x})$ of $\vec{x}$ onto $\Delta_{n,r}$. An algorithm for this task is given in Appendix~\ref{alg:proj_I_r} along with the analysis.

For our algorithm, we also require step size parameters $\beta_t$ for $t \in \{n+1,n+2,\dots\}$ satisfying $\beta_t \geq \beta_{t+1}$ for all $t \in \{n+1,n+2,\dots\}$.
\begin{algorithm}
\label{algo:2}
\SetAlgoLined
\DontPrintSemicolon

\For{each time slot $t \in [1:n]$}{
   Set $\mathcal{A}_1(t)$ to be an arbitrary action set with $\mathcal{A}_1(t)\subset [1:n]$ satisfying $|\mathcal{A}_1(t)| = r$ and $t \in \mathcal{A}_1(t)$.\;
   Receive feedback $\{W_k(t); 1\leq k \leq n, k \in \mathcal{A}_1(t)\}$.\;
}
Initialize $\vec{p}(n+1) \in \Delta_{n,r}$.\;
\For{each time slot $t \in \{n+1,n+2,\dots,\}$}{
Sample an action set $\mathcal{A}_1(t) \subset [1:n]$ using the Madow's sampling technique such that $|\mathcal{A}_1(t)| = r$, and  $p_k(t) = \mathbb{E}\{\mathbbm{1}_{[k \in \mathcal{A}_1(t)]}|\vec{p}(t)\}$ for each $k \in [1:n]$. In particular, given $\vec{p}(t)$, we sample the above action set independent of the past $\mathcal{H}(t)$ (see Appendix~\ref{app:madows} for the implementation).\; 
Receive feedback $\{W_k(t); 1\leq k \leq n, k \in \mathcal{A}_1(t)\}$.\;
Find $\bar{E}_k(t)$, and $\tilde{E}_k(t)$ for each $k \in [1:n]$ using \eqref{eqn:bar_e} and \eqref{eqn:til_e}, respectively.\; 
Find $\vec{x}(t)$ by solving,
\begin{align}\label{eqn:intem_prob}
    \vec{x}(t) \in \arg\min_{\vec{x}\in \mathcal{J}} f_t(\vec{p}(t),\vec{x})
\end{align}
using Algorithm~\ref{algo:0}, where $f_t$ is defined in \eqref{eqn:def_f_t}.\;
Obtain $\vec{p}(t+1)$ by using,
\begin{align}\label{eqn:p_next}
    \vec{p}(t+1) = \Pi_{\Delta_{n,r}}\left(\vec{p}(t) + \beta_t\nabla_{\vec{p}} f_t(\vec{p}(t),\vec{x}(t)) \right),
\end{align}
where $\Pi_{\Delta_{n,r}}(\vec{y})$ denotes the projection of $\vec{y}$ onto $\Delta_{n,r}$ (See Appendix~\ref{alg:proj_I_r} for an algorithm) and $\beta_t$ is the step size parameter.
}
\caption{UCB based algorithm for worst-case maximization}
\end{algorithm}

\subsection{Analysis of the Algorithm}
In this section, we focus on establishing performance of Algorithm~\ref{algo:2}.
\begin{theorem}\label{lemma:UCBr1_regret_bound}
 Fix $T \in \{n+2,n+3,\dots\}$. 
 \begin{enumerate}
     \item[(a)]  Running the UCB based worst-case maximization algorithm in Algorithm~\ref{algo:2} for $T$ time slots with $\beta_t>0$ such that $\beta_t \geq \beta_{t+1}$ and $\delta_t \in (0,1)$ such that $\delta_t \geq \delta_{t+1}$ for all $t \in \{n+1,n+2,\dots\}$ yields
\begin{align}
    f^{\text{worst},*}-R^{\text{worst}}(T) &\leq   \frac{n}{2\beta_T T} + \frac{ n rC}{T}+  \frac{n D_{T}^2\sum_{t=n+1}^T\beta_t}{2T}   + 4\sqrt{\frac{2nr \log{\frac{T(T+1)}{\delta_T}}}{T}}\nonumber\\&\ \ \ \ +\frac{1}{T}\sum_{t=n+1}^T\left(2rC+\frac{n}{\beta_t}\right)n\delta_t,
\end{align}
where $R^{\text{worst}}(T)$ is the time-average worst case expected reward achieved by the algorithm (See \eqref{eqn:r_worst}), $C$ is defined in \eqref{eqn:def_C}, and $D_T$ is defined in \eqref{eqn:d_def}. Notice that the algorithm does not require the knowledge of $T$. Hence, the algorithm can be implemented in a setting where the time horizon is unknown.
\item [(b)] Running the UCB based worst-case maximization algorithm for $T$ time slots with  $\delta_t = \Theta(1/t)$ and $\beta_t = \Theta(1/\sqrt{t})$ for all $t \in \{n+1,n+2,\dots\}$, we have that
\begin{align}
     f^{\text{worst},*}-R^{\text{worst}}(T) \leq  \Theta\left(\frac{\log(T)}{\sqrt{T}}\right).
\end{align}
 \end{enumerate}
\begin{proof}
We will first prove part-(a). 

\noindent
(a) 
Fix any $t \in \{n+1,\dots,T\}$ and assume the ``good" event $G_t$ holds. Lemma~\ref{lemma:prel_lemma} implies $\lVert \nabla_{\vec{p}}f_t(\vec{p}(t),\vec{x}(t))\lVert^2 \leq nD_t^2 \leq nD_T^2$, where the first inequality follows from Lemma~\ref{lemma:prel_lemma}-(b) and the second inequality follows since $D_t \leq D_T$ for all $t \in \{n+1,\dots,T\}$ (see the definition of $D_t$ in \eqref{eqn:d_def} and use the fact that $\delta_t \geq \delta_T$ for all $t \in \{n+1,\dots,T\}$). Also, we have
\begin{align}\label{eqn:ftx_ft_xst}
    f_t(\vec{p}^*,\vec{x}(t)) \geq f(\vec{p}^*,\vec{x}^*) = f^{\text{worst},*},
\end{align}
where $\vec{x}(t)$ is defined in \eqref{eqn:intem_prob}, $f_t$ is defined in \eqref{eqn:def_f_t}, the first inequality follows from Lemma~\ref{lemma:prel_lemma}-(a) and the last equality follows from \eqref{eqn:f_worst_star_rem}. Define $\tilde{\vec{x}}(t) \in \arg\min_{\vec{x} \in \mathcal{J}} f(\vec{p}(t),\vec{x})$. Thus
\begin{align}\label{eqn:ftilx_f_worst}
    f(\vec{p}(t),\tilde{\vec{x}}(t) ) = f^{\text{worst}}(\vec{p}(t))
\end{align}
by the definition of $f^{\text{worst}}$ in \eqref{eqn:def_g}. Due to the definition of $\vec{x}(t)$ in \eqref{eqn:intem_prob}, we have
\begin{align}\label{eqn:ftx_fttilx}
    f_t(\vec{p}(t),\vec{x}(t)) \leq  f_t(\vec{p}(t),\tilde{\vec{x}}(t)).
\end{align}
Also, notice that
\begin{align}\label{eqn:first_chain_1}
    f_t(\vec{p}(t),\vec{\tilde{x}}(t)) &=_{(a)} \sum_{k=1}^n \frac{\tilde{E}_k(t)p_k(t)}{1+\tilde{x}_k(t)} = \sum_{k=1}^n \frac{E_kp_k(t)}{1+\tilde{x}_k(t)} + \sum_{k=1}^n \frac{[\tilde{E}_k(t)-E_k]p_k(t)}{1+\tilde{x}_k(t)}\nonumber\\&=_{(b)} f^{\text{worst}}(\vec{p}(t)) +\sum_{k=1}^n \frac{[\tilde{E}_k(t)-E_k]p_k(t)}{1+\tilde{x}_k(t)} \nonumber\\&\leq_{(c)}f^{\text{worst}}(\vec{p}(t)) + \sum_{k=1}^n\left(\frac{2p_k(t)}{1+\tilde{x}_k(t)}\sqrt{\frac{2\log{\frac{n_k(t)(n_k(t)+1)}{\delta_t}}}{n_k(t)}}\right)\nonumber\\&\leq_{(d)}f^{\text{worst}}(\vec{p}(t)) + 2\sum_{k=1}^n\left(p_k(t)\sqrt{\frac{2\log{\frac{T(T+1)}{\delta_T}}}{n_k(t)}}\right)
\end{align}
where (a) follows from the definition of $f_t$ in \eqref{eqn:def_f_t}; (b) follows from \eqref{eqn:ftilx_f_worst}; (c) follows since we assume the ``good" event $G_t$ holds (hence, the inequality \eqref{eqn:good_ev_2} is true); (d) follows since $n_k(t) \leq T$, $\delta_t \geq \delta_{T}$ for all $t \in \{n+1,\dots,T\}$, and $\tilde{x}_k(t) \geq 0$ for all $k \in [1:n]$. 
Since $\vec{p}(t+1)$ is defined in \eqref{eqn:p_next} as the projection of $\vec{p}(t) + \beta_t\nabla_{\vec{p}} f_t(\vec{p}(t),\vec{x}(t))$ onto the convex set $\Delta_{n,r}$, we have that,
\begin{align}\label{eqn:before_sum}
    &\lVert \vec{p}(t+1) - \vec{p}^*\rVert^2  \leq_{(a)}  \lVert\vec{p}(t) + \beta_t\nabla_{\vec{p}} f_t(\vec{p}(t),\vec{x}(t)) - \vec{p}^* \rVert^2 \nonumber\\& \leq \lVert \vec{p}(t) - \vec{p}^* \rVert^2 + \beta_t^2 \lVert \nabla_{\vec{p}}f_t(\vec{p}(t),\vec{x}(t))\lVert^2 - 2\beta_t(\vec{p}^*-\vec{p}(t))^{\top}\nabla_{\vec{p}}f_t(\vec{p}(t),\vec{x}(t)) \nonumber\\& =_{(b)}  \lVert \vec{p}(t) - \vec{p}^* \rVert^2 + \beta_t^2 \lVert \nabla_{\vec{p}}f_t(\vec{p}(t),\vec{x}(t))\lVert^2 - 2\beta_t(f_t(\vec{p}^*,\vec{x}(t))-f_t(\vec{p}(t),\vec{x}(t))) \nonumber\\& \leq_{(c)}  \lVert \vec{p}(t) - \vec{p}^* \rVert^2 + n\beta_t^2D_T^2 - 2\beta_tf^{\text{worst},*}+2\beta_tf_t(\vec{p}(t),\vec{\tilde{x}}(t))) \nonumber\\& \leq_{(d)}\lVert \vec{p}(t) - \vec{p}^* \rVert^2 + n\beta_t^2D_T^2 +4\beta_t \sum_{k=1}^n\left\{p_k(t)\sqrt{\frac{2\log{\frac{T(T+1)}{\delta_T}}}{n_k(t)}}\right\}\nonumber\\&\ \ \ \  - 2\beta_t f^{\text{worst},*}+2\beta_t f^{\text{worst}}(\vec{p}(t))
\end{align}
where (a) follows since projection onto the convex set $\Delta_{n,r}$ reduces the distance to any point in the set, (b) follows from the subgradient equality for the linear function $f_t(\cdot,\vec{x}(t))$, (c) follows from~\eqref{eqn:ftx_ft_xst} and \eqref{eqn:ftx_fttilx}, and (d) follows from \eqref{eqn:first_chain_1}.

Hence, we have that for all $t \in \{n+1,\dots,T\}$, given that the ``good" event $G_t$ is true
\begin{align}\label{eqn:before expec}
    &2 f^{\text{worst},*}-2 f^{\text{worst}}(\vec{p}(t)) -\frac{1}{\beta_t} \lVert \vec{p}(t) - \vec{p}^* \rVert^2 + \frac{1}{\beta_t}\lVert \vec{p}(t+1) - \vec{p}^*\rVert^2\nonumber\\&  \leq n\beta_t D_T^2+ 4 \sum_{k=1}^n\left\{p_k(t)\sqrt{\frac{2\log{\frac{T(T+1)}{\delta_T}}}{n_k(t)}}\right\}
\end{align}

Notice that
\begin{align}\label{eqn:chain_44}
    p_k(t) &=_{(a)} \mathbb{E}\{\mathbbm{1}_{[k \in \mathcal{A}_1(t)]}|\vec{p}(t)\} =_{(b)} \mathbb{E}\{\mathbbm{1}_{[k \in \mathcal{A}_1(t)]}|\vec{p}(t),\mathcal{H}(t)\} \nonumber\\&=_{(c)} \mathbb{E}\{\mathbbm{1}_{[k \in \mathcal{A}_1(t)]}|\mathcal{H}(t)\},
\end{align}
where (a) follows due to the sampling of the set $\mathcal{A}_1(t)$ in line 7 of Algorithm~\ref{algo:2}, (b) follows because the action set $\mathcal{A}_1(t)$ is sampled independent of the history $\mathcal{H}(t)$ given $\vec{p}(t)$ (see line 7 of Algorithm~\ref{algo:2}), and (c) follows since $\vec{p}(t)$ is $\mathcal{H}(t)$-measurable.

Now we take the expectation (Conditioned on the event $G_t$) of both sides of \eqref{eqn:before expec} which gives,
\begin{align}\label{eqn:first_chain}
    &\mathbb{E}\left\{2 f^{\text{worst},*}-2 f^{\text{worst}}(\vec{p}(t)) - \frac{1}{\beta_t}\lVert \vec{p}(t) - \vec{p}^* \rVert^2 + \frac{1}{\beta_t}\lVert \vec{p}(t+1) - \vec{p}^*\rVert^2\Bigg{|}G_t\right\} \nonumber\\&\leq  n\beta_t D_T^2+4 \mathbb{E}\left\{\sum_{k=1}^np_k(t)\sqrt{\frac{2\log{\frac{T(T+1)}{\delta_T}}}{n_k(t)}} \Bigg{|} G_t\right\} \nonumber\\&\leq n\beta_tD_T^2+\frac{4}{\mathbb{P}\{G_t\}}\mathbb{E}\left\{\sum_{k=1}^np_k(t)\sqrt{\frac{2\log{\frac{T(T+1)}{\delta_T}}}{n_k(t)}}\right\} \nonumber\\&=_{(a)}n\beta_t D_T^2+\frac{4}{\mathbb{P}\{G_t\}}\mathbb{E}\left\{\sum_{k=1}^n\mathbb{E}\{\mathbbm{1}_{[k \in \mathcal{A}_1(t)]}|\mathcal{H}(t)\}\sqrt{\frac{2\log{\frac{T(T+1)}{\delta_T}}}{n_k(t)}}\right\} \nonumber\\&=_{(b)}n\beta_tD_T^2+\frac{4}{\mathbb{P}\{G_t\}}\mathbb{E}\left\{\mathbb{E}\left\{\sum_{k=1}^n\mathbbm{1}_{[k \in \mathcal{A}_1(t)]}\sqrt{\frac{2\log{\frac{T(T+1)}{\delta_T}}}{n_k(t)}}\Bigg{|}\mathcal{H}(t)\right\}\right\} \nonumber\\&=n\beta_tD_T^2+\frac{4}{\mathbb{P}\{G_t\}}\mathbb{E}\left\{\mathbb{E}\left\{\sum_{{j:j \in \mathcal{A}_1(t)}}\sqrt{\frac{2\log{\frac{T(T+1)}{\delta_T}}}{n_{j}(t) }}\Bigg{|}\mathcal{H}(t)\right\}\right\} \nonumber\\&=n\beta_tD_T^2+\frac{4}{\mathbb{P}\{G_t\}}\mathbb{E}\left\{\sum_{j:j \in \mathcal{A}_1(t)}\sqrt{\frac{2\log{\frac{T(T+1)}{\delta_T}}}{n_{j}(t) }}\right\}
\end{align}
where (a) follows from \eqref{eqn:chain_44} and (b) follows since $n_k(t)$ is $\mathcal{H}(t)$-measurable. Hence, we have that for $t \in \{n+1,\dots,T\}$
\begin{align}\label{eqn:chain_2}
    &\mathbb{E}\left\{2f^{\text{worst},*}-2f^{\text{worst}}(\vec{p}(t))|G_t\right\} \leq \frac{1}{\beta_t}\mathbb{E}\{\lVert \vec{p}(t) - \vec{p}^*\rVert^2|G_t\} - \frac{1}{\beta_t}\mathbb{E}\{\lVert \vec{p}(t+1) - \vec{p}^*\rVert^2|G_t\} \nonumber\\&\ \ \ + n\beta_tD_T^2+\frac{4}{\mathbb{P}\{G_t\}}\mathbb{E}\left\{\sum_{j:j \in \mathcal{A}_1(t)}\sqrt{\frac{2\log{\frac{T(T+1)}{\delta_T}}}{n_{j}(t) }}\right\}.
\end{align}
Now, notice that
\begin{align}\label{eqn:chain_3}
    \mathbb{E}\{\lVert \vec{p}(t+1) - \vec{p}^*\rVert^2|G_t\}\mathbb{P}\{G_t\} &= \mathbb{E}\{\lVert \vec{p}(t+1) - \vec{p}^*\rVert^2\} - \mathbb{E}\{\lVert \vec{p}(t+1) - \vec{p}^*\rVert^2|G_t^c\}\mathbb{P}\{G_t^c\}  \nonumber\\&\geq  \mathbb{E}\{\lVert \vec{p}(t+1) - \vec{p}^*\rVert^2\} - n\mathbb{P}\{G_t^c\},
\end{align}
where the last inequality follows from $\lVert \vec{p}(t+1) - \vec{p}^*\rVert^2 \leq n$ (since $\vec{p}(t+1),\vec{p}^* \in \Delta_{n,r}$). Next, notice that
\begin{align}\label{eqn:f_worst_bound}
    f^{\text{worst},*} = \sum_{k=1}^n \frac{p^*_kE_k}{1+x^*_k} \leq \sum_{k=1}^n p^*_kC = rC,
\end{align}
where the first equality follows from \eqref{eqn:f_worst_star_rem}, the inequality follows from the definition of $C$ in \eqref{eqn:def_C} and the fact that $x^*_k \geq 0$ for all $k \in [1:n]$, and the last equality follows since $\vec{p}^* \in \Delta_{n,r}$ (see \eqref{eqn:def_star_p_1}). Hence,
\begin{align}\label{eqn:comp_evet}
    \mathbb{E}\left\{2f^{\text{worst},*}-2f^{\text{worst}}(\vec{p}(t))|G_t^c\right\} \leq 2f^{\text{worst},*} \leq 2rC,
\end{align}
where the last inequality follows from \eqref{eqn:f_worst_bound}. Notice that,
\begin{align}\label{eqn:11111}
    &\mathbb{E}\{2 f^{\text{worst},*}-2 f^{\text{worst}}(\vec{p}(t))\} \nonumber\\&= \mathbb{E}\{2 f^{\text{worst},*}-2 f^{\text{worst}}(\vec{p}(t))|G_t\}\mathbb{P}\{G_t\}+ \mathbb{E}\{2 f^{\text{worst},*}-2 f^{\text{worst}}(\vec{p}(t))|G_t^c\}\mathbb{P}\{G_t^c\} \nonumber\\& \leq_{(a)} \frac{1}{\beta_t}\mathbb{E}\{\lVert \vec{p}(t) - \vec{p}^* \rVert^2|G_t\}\mathbb{P}\{G_t\} - \frac{1}{\beta_t}\mathbb{E}\{\lVert \vec{p}(t+1) - \vec{p}^*\rVert^2|G_t\}\mathbb{P}\{G_t\}  +n\beta_t D_T^2\mathbb{P}\{G_t\}\nonumber\\&\ \   + 4\mathbb{E}\left\{\sum_{j:j \in \mathcal{A}_1(t)}\sqrt{\frac{2\log{\frac{T(T+1)}{\delta_T}}}{n_{j}(t) }}\right\} +2rC\mathbb{P}\{G_t^c\}\nonumber\\& \leq_{(b)} \frac{1}{\beta_t}\mathbb{E}\{\lVert \vec{p}(t) - \vec{p}^* \rVert^2\} - \frac{1}{\beta_t}\mathbb{E}\{\lVert \vec{p}(t+1) - \vec{p}^*\rVert^2\} + \frac{n}{\beta_t}\mathbb{P}\{G_t^c\}+ n\beta_tD_T^2 \nonumber\\& \ \ + 4\mathbb{E}\left\{\sum_{j:j \in \mathcal{A}_1(t)}\sqrt{\frac{2\log{\frac{T(T+1)}{\delta_T}}}{n_{j}(t) }}\right\}  + 2rC\mathbb{P}\{G_t^c\}\nonumber\\&\leq_{(c)} \frac{1}{\beta_t}\mathbb{E}\{\lVert \vec{p}(t) - \vec{p}^* \rVert^2\} - \frac{1}{\beta_t}\mathbb{E}\{\lVert \vec{p}(t+1) - \vec{p}^*\rVert^2\} + n\beta_tD_T^2 \nonumber\\& \ \  + 4\mathbb{E}\left\{\sum_{j:j \in \mathcal{A}_1(t)}\sqrt{\frac{2\log{\frac{T(T+1)}{\delta_T}}}{n_{j}(t) }}\right\} +2\left(2rC+\frac{n}{\beta_t}\right)n\delta_t ,
\end{align}
where (a) follows from \eqref{eqn:chain_2} and \eqref{eqn:comp_evet}, (b) follows since $\mathbb{E}\{X|Y\}\mathbb{P}\{Y\} \leq \mathbb{E}\{X\}$ for a positive valued random variable $X$ and \eqref{eqn:chain_3}, and (c) follows from \eqref{eqn:complment_ineq}. Now, we sum \eqref{eqn:11111} for $t \in \{n+1,\dots,T\}$ to get
\begin{align}\label{eqn:from_n}
    &\mathbb{E}\left\{2 (T-n) f^{\text{worst},*}-2\sum_{t=n+1}^{T} f^{\text{worst}}(\vec{p}(t))\right\}\nonumber\\&\leq \frac{\mathbb{E}\{\lVert \vec{p}(n+1) - \vec{p}^* \rVert^2\}}{\beta_{n+1}}+ \sum_{t=n+2}^T\left[\frac{1}{\beta_t} - \frac{1}{\beta_{t-1}}\right]\mathbb{E}\{\lVert \vec{p}(t) - \vec{p}^*\rVert^2\} -\frac{\mathbb{E}\{\lVert \vec{p}(T+1) - \vec{p}^* \rVert^2\}}{\beta_{T+1}} \nonumber\\&\ \ + nD_T^2\sum_{t=n+1}^{T}\beta_t+ 4\sum_{t=n+1}^{T}\mathbb{E}\left\{\sum_{j:j \in \mathcal{A}_1(t)}\sqrt{\frac{2\log{\frac{T(T+1)}{\delta_T}}}{n_{j}(t) }}\right\}   + \sum_{t=n+1}^T2\left(2rC+\frac{n}{\beta_t}\right)n\delta_t \nonumber\\&\leq_{(a)} \frac{n}{\beta_{n+1}}+ \sum_{t=n+2}^Tn\left[\frac{1}{\beta_t} - \frac{1}{\beta_{t-1}}\right] + nD_T^2\sum_{t=n+1}^{T}\beta_t+ 4\sum_{t=n+1}^{T}\mathbb{E}\left\{\sum_{j:j \in \mathcal{A}_1(t)}\sqrt{\frac{2\log{\frac{T(T+1)}{\delta_T}}}{n_{j}(t) }}\right\} \nonumber\\&\ \   + \sum_{t=n+1}^T2\left(2rC+\frac{n}{\beta_t}\right)n\delta_t\nonumber\\&= \frac{n}{\beta_{T}} +  nD_{T}^2\sum_{t=n+1}^T\beta_t  + 4\sum_{t=n+1}^{T}\mathbb{E}\left\{\sum_{j:j \in \mathcal{A}_1(t)}\sqrt{\frac{2\log{\frac{T(T+1)}{\delta_T}}}{n_{j}(t) }}\right\}  +\sum_{t=n+1}^T2\left(2rC+\frac{n}{\beta_t}\right)n\delta_t 
\end{align}
where (a) follows since $1/\beta_t - 1/\beta_{t-1} \geq 0$ for all $t \in \{n+2,\dots,T\}$ and $\lVert \vec{p}(t) - \vec{p}^* \rVert^2 \leq n$ for all $t \in \{n+1,\dots,T\}$ (since $\vec{p}(t),\vec{p}^* \in \Delta_{n,r}$).
Now, notice that
\begin{align}
  \sum_{t=n+1}^{T}&\mathbb{E}\left\{\sum_{j:j \in \mathcal{A}_1(t)}\sqrt{\frac{2\log{\frac{T(T+1)}{\delta_T}}}{n_{j}(t) }}\right\}= \mathbb{E}\left\{\sum_{k=1}^n\sum_{\substack{t=n+1\\ k:k \in \mathcal{A}_1(t)}}^{T}\sqrt{\frac{2\log{\frac{T(T+1)}{\delta_T}}}{n_{k}(t) }}\right\} \nonumber\\&= \mathbb{E}\left\{\sum_{k=1}^n\sum_{j=n_k(n+1)}^{n_k(T)}\sqrt{\frac{2\log{\frac{T(T+1)}{\delta_T}}}{j}}\right\}\leq \mathbb{E}\left\{\sum_{k=1}^n\sum_{j=1}^{n_k(T)}\sqrt{\frac{2\log{\frac{T(T+1)}{\delta_T}}}{j}}\right\}\nonumber\\&\leq_{(a)}2\mathbb{E}\left\{\sum_{k=1}^n\sqrt{2n_k(T) \log{\frac{T(T+1)}{\delta_T}}}\right\} \leq_{(b)}2\sqrt{2n \log{\frac{T(T+1)}{\delta_T}}}\sqrt{ \sum_{k=1}^n n_k(T) }\nonumber\\& \leq_{(c)} 2\sqrt{2nrT \log{\frac{T(T+1)}{\delta_T}}},
\end{align}
where (a) follows from $\sum_{j=1}^l \sqrt{j}^{-1} \leq 2\sqrt{l}$, (b) follows since $\sum_{k=1}^n \sqrt{n_k(T)} \leq \sqrt{n\sum_{k=1}^n n_k(T)}$, and (c) follows since  $\sum_{k=1}^n n_k(T) = r(T-1) \leq rT$. Substituting above in \eqref{eqn:from_n}, we have that
\begin{align}\label{eqn:from_n_1}
    &\mathbb{E}\left\{2 (T-n) f^{\text{worst},*}-2\sum_{t=n+1}^{T} f^{\text{worst}}(\vec{p}(t))\right\}\nonumber\\&\leq \frac{n}{\beta_{T}} +  nD_{T}^2\sum_{t=n+1}^T\beta_t  + 8\sqrt{2nrT \log{\frac{T(T+1)}{\delta_T}}}  +\sum_{t=n+1}^T2\left(2rC+\frac{n}{\beta_t}\right)n\delta_t 
\end{align}

For $t \in [1:n]$, define $\vec{p}(t)$ as $p_k(t) = \mathbbm{1}_{[k \in \mathcal{A}_1(t)]}$. This definition is consistent with the definition of $\vec{p}(t)$ for $t \in \{n+1,\dots\}$, since $\mathcal{A}_1(t)$ is deterministic for $t \in [1:n]$ (see lines 1-4 of Algorithm~\ref{algo:2}). Hence,
\begin{align}\label{eqn:upto_n}
    &\mathbb{E}\left\{2 n f^{\text{worst},*}-2\sum_{t=1}^{n} f^{\text{worst}}(\vec{p}(t))\right\} \leq 2 n f^{\text{worst},*} \leq 2 n rC,
\end{align}
where the last inequality follows from \eqref{eqn:f_worst_bound}. Adding \eqref{eqn:from_n_1} and \eqref{eqn:upto_n}, and dividing by $2 T $, we have that
\begin{align}\label{eqn:up_bound}
   &\mathbb{E}\left\{ f^{\text{worst},*}-\frac{1}{T}\sum_{t = 1}^{T} f^{\text{worst}}(\vec{p}(t))\right\} \leq \frac{n}{2\beta_T T} + \frac{ n rC}{T}+  \frac{n D_{T}^2\sum_{t=n+1}^T\beta_t}{2T}  \nonumber\\&\ \ \ \ + 4\sqrt{\frac{2nr \log{\frac{T(T+1)}{\delta_T}}}{T}} +\frac{1}{T}\sum_{t=n+1}^T\left(2rC+\frac{n}{\beta_t}\right)n\delta_t.
\end{align}
 Using the definition of $R^{\text{worst}}(T)$ defined in \eqref{eqn:r_worst} in the above, we are done.

\noindent
(b) To prove the (b), consider $\delta_t = \Theta(1/t)$ and $\beta_t = \Theta(1/\sqrt{t})$ for all $t \in \{n+1,n+2,\dots\}$. We analyze each term in the right hand side of the bound obtained in part-(a).  Notice that $\frac{n}{2\beta_T T} $ is $\Theta(1/\sqrt{T})$, $\frac{ n rC}{T}$ is $\Theta(1/T)$, and  $\sqrt{\frac{2nr \log{\frac{T(T+1)}{\delta_T}}}{T}}$ is $\Theta(\sqrt{\log(T)/T})$.  We will analyze the remaining two terms separately. For simplicity, we will use $\delta_t = 1/t$ and $\beta_t = 1/\sqrt{t}$. First,
\begin{align}
    \frac{n D_{T}^2\sum_{t=n+1}^T\beta_t}{2T} &=_{(a)} \frac{n}{2T} \left(C + 2\sqrt{2\log T^2(T+1)}\right)^2\sum_{t=n+1}^T \frac{1}{\sqrt{t}} \nonumber\\&\leq_{(b)} \frac{n}{\sqrt{T}} \left(C + 2\sqrt{2\log T^2(T+1)}\right)^2 = \Theta\left(\frac{\log(T)}{\sqrt{T}}\right),
\end{align}
where (a) follows from the definition of $D_T$ in \eqref{eqn:d_def} and for (b) we have used $\sum_{k=1}^l 1/\sqrt{k} \leq 2 \sqrt{l}$.
Next,
\begin{align}
    \frac{1}{T}\sum_{t=n+1}^T\left(2rC+\frac{n}{\beta_t}\right)n\delta_t &= \frac{1}{T}\sum_{t=n+1}^T\left(\frac{2rnC}{t}+\frac{n^2}{\sqrt{t}}\right) \leq \frac{1}{T}\sum_{t=1}^T\left(\frac{2rnC+n^2}{\sqrt{t}}\right)\nonumber\\&\leq_{(a)} \frac{4rnC+2n^2}{\sqrt{T}}=  \Theta\left(\frac{1}{\sqrt{T}}\right),
\end{align}
where for (a) we have used $\sum_{k=1}^l 1/\sqrt{k} \leq 2 \sqrt{l}$. Combining the terms, we are done.
\end{proof}
\end{theorem}
\section{Finding $f^{\text{worst},*}$ with known $\vec{E}$}\label{sec:singlton}
If $\vec{E}$ is known, given $\vec{p} \in \Delta_{n,r}$, the problem of finding $f^{\text{worst}}(\vec{p})$ has been well studied in the literature. In particular, we can find $\vec{x}^* \in \arg\min_{\vec{x} \in \mathcal{J}}f(\vec{p},\vec{x})$. In Appendix~\ref{app:solving_selection}, we provide the algorithm for completeness. Hence, we can use standard min-max optimization techniques such as min-oracle algorithm~\cite{pmlr-v119-jin20e} to find $f^{\text{worst},*}$ and $\vec{p}^*$. Also, since $\mathcal{J}$ is a finite set, and the function $f$ is concave in the first argument, from the Danskin's theorem~\cite{Bertsekas_99}, we can easily calculate a subgradient of $f^{\text{worst}}$ at $\vec{p} \in \Delta_{n,r}$ as
\begin{align}
    \nabla_{\vec{p}}f^{\text{worst}}(\vec{p}) = \nabla_{\vec{p}}f(\vec{p},\vec{x}^*),
\end{align}
where $\vec{x}^* \in \arg\min_{\vec{x} \in \mathcal{J}}f(\vec{p},\vec{x})$. Hence, we can also use standard subgradient descent with Euclidean projections onto $\Delta_{n,r}$ (see Algorithm~\ref{alg:proj_I_r} to project onto $\Delta_{n,r}$) to find $f^{\text{worst},*}$.

The work of \cite{wijewardena2023twoplayer_1} finds $f^{\text{worst},*}$ and $\vec{p}^*$ explicitly for the case $m=2,r=1$. We discuss the solution of this case in Section~\ref{subsec:m2r1}. In Section~\ref{subsec:solving_a0b0}, we extend this to the case $m=3,r=1$. These explicit solutions provide a fast way to find $f^{\text{worst},*}$ and provide insight into the structure of optimal $\vec{p}^*$. We also, provide a partial solution for the case $m=2$ with general $n , r$ in Section~\ref{subsec:solving_a0b0_non_singleton}. In this section, we assume, without loss of generality, that $E_k > 0$ for all $1 \leq k \leq n$ since otherwise, we can transform the problem into a lower dimensional version. Without loss of generality, we also assume that $E_k \geq E_{k+1}$ for $1 \leq k \leq n-1$. Before moving on to the cases, we state the well-known Lagrange multiplier lemma, which will be useful in constructing the solution for both cases.
We first state a lemma that is useful for the proof.
\begin{lemma}\label{lemma:lagrange_lemma}
     Consider the following constrained optimization problem,
     \begin{maxi}
  {\vec{x}}{z_0(\vec{x})}{}{}
     \addConstraint{z_i(\vec{x})}{\geq 0 \ \text{ for } i \in \{1,2,\dots,k\}}  
     \addConstraint{\vec{x}}{\in \mathcal{Y}},
\end{maxi}
where $z_i:\mathbb{R}^n \to \mathbb{R}$ for $i \in \{0,1,2,\dots,k\}$, and $\mathcal{Y} \subset \mathbb{R}^n$. Consider the following unconstrained problem for some $\vec{\mu} \geq 0$.
\begin{maxi}
  {\vec{x}}{z_0(\vec{x})+\sum_{i=1}^k\mu_iz_i(\vec{x})}{}{} 
  \addConstraint{\vec{x}}{\in \mathcal{Y}}.
\end{maxi}
Let $\vec{x}^*$ be a solution to the unconstrained problem. Assume $\vec{x}^*$ satisfies for all $i \in \{1,2,\dots,k\}$,
\begin{enumerate}
    \item[(a)] $z_i(\vec{x}^*) \geq 0$ (That is $\vec{x}^*$ is feasible for the constrained problem)
    \item[(b)] $\mu_i > 0$ implies $z_i(\vec{x}^*) = 0$. 
\end{enumerate}
Then $\vec{x}^*$ is optimal for the constrained problem.
\begin{proof}
    The proof of the lemma is immediate and omitted for brevity.
\end{proof}
\end{lemma}
\subsection{$r=1$, $m=2$}\label{subsec:m2r1}
For this section, we use the notation $\Delta_n$ for $\Delta_{n,1}$.
This is solved in~\cite{wijewardena2023twoplayer}. The solution is given by,
$\vec{p}^*$ where,
\begin{align}\label{eqn:wrst_case_plcy}
    p^*_k = \begin{cases}
        \frac{1}{E_k\left(\sum_{j=1}^{v} \frac{1}{E_j}\right)} & \text{ if } k \leq v,\\
        0 & \text{ otherwise, }
    \end{cases}
\end{align}
and,
\begin{align}\label{eqn:v_def__}
 v = \arg\max_{1 \leq k \leq n} \frac{k - \frac{1}{2}}{\sum_{j=1}^k \frac{1}{E_j}},
\end{align}
See~\cite{wijewardena2023twoplayer} for the proof.

Notice that $\mathcal{J}$ is the set of standard unit vectors. Assume player 1 follows the worst-case policy given by $\vec{p}^*$ in \eqref{eqn:wrst_case_plcy}. It can be shown that $\vec{x}^*$ given by $x^*_1 = 1$ and $x^*_i = 0$ for all $i \in [2:n]$ satisfies $\vec{x}^* \in \arg\min_{\vec{x} \in \mathcal{J}} f(\vec{p}^*,\vec{x})$. Lemma~\ref{lemma:kind_of_nash_} shows the $\vec{x}^*$ is also the best response of Player 2 for Player 1. Hence, player 2 has no incentive to deviate from the policy $\vec{x}^*$, given that player 1 uses $\vec{p}^*$. Notice that this may not be a Nash equilibrium since $\vec{p}^*$ may not be the best response of player 1 to $\vec{x}^*$. However, this property incentivizes player 2 to use $\vec{x}^*$ as the strategy, even if they do not care about hurting player 1. For general $m,r$, given that player 1 is using strategy $\vec{p}^*$, the above raises the question of whether the strategies of players $[2:m]$ that gives the worst-case to $\vec{p}^*$ are also the best responses to the other players. It turns out this is not true in general. Particularly, when $m >2$, players $[2:m]$ will focus on increasing the congestion of resources with high mean rewards to reduce the expected reward of player 1. In such a scenario, a player in $[2:m]$ may increase their reward by switching to a resource with a less mean reward and less congestion. However, there are special cases in which the above property is true even when $m>2$. One such example is discussed in the case $m = 3, r=1$.

\begin{lemma}\label{lemma:kind_of_nash_}
Consider the special case $m=2,r=1$ with $n \in \mathbb{N}$ satisfying $n \geq 2$. Without loss of generality, assume that $\vec{E}$ satisfies $E_k \geq E_{k+1}$ for all $k \in [1:n-1]$. Let $\vec{p}^* \in \Delta_{n}$ denote the worst-case strategy of player 1 defined in \eqref{eqn:wrst_case_plcy}. Then $\vec{x}^*$ defined by,
\begin{align}
    x^*_i = \begin{cases}
        1 & \text{ if } i = 1\\
        0 & \text{ otherwise}.
    \end{cases}
\end{align}
satisfies $\vec{x}^* \in \arg\min_{\vec{x} \in \mathcal{J}} f(\vec{p}^*,\vec{x})$. Hence, given that player 1 uses strategy $\vec{p}^*$, player 2 can use $\vec{x}^*$ to enforce the worst-case for player 1. Additionally the strategy profile $(\vec{p}^*,\vec{x}^*)$ has the property that player 2 cannot increase their reward by unilaterally switching to a different strategy $\vec{q} \in \Delta_{n}$.
\begin{proof}
Define $C = \frac{1}{\sum_{j=1}^v \frac{1}{E_j}}$,
where $v$ is defined in~\eqref{eqn:v_def__}.  It can be easily shown that the worst-case for player 1 occurs when player 2 chooses resource 1 with probability 1 (player 2 uses strategy $\vec{x}^*$). The expected reward of player 2 under this profile is $\frac{p^*_1 E_1}{2} + (1-p^*_1)E_1 = E_1 - \frac{C}{2}$.

Consider the scenario where player 2 switches to a different policy $\vec{q} \in \Delta_{n}$. Under the profile $(\vec{p}^*,\vec{q}$), the expected reward of player 2 is
\begin{align}
   h(\vec{q}) &=  \sum_{j=1}^v q_j \left(\frac{p^*_j E_j}{2} + (1-p^*_j)E_j\right)+ \sum_{j=v+1}^n q_j E_j \nonumber \\&= \sum_{j=1}^v q_j \left(E_j -\frac{C}{2}\right)+ \sum_{j=v+1}^n q_j E_j.
\end{align}
We establish that $h(\vec{q}) \leq h(\vec{x}^*)$ for all $\vec{q} \in \Delta_{n}$. 
% Although the above is not true for general $m,r$, it should be noted that the policies of players $[2:m]$ that gives the worst-case for player 1 also 

Throughout the proof, let us call the property $E_k \geq E_{k+1}$ for $k \in [1:n-1]$ to be the nondecreasing property. We consider two cases. First, if $v = n$, notice that
\begin{align}
   h(\vec{q}) = \sum_{j=1}^n q_j \left(E_j -\frac{C}{2}\right) \leq \sum_{j=1}^n q_j \left(E_1 -\frac{C}{2}\right) = \left(E_1 -\frac{C}{2}\right) = h(\vec{x}^*),
\end{align}
where the inequality follows from the nondecreasing property.

Hence, we are done. Now, consider the case $v < n$. Hence, from the definition of $v$ in \eqref{eqn:v_def__}, we have $\frac{v-\frac{1}{2}}{\sum_{k=1}^{v}\frac{1}{E_k}} \geq \frac{v+\frac{1}{2}}{\sum_{k=1}^{v+1}\frac{1}{E_k}}$. This inequality simplifies to
\begin{align}\label{eqn:1010102}
    \frac{\left(v - \frac{1}{2}\right)}{E_{v+1}} \geq \sum_{k=1}^{v}\frac{1}{E_k} \geq \frac{v}{E_1}.
\end{align}
where the last inequality follows from the nondecreasing property.
We also have
\begin{align}\label{eqn:1010103}
    C = \frac{1}{\sum_{k=1}^{v}\frac{1}{E_k}}  \leq \frac{E_1}{v}.
\end{align}
where the inequality follows from the nondecreasing property.
Hence, 
\begin{align}
   h(\vec{q}) &=  \sum_{j=1}^v q_j \left(E_j -\frac{C}{2}\right)+ \sum_{j=v+1}^n q_j E_j \leq \left(\sum_{j=1}^v q_j\right)\left(E_1 -\frac{C}{2}\right)+\left(\sum_{j=v+1}^n q_j\right)E_{v+1} \nonumber\\&= \left(\sum_{j=1}^v q_j\right)\left(E_1 -\frac{C}{2}\right)+\left(\sum_{j=v+1}^n q_j\right)\left(E_{v+1} - \frac{C}{2} + \frac{C}{2}\right)\nonumber\\&\leq_{(a)} \left(\sum_{j=1}^v q_j\right)\left(E_1 -\frac{C}{2}\right)+\left(\sum_{j=v+1}^n q_j\right)\left(\frac{E_1\left(v - \frac{1}{2}\right)}{v}  - \frac{C}{2} + \frac{E_1}{2v}\right) = h(\vec{x}^*), \nonumber
\end{align}
where (a) follows due to \eqref{eqn:1010102} and \eqref{eqn:1010103}. 
\end{proof}
\end{lemma}
\subsection{$r=1$, $m=3$}\label{subsec:solving_a0b0} 
This section finds $\vec{p}^*$ for the case $m = 3,r=1$ where $n$ is a positive integer and $\vec{E} = [E_1,\dots,E_n]$ is known. Define
\begin{align}
    \vec{p}^*  \in \arg\min_{\vec{p} \in \Delta_{n}} f^{\text{worst}}(\vec{p}),
\end{align}
\begin{align}
    f^{\text{worst}}(\vec{p}) = \min_{\vec{x} \in \mathcal{J}}f(\vec{p},\vec{x}),
\end{align}
and
\begin{align}
    f(\vec{p},\vec{x}) = \sum_{k =1}^n \frac{p_kE_k}{1+x_k}.
\end{align}
 Notice that in this case we have
\begin{align}
\mathcal{J} = \left\{\vec{x} \in \{0,1,2\}^n \Bigg{|} \sum_{j=1}^n x_j = 2 \right\}. 
\end{align}
We will use the notation $\Delta_n$ for $\Delta_{n,1}$.
 We first focus on solving the problem $\min_{\vec{x} \in \mathcal{J}} f(\vec{p},\vec{x})$ for given $\vec{p} \in \Delta_{n}$. 
\begin{lemma}\label{lemma:worst_case}
Consider the special case $m=3,r=1$ with $n \in \mathbb{N}$ satisfying $n \geq 2$, and a fixed $\vec{p} \in \Delta_{n}$. Let $a = \arg\max_{1 \leq i \leq n} E_ip_i$, and $b = \arg\max_{1 \leq i \leq n, i \neq a} E_ip_i$, where we assume that $\arg\max$ returns the least index in the case of ties. Define the two vectors $\vec{x}^1,\vec{x}^2 \in \mathcal{J}$, where
\begin{align}\label{eqn:first_assign}
x^1_k = \begin{cases}
    2 & \text{ if } k = a,\\
    0 &  \text{ otherwise, }
\end{cases}
\end{align}
and
\begin{align}\label{eqn:second_assign}
x^2_k = \begin{cases}
    1 & \text{ if } k \in \{a,b\},\\
    0 &  \text{ otherwise. }
\end{cases}
\end{align}
Then $\vec{x}^* \in \arg\min_{\vec{x} \in \mathcal{J}}$ $f(\vec{p},\vec{x})$ can be given in two cases.

\noindent
\textbf{Case 1:} $E_a p_a \geq 3E_bp_b$: We have $\vec{x}^* = \vec{x}^1$.

\noindent
\textbf{Case 2:} $E_a p_a < 3E_bp_b$: We have $\vec{x}^* = \vec{x}^2$.

\begin{proof}
Since $\vec{x}^* \in \mathcal{J}$, we know $\vec{x}^*$ has nonnegative components that sum to $2$. If $\vec{x}^*$ has only one nonzero component at some index $k \in \{1,\dots,n\}$, then $x^*_k = 2$ and $f$ is minimized by choosing $k = a$, so assignment \eqref{eqn:first_assign} holds.

Else, $\vec{x}^*$ has exactly two nonzero components at indices $k,j \in \{1,2,\dots,n\}$ ($k\neq j$) and $f$ is minimized by choosing $k = a$ and $j = b$, so assignment \eqref{eqn:second_assign} holds. It remains to compare \eqref{eqn:first_assign} and \eqref{eqn:second_assign}.
\\
\noindent
Under \eqref{eqn:first_assign}:
\begin{align}
    f(\vec{p},\vec{x}) = \frac{p_aE_a}{3} + p_bE_b + \sum_{k \not \in \{a,b\}} p_kE_k = \sum_{k=1}^n p_kE_k - \frac{2p_aE_a}{3}.
\end{align}\\
Under \eqref{eqn:second_assign}: 
\begin{align}
    f(\vec{p},\vec{x}) = \frac{p_aE_a}{2} + \frac{p_bE_b}{2} + \sum_{k \not \in \{a,b\}} p_kE_k = \sum_{k=1}^n p_kE_k - \frac{p_aE_a}{2}- \frac{p_bE_b}{2}
\end{align}
Comparing the two cases, we have that for assignment~\eqref{eqn:first_assign}, we require $E_ap_a \geq 3E_bp_b$ and for assignment~\eqref{eqn:second_assign}, we require $E_ap_a < 3E_bp_b$. Hence, we are done.
\end{proof}
\end{lemma}
Now state the solution of $m =3 , r = 1$ as a Theorem, after which we move on to the proof.
\begin{theorem}\label{theorem:a0b0explicit}
Consider the special case $m=3,r=1$ with $n \in \mathbb{N}$ satisfying $n \geq 2$. Without loss of generality, assume that $\vec{E}$ satisfies $E_k \geq E_{k+1}$ for all $k \in [1:n-1]$. Define the two sequences $(U_i; 1\leq i \leq n)$ and $(V_i; 2\leq i \leq n)$ according to
\begin{align}\label{eqn:def_r_i_1}
    U_i = \frac{i}{\frac{3}{E_1} + \sum_{k=2}^{i}\frac{1}{E_k}}
\end{align}
and
\begin{align}\label{eqn:def_S_i_1}
    V_i = \frac{i-1}{\sum_{k=1}^{i}\frac{1}{E_k}}.
\end{align}
Let $u = \arg\max_{1 \leq i \leq n} U_i$, and $v = \arg\max_{2 \leq i \leq n} V_i$, where $\arg\max$ returns the least index in the case of ties. Then, $\vec{p}^*$ can be described under two cases.

\noindent
\textbf{Case 1: } If $V_v > U_u$, 
\begin{align}\label{eqn:c1_p2}
   p^*_k = \begin{cases}
       \frac{\frac{1}{E_k}}{\sum_{j=1}^v \frac{1}{E_j}} &\text{ if } 1\leq k \leq v\\
       0 & \text{ otherwise.}
   \end{cases}
\end{align}

\noindent
\textbf{Case 2: } If $U_u \geq V_v$,

\begin{align}\label{eqn:c2_p2}
   p^*_k = \begin{cases}
       \frac{\frac{3}{E_1}}{\frac{3}{E_1}+\sum_{j=2}^u \frac{1}{E_j}} &\text{ if }  k = 1\\
       \frac{\frac{1}{E_k}}{\frac{3}{E_1}+\sum_{j=2}^u \frac{1}{E_j}} &\text{ if } 2\leq k \leq u\\
       0 & \text{ otherwise.}
   \end{cases}
\end{align}
\end{theorem}
Before moving on to the proof of the theorem, we discuss the result. It is interesting to note the variation of choice probabilities in both cases of the theorem. In the first scenario, while choosing a collection of resources with the highest mean rewards with nonzero probability is optimal, one chooses resources with lower mean rewards with higher probability within the collection. In particular, player 1 first chooses a set of resources $[1:v]$ to be chosen with nonzero probability. Then player 1 assigns probability $p^*_k$ for $k \in [1:v]$ such that the $p^*_kE_k = C$ for some constant $C$. This behavior can be explained as follows. First, player $1$ never chooses resources with mean rewards below a certain threshold. Second, within the collection of resources with relatively high mean rewards, player $1$ is tempted to choose resources with lower mean rewards with high probability since, in the worst case, opponents choose the rewards with the highest mean rewards.

In Case 2, a similar behavior can be observed. Player 1 first chooses a set of resources $[1:u]$ to be chosen with nonzero probability. However, now player 1 assigns probability $p^*_k$ for $k \in [1:u]$ such that $p^*_kE_k = D$ for each $k \in [2:u]$ and $p^*_1E_1 = 3D$ for some constant $D$. In particular, player 1 chooses the first resource with a higher probability. To see this clearly, consider the two scenarios in Fig.~\ref{fig:201}, where we consider two possibilities of $\vec{E}$ for $n = 10$  (Scenario 1 and Scenario 2). Fig.~\ref{fig:201}-Left denotes the plot of $\vec{E}$ for the two scenarios. Although in the two scenarios, $\vec{E}$ is different only in $E_1$ by $0.1$, Scenario 1 belongs to Case 1 of Theorem~\ref{theorem:a0b0explicit}, whereas Scenario 2 belongs to Case 2. Fig.~\ref{fig:201}-Right shows the higher choice probability of resource 1 in Scenario 2. Here, the mean reward of the first resource is
high enough to give a high per-player reward even if many players select it. However, the mean reward is insufficient for player 1 to choose it with probability 1.  

Notice that in both scenarios, we have $1 \in \arg\max_{1 \leq i \leq n} E_ip^*_i$, and $2 \in \arg\max_{1 \leq i \leq n, i \neq 1} E_ip^*_i$. Also, since Scenario 1 belongs to Case 1 of Theorem~\ref{theorem:a0b0explicit}, we have $p^*_1 E_1 < 3p^*_2E_2$. Hence, from Lemma~\ref{lemma:worst_case}, we have $\vec{x}^* \in \mathcal{J}$ given by $x^*_i = 1$ if $i \in \{1,2\}$ and $x^*_i = 0$ if $i \in [3:n]$ satisfies $ \vec{x}^* \in \arg\min_{\vec{x} \in \mathcal{J}} f(\vec{p}^*,\vec{x})$. Hence, the worst-case for player 1 occurs when player 2 always chooses resource 1 and player 3 always chooses resource 2 (or vice versa). Using a similar argument, one can establish that in Scenario 2, the worst-case for player 1 occurs when player 2 and player 3 always choose resource 1. 

Assuming player 1 plays the strategy $\vec{p}^*$, (worst-case strategy) and the other two players play the strategies that give the worst-case to $\vec{p}^*$, the expected reward vector of the three players in Scenario 1 is $(4.52, 7.87, 5.57)$. In the above strategy profile, the strategies of players 2 and 3 are the best responses for the other two players. In fact this property holds more generally ($m =3,r = 1$, and the solution comes from Case 1 of Theorem~\ref{theorem:a0b0explicit}) as proven in Lemma~\ref{lemma:kind_of_nash}. Hence, it makes sense for players 2 and 3 to play the above strategy even if they do not intend to penalize player 1. However, in Scenario 2, the same vector is $(4.54, 3.79, 3.79)$. Here, players 2 and 3 can improve their expected reward by unilaterally deviating. Hence, in this case, the only motivation for players 2 and 3 to play the above strategy is to hurt player 1. Also, in Scenario 2, when all three players play the worst-case strategy $\vec{p}^*$,  each can increase the expected reward to $4.98$.
\begin{figure*}
\centering
\input{opt_sol_m3r1}
\end{figure*} 

\begin{lemma}\label{lemma:kind_of_nash}
Consider the special case $m=3,r=1$ with $n \in \mathbb{N}$ satisfying $n \geq 2$. Without loss of generality, assume that $\vec{E}$ satisfies $E_k \geq E_{k+1}$ for all $k \in [1:n-1]$. Consider the two sequences $(U_i; 1\leq i \leq n)$ and $(V_i; 2\leq i \leq n)$, and the indices $v,u$ defined in Theorem~\ref{theorem:a0b0explicit} and assume $V_v > U_u$. Let $\vec{p}^* \in \Delta_{n}$ denote the worst-case strategy of Player 1 defined in \eqref{eqn:c1_p2}. Then $\vec{x}^2,\vec{x}^3$ defined by,
\begin{align}
    x^2_i = \begin{cases}
        1 & \text{ if } i = 1\\
        0 & \text{ otherwise}.
    \end{cases}
\end{align}
and 
\begin{align}
    x^3_i = \begin{cases}
        1 & \text{ if } i = 2\\
        0 & \text{ otherwise}.
    \end{cases}
\end{align}
satisfies $\vec{x}^2 + \vec{x}^3 \in \arg\min_{\vec{x} \in \mathcal{J}} f(\vec{p}^*,\vec{x})$. Hence, given that Player 1 uses strategy $\vec{p}^*$, Players 2 and Players 3 can use decisions $\vec{x}^2,\vec{x}^3$ to enforce the worst-case for Player 1. Additionally the strategy profile $(\vec{p}^*,\vec{x}^2,\vec{x}^3)$ has the following property. Players $j \in \{2,3\}$ cannot increase their reward by unilaterally switching to a different strategy $\vec{q} \in \Delta_{n}$.
\begin{proof}
    The first part follows by applying Lemma~\ref{lemma:worst_case} (recall that we assumed the solution comes from case 2 of Theorem~\ref{theorem:a0b0explicit}). We move on to the second part. Define $C = \frac{1}{\sum_{k=1}^v \frac{1}{E_k}}$, where $v$ is defined in Theorem~\ref{theorem:a0b0explicit}. We first prove five claims that are useful in the solution.

    Throughout the proof, let us call the property $E_k \geq E_{k+1}$ for $k \in [1:n-1]$ to be the nondecreasing property.
    
    \noindent
    \textbf{Claim 1: } $C \leq \frac{E_1}{v}$
    \begin{proof}
        Notice that
    \begin{align}
        C = \frac{1}{\sum_{k=1}^v \frac{1}{E_k}} \leq \frac{1}{\sum_{k=1}^v \frac{1}{E_1}}  =   \frac{E_1}{v}
    \end{align}
    where the inequality follows from the nondecreasing property.
    \end{proof}

    \noindent
    \textbf{Claim 2: } If $v<n$, $\frac{E_1(v-1)}{v} \geq E_{v+1}$
    \begin{proof}
         Notice that from the definition of $v$, we have that $V_{v} \geq V_{v+1}$ which is,
    \begin{align}
        \frac{v-1}{\sum_{k=1}^v \frac{1}{E_k}} \geq \frac{v}{\sum_{k=1}^{v+1} \frac{1}{E_k}}.
    \end{align}
    Simplifying this gives
    \begin{align}
        \frac{v-1}{E_{v+1}} \geq \sum_{k=1}^v \frac{1}{E_k} \geq \frac{v}{E_1},
    \end{align}
    where the last inequality follows from the nondecreasing property.
    \end{proof}
   
    \noindent
     \textbf{Claim 3: } $\frac{E_1(v-1)}{2v-3} \leq E_{2}$
     \begin{proof}
          Notice that from the definition of $v$, we have that $V_{v} \geq U_{v}$ which is,,
    \begin{align}
        \frac{v-1}{\sum_{k=1}^v \frac{1}{E_k}} \geq \frac{v}{\frac{3}{E_1}+\sum_{k=2}^{v} \frac{1}{E_k}}.
    \end{align}
    Simplifying this gives
    \begin{align}
        \frac{2v-3}{E_1} \geq \sum_{k=2}^{v} \frac{1}{E_k} \geq \frac{v-1}{E_2}
    \end{align}
    where the last inequality follows from the nondecreasing property.
     \end{proof}

     \noindent
     \textbf{Claim 4:} $\frac{E_2(2v-3)}{2v-2} \geq E_{v+1}$.

     \begin{proof}
         Notice that from the definition of $v$, we have that $V_{v} \geq V_{v+1}$ which is,
    \begin{align}
        \frac{v-1}{\sum_{k=1}^v \frac{1}{E_k}} \geq \frac{v}{\sum_{k=1}^{v+1} \frac{1}{E_k}}.
    \end{align}
    Simplifying this gives
    \begin{align}
        \frac{v-1}{E_{v+1}} \geq \sum_{k=1}^v \frac{1}{E_k} \geq_{(a)} \frac{1}{E_1} + \frac{v-1}{E_2} \geq_{(b)} \frac{1}{E_2}\left(\frac{v-1}{2v-3}+ v-1\right)
    \end{align}
    where (a) follows from the nondecreasing property, and (b) follows from Claim 3. Simplifying the above ineqaulity, we have the result.
     \end{proof}

     \noindent
     \textbf{Claim 5:} $\frac{E_2}{v-1} \geq C$.

     \begin{proof}
          Notice that
    \begin{align}
        C = \frac{1}{\sum_{k=1}^v \frac{1}{E_k}} \leq \frac{1}{\sum_{k=2}^v \frac{1}{E_k}}\leq \frac{1}{\sum_{k=2}^v \frac{1}{E_2}} = \frac{E_2}{v-1}
    \end{align}
    where the last inequality follows from the nondecreasing property.
    \end{proof}
     
     Now, we are ready to prove the result. First consider the scenario where the three players use the strategy profile $(\vec{p}^*,\vec{q},\vec{x}^3)$, where $\vec{p}^*,\vec{x}^3$ are defined in the statement of the lemma and $\vec{q} \in \Delta_{n}$. Then the expected reward $h^2(\vec{q})$ of player 2 is 
    \begin{align}
        &h^2(\vec{q}) \nonumber\\&= q_1\left(\frac{p^*_1E_1}{2}+ (1-p^*_1)E_1\right)+ q_2\left(\frac{p^*_2E_2}{3}+ \frac{(1-p^*_2)E_2}{2}\right) + \sum_{k=3}^v q_k\left(\frac{p^*_kE_k}{2}+ (1-p^*_k)E_k\right)\nonumber\\&\ \ \ \  + \sum_{k = v+1}^n q_k E_k \nonumber\\& = q_1\left(E_1 - \frac{C}{2}\right)+ q_2\left(\frac{E_2}{2}- \frac{C}{6}\right) + \sum_{k=3}^v q_k\left(E_k- \frac{C}{2}\right) + \sum_{k = v+1}^n q_k E_k,
    \end{align}
    where the last equality follows from the definition of $\vec{p}^*$ in \eqref{eqn:c1_p2}.
    Notice that we require proving $h^2(\vec{q}) \leq h^2(\vec{x}^2)$ for all $\vec{q} \in \Delta_{n}$. Due to the nondecreasing property, this reduces to proving
    \begin{align}
        E_1 - \frac{C}{2}  \geq E_{v+1} , \text{ and } E_1 - \frac{C}{2}  \geq \frac{E_2}{2}- \frac{C}{6}
    \end{align}

    For the first inequality notice that,
    \begin{align}
        E_{v+1} = E_{v+1} - \frac{C}{2} + \frac{C}{2}  \leq_{(a)}  \frac{E_1(v-1)}{v} - \frac{C}{2} +  \frac{E_1}{2v} = E_1 - \frac{C}{2} - \frac{E_1}{2v} \leq E_1 - \frac{C}{2},
    \end{align}
    where (a) follows from Claim 1 and Claim 2. For the second inequality, notice that 
    \begin{align}
         \frac{E_2}{2} + \frac{C}{3} \leq_{(a)} \frac{E_1}{2}+ \frac{E_1}{3v} \leq E_1
    \end{align}
    where (a) follows from the nondecreasing property and Claim 1. Rearranging above we have the requirred inequality. Hence, we are done.

    Now consider the scenario where the three players use the strategy profile $(\vec{p}^*,\vec{x}^2,\vec{q})$, where $\vec{p}^*,\vec{x}^2$ are defined in the statement of the lemma and $\vec{q} \in \Delta_{n}$. Then the expected reward $h^3(\vec{q})$ of player 2 is 
    \begin{align}
        &h^3(\vec{q}) \nonumber\\&= q_1\left(\frac{p^*_1E_1}{3}+ \frac{(1-p^*_1)E_1}{2}\right)+ \sum_{k=2}^v q_k\left(\frac{p^*_kE_k}{2}+ (1-p^*_k)E_k\right) + \sum_{k = v+1}^n q_k E_k \nonumber\\& = q_1\left(\frac{E_1}{2} - \frac{C}{6}\right)+ \sum_{k=2}^v q_k\left(E_k- \frac{C}{2}\right) + \sum_{k = v+1}^n q_k E_k,
    \end{align}
    where the last equality follows from the definition of $\vec{p}^*$ in \eqref{eqn:c1_p2}.
    Notice that we require proving $h^3(\vec{q}) \leq h^3(\vec{x}^3)$ for all $\vec{q} \in \Delta_{n}$. Due to the nondecreasing property, this reduces to proving
    \begin{align}
        E_2 - \frac{C}{2}  \geq E_{v+1} , \text{ and } E_2 - \frac{C}{2}  \geq \frac{E_1}{2}- \frac{C}{6}
    \end{align}

    For the first inequality, notice that,
    \begin{align}
        E_2 - E_{v+1} \geq_{(a)} E_2 - \frac{2v-3}{2v-2}E_2 = \frac{E_2}{2(v-1)} \geq \frac{C}{2}
    \end{align}
    where (a) follows from Claim 4 and the last inequality follows from Claim 5. 

    For the second inequality, notice that,
    \begin{align}
        E_2 - \frac{E_1}{2} \geq_{(a)} E_2 - \frac{2v-3}{2v-2}E_2 = \frac{E_2}{2(v-1)} \geq_{(b)} \frac{C}{2} \geq \frac{C}{3}
    \end{align}
    where (a) follows from Claim 3 and (b) follows from Claim 5. Rearranging, we are done. Hence, we are done with the proof.
\end{proof}
\end{lemma}
Now, we prove Theorem~\ref{theorem:a0b0explicit} step by step.
Recall that we assumed without loss of generality that $\vec{E}$ is sorted as $E_k \geq E_{k+1}$ for all $k \in \{1,2,\dots,n-1\}$. Notice that from Lemma~\ref{lemma:worst_case} we have,
\begin{align}\label{def:f_worst_m3}
f^{\text{worst}}(\vec{p}) = \begin{cases}
    \sum_{k =1}^n p_kE_k - \frac{2}{3}\Gamma_1(\vec{p}) &\text{ if } \Gamma_1(\vec{p})>3\Gamma_2(\vec{p})\\
    \sum_{k =1}^n p_kE_k - \frac{1}{2}\Gamma_1(\vec{p})- \frac{1}{2}\Gamma_2(\vec{p}) &\text{ if } \Gamma_1(\vec{p})\leq 3\Gamma_2(\vec{p})\\
\end{cases}
\end{align}
where $\Gamma_1(\vec{p})$, $\Gamma_2(\vec{p})$ are the largest and the second largest elements of the set $\{p_kE_k; 1 \leq k \leq n \}$, respectively. Observe that if $\Gamma_1(\vec{p}) = 3\Gamma_2(\vec{p})$, then $\frac{2}{3}\Gamma_1(\vec{p})= \frac{1}{2}\Gamma_1(\vec{p})+\frac{1}{2}\Gamma_1(\vec{p})$. In particular, the function $f^{\text{worst}}(\vec{p})$ is continuous and so it has a maximizer $\vec{p}^*$ over the compact set $\Delta_{n}$. By considering the case $\Gamma_1(\vec{p}^*) \geq 3\Gamma_2(\vec{p}^*)$ and a particular index $i \in \{1,\dots,n\}$ achieves $p^*_iE_i = \Gamma_1(\vec{p}^*)$, and the case $\Gamma_1(\vec{p}^*) \leq 3\Gamma_2(\vec{p}^*)$ and particular indices $i \neq j$ achieve $p^*_iE_i = \Gamma_1(\vec{p}^*)$, $p^*_jE_j = \Gamma_2(\vec{p}^*)$, we notice that $\vec{p}^*$ is the solution of the problem with the maximal optimal objective out of the $n^2$ linear programs,
\begin{maxi}
  {}{\sum_{k =1}^n p_kE_k - \frac{2p_iE_i}{3} }{}{\text{(P1-}i\text{)}:}
  \label{prob:p1_i}
  \addConstraint{\vec{p}\in}{ \Delta_{n}}
  \addConstraint{p_iE_i}{ \geq 3p_kE_k \text{ } \forall 1 \leq k \leq n},
\end{maxi}
and 
\begin{maxi}
  {}{ \sum_{k =1}^n p_kE_k - \frac{p_iE_i}{2}-\frac{p_jE_j}{2} }{}{\text{(P1-}(i,j)\text{)}:}
  \label{prob:p1_i_j}
  \addConstraint{\vec{p} \in \Delta_{n},}{\ p_iE_i \leq 3p_jE_j, \ p_iE_i \geq p_jE_j }
  \addConstraint{p_jE_j}{ \geq p_kE_k \text{ } \forall 1 \leq k \leq n , k \neq i},
\end{maxi}
where $i,j \in [1:n]$ and $i\neq j$. To solve $\text{(P1-}i\text{)}$, and $\text{(P1-}(i,j)\text{)}$, it shall be useful to re-index to associate $i$ with 1, and $(i,j)$ with 1 and 2. Hence, we define the two problems.
\begin{maxi}
  {}{f_1(\vec{p}) =   \sum_{k =1}^n p_kF_k-\frac{2p_1F_1}{3}}{}{\text{(P1-1)}:}
  \addConstraint{\vec{p}}{\in \Delta_{n}}
  \addConstraint{p_1F_1}{ \geq 3p_{k+1}F_{k+1} \text{ } \forall k \in \{1,\dots,n-1\}},
\end{maxi}
and 
\begin{maxi}
  {}{f_2(\vec{p}) = \sum_{k =1}^n p_kF_k-\frac{p_1F_1}{2} -\frac{p_2F_2}{2}}{}{\text{(P1-2)}:}
  \addConstraint{\vec{p} \in \Delta_{n},}{\ p_1F_1 \leq  3p_2F_2, \ p_1F_1 \geq p_2F_2}
  \addConstraint{p_2F_2}{ \geq p_kF_k \text{ } \forall 3 \leq k \leq n },
\end{maxi}
where for (P1-1), without loss of generality $\vec{F} \in \mathbb{R}^n$ is assumed to a positive vector such that $F_k \geq F_{k+1}$ for $k \in [2:n-1]$, and for (P1-2), $\vec{F} \in \mathbb{R}^n$ is assumed to a positive vector such that $F_k \geq F_{k+1}$ for $k \in [3:n-1]$. It should be noted that the $F_k$ values are just the $E_k$ values under more convenient indexing. Solving the above two problems immediately solves each of the previously defined $n^2$ problems. Define the two sequences $(U_i; 1\leq i \leq n)$, and  $(V_i; 2\leq i \leq n)$ by,
\begin{align}\label{eqn:def_r_i}
    U_i = \frac{i}{\frac{3}{F_1} + \sum_{k=2}^{i}\frac{1}{F_k}},
\end{align}
and,
\begin{align}\label{eqn:def_S_i}
    V_i = \frac{i-1}{\sum_{k=1}^{i}\frac{1}{F_k}}.
\end{align}
These two sequences are useful when constructing the solutions to (P1-1) and (P1-2).

\subsubsection{Solving (P1-1)}\label{sec:solving_P2_1}
Consider the problem (P1-1):
\begin{maxi}
  {}{f_1(\vec{p})}{}{\text{(P1-1)}:}
  \addConstraint{\vec{p}}{\in \Delta_{n}}
  \addConstraint{p_1F_1}{\geq 3p_{k+1}F_{k+1}  \text{ } \forall k \in \{1,2,\dots,n-1\}},
\end{maxi}
where the function $f_1$ is defined by
\begin{align}\label{eqn:f_1_def}
    f_1(\vec{p}) =   \sum_{k =1}^n p_kF_k-\frac{2p_1F_1}{3}.
\end{align}
Let us define 
\begin{align}\label{eqn:u_def}
    u = \arg\max_{1 \leq i \leq n} U_i,
\end{align}
where the sequence $(U_i; 1\leq i \leq n)$ is defined in \eqref{eqn:def_r_i} and $\arg\max$ returns the least index in the case of ties. We establish that the solution to (P1-1) is $\vec{\tilde{p}}^*$, where
\begin{align}\label{eqn:p2-2_sol}
\tilde{p}^*_k = \begin{cases}
   \frac{\frac{3}{F_1}}{\frac{3}{F_1}+\sum_{j=2}^{u} \frac{1}{F_j}} &\text{ if }  k = 1\\
   \frac{\frac{1}{F_k}}{\frac{3}{F_1}+\sum_{j=2}^{u} \frac{1}{F_j}} &\text{ if } 2\leq k \leq {u}\\
   0 & \text{ otherwise,}
\end{cases}
\end{align}
with optimal objective value $U_u$.

Consider the vector $\vec{\tilde{\mu}}^{*} \in \mathbb{R}^{n-1}$ defined by
\begin{align}\label{eqn:def_mu_p2-1}
\tilde{\mu}^*_k = \begin{cases}
    \frac{1}{3}\left(1 - \frac{1}{F_{k+1}}\frac{u}{\frac{3}{F_1}+\sum_{j=2}^{u}\frac{1}{F_j}}\right) & \text{ if } 1 \leq k \leq u-1\\
    0 & \text{ otherwise, }
\end{cases}
\end{align}
where $u$ is defined in \eqref{eqn:u_def}. In the subsequent analysis, we establish that $\vec{\tilde{\mu}}^{*}$ defined above is a valid Lagrange multiplier ($\tilde{\mu}_k^* \geq 0$ for all $k \in [1:n-1]$) and  $(\vec{\tilde{p}}^*, \vec{\tilde{\mu}}^*)$ satisfy the conditions of Lemma~\ref{lemma:lagrange_lemma}, where for $k \in [1:n-1]$, $\tilde{\mu}_k^*$ corresponds to the constraint $p_1F_1 \geq 3p_{k+1}F_{k+1}$ of (P1-1). This establishes that $\vec{\tilde{p}}^*$ solves (P1-1). It can be easily checked by substitution that the objective value of (P1-1) for $\vec{\tilde{p}}^*$ is $U_u$. Hence, the steps of the proof can be summarized as:
\begin{enumerate}
    \item $\tilde{\mu}_k^* \geq 0$ for all $k \in [1:n-1]$.
    \item $\vec{\tilde{p}}^{*}$ is feasible for (P1-1). In particular, we have that $\vec{\tilde{p}}^{*} \in \Delta_{n}$ and $\tilde{p}^{*}_1F_1 \geq 3\tilde{p}^{*}_{k+1}F_{k+1}$ for $k \in \{1,\dots,n-1\}$.
    \item $\vec{\tilde{p}}^{*}$ solves the unconstrained problem with Lagrange multiplier vector $\vec{\tilde{\mu}}^{*}$ (See Lemma~\ref{lemma:lagrange_lemma} for the construction of the unconstrained problem).
    \item For $k \in \{1,\dots,n-1\}$, $\tilde{\mu}^*_k>0$ implies the corresponding constraint of (P1-1) is met with equality.
\end{enumerate}
Notice that step 2 above can be checked by direct substitution from \eqref{eqn:p2-2_sol}. Also, for step 4, notice that from the definition of $\vec{\tilde{\mu}}^{*}$ in \eqref{eqn:def_mu_p2-1}, $\tilde{\mu}^*_k>0$  implies that $k \in \{1,\dots,u-1\}$. By substitution from the definition of $\vec{\tilde{p}}^*$ in \eqref{eqn:p2-2_sol}, it follows that $\tilde{p}^{*}_1F_1 = 3\tilde{p}^{*}_{k+1}F_{k+1}$ for $k \in \{1,\dots,u-1\}$. Hence, we are only required to establish steps 1 and 3. We establish step 1 along with two other results that will be useful for step 3 in Lemma~\ref{lemma:p2-1-mu} below, after which we establish step 3 in Lemma~\ref{eqn:uncons_solve_lemma}.
\begin{lemma}\label{lemma:p2-1-mu}
Consider the $\vec{\tilde{\mu}}^*$ defined in \eqref{eqn:def_mu_p2-1}. We have that
\begin{itemize}
    \item[(a)] $\tilde{\mu}^*_k \geq 0$ for all $k$ such that $1 \leq k \leq n-1$.
    \item[(b)] $F_k(1-3\tilde{\mu}^*_{k-1}) = \frac{u}{\frac{3}{F_1}+\sum_{i=2}^{u}\frac{1}{F_i}} \text{ for } 2 \leq k \leq u$ and
    $F_1\left(\frac{1}{3} + \sum_{i=1}^{u-1}\tilde{\mu}^*_i\right)   = \frac{u}{\frac{3}{F_1}+\sum_{i=2}^{u}\frac{1}{F_i}}.$
    \item[(c)] $F_k  \leq  \frac{u}{\frac{3}{F_1}+\sum_{j=2}^{u}\frac{1}{F_j}}$ for $u+1 \leq k \leq n$.
\end{itemize}
\begin{proof}
 Notice that since $u = \arg\max_{1 \leq i \leq n} U_i$, we have that
\begin{align}\label{eqn:u_eq}
    U_u \geq U_j \text{ for all $j \in [1:n]$.}
\end{align}
\begin{itemize}
\item[(a)] Notice that  when $k>u-1$, by definition of $\vec{\tilde{\mu}}^*$ in \eqref{eqn:def_mu_p2-1}, we have that $\tilde{\mu}^*_k = 0$. Now suppose $k \leq u-1$. Hence, we can assume $u \geq 2$. From the definition of $\vec{\tilde{\mu}}^*$ in \eqref{eqn:def_mu_p2-1}, we are required to prove,
\begin{align}
    F_{k+1} \geq \frac{u}{\frac{3}{F_1}+\sum_{j=2}^{u}\frac{1}{F_j}},
\end{align}
for all $k \in \{1,2,\dots,u-1\}$. It is enough to prove the above for $k = u-1$, since $F_k \geq F_{k+1}$ for $k \geq 2$. Notice that from \eqref{eqn:u_eq} we have that $U_u \geq U_{u-1}$ (recall that $u\geq 2$). Substituting from \eqref{eqn:def_r_i}, $U_u \geq U_{u-1}$ translates to,
\begin{align}
    \frac{u}{\frac{3}{F_1} + \sum_{j=2}^{u}\frac{1}{F_j}} \geq \frac{u-1}{\frac{3}{F_1} + \sum_{j=2}^{u-1}\frac{1}{F_j}}.
\end{align}
Simplifying the above gives
\begin{align}
    F_{u} \geq \frac{u}{\frac{3}{F_1}+\sum_{j=2}^{u}\frac{1}{F_j}}
\end{align}
as desired.
\item[(b)] Substituting from the definition of $\tilde{\mu}^*_k$ in \eqref{eqn:def_mu_p2-1} and simplifying yields the result. 

\item[(c)] If $u = n$, there is nothing to prove. Hence, we can assume $u<n$. Notice that it is enough to prove the result for $k = u+1$, since $F_k \geq F_{k+1}$ for $k \geq 2$. From \eqref{eqn:u_eq}, we have that $U_u \geq U_{u+1}$ (recall that $u < n$). Substituting from \eqref{eqn:def_r_i}, $U_u \geq U_{u+1}$ translates to,
\begin{align}
    \frac{u}{\frac{3}{F_1} + \sum_{j=2}^{u}\frac{1}{F_j}} \geq \frac{u+1}{\frac{3}{F_1} + \sum_{j=2}^{u+1}\frac{1}{F_j}} 
\end{align}
Simplifying the above we have,
\begin{align}
    F_{u+1} \leq \frac{u}{\frac{3}{F_1}+\sum_{j=2}^u\frac{1}{F_j}}
\end{align}
as desired.
\end{itemize}
\end{proof}
\end{lemma}
\begin{lemma}\label{eqn:uncons_solve_lemma}
The vector $\vec{\tilde{p}}^{*}$ defined in \eqref{eqn:p2-2_sol} solves unconstrained problem with Lagrange multiplier vector $\vec{\tilde{\mu}}^*$ defined in \eqref{eqn:def_mu_p2-1} (See Lemma~\ref{lemma:lagrange_lemma} for the construction of the unconstrained problem). In particular, $\vec{\tilde{p}}^{*}$ solves
\begin{maxi}
  {}{f_1(\vec{p}) +  \sum_{k=1}^{n-1} \tilde{\mu}^*_k (p_1F_1-3p_{k+1}  F_{k+1})}{}{}
  \label{prob:p1_1_m}
  \addConstraint{\vec{p}}{ \in \Delta_{n}},
\end{maxi}
where the function $f_1$ is defined in \eqref{eqn:f_1_def}.
\begin{proof}
Noticing from the definition of $\vec{\tilde{\mu}}^*$ in \eqref{eqn:def_mu_p2-1} that $\tilde{\mu}_k^* = 0$ for $k>u$, and using the definition of function $f_1$ in \eqref{eqn:f_1_def}, the objective of the above unconstrained problem simplifies as
\begin{align}
    &f_1(\vec{p}) +  \sum_{k=1}^{n-1} \tilde{\mu}^*_k (p_1F_1-3p_{k+1}  F_{k+1}) \nonumber\\&= p_1F_1\left(\frac{1}{3} + \sum_{i=1}^{u-1}\tilde{\mu}^*_i\right) + \sum_{k=2}^u p_kF_k(1-3\tilde{\mu}^*_{k-1}) + \sum_{k=u+1}^np_kF_k \nonumber\\&= \sum_{i=1}^u p_iC + \sum_{k=u+1}^np_kF_k,
\end{align}
where 
\begin{align}\label{eqn:C_1}
     C = \frac{u}{\frac{3}{F_1}+\sum_{i=2}^u\frac{1}{F_i}}
\end{align}
and the last inequality follows from Lemma~\ref{lemma:p2-1-mu}-(b). Also, notice that from Lemma~\ref{lemma:p2-1-mu}-(c), we have that $C \geq F_k$ for all $k \in \{u+1,\dots,n\}$. Hence, the optimal solution to the above defined unconstrained problem is any $\vec{p} \in \Delta_{n}$ such that $p_k = 0$ for all $k \in \{u+1,\dots,n\}$. In particular, $\vec{\tilde{p}}^*$ given in \eqref{eqn:p2-2_sol} is a solution to the unconstrained problem. 
\end{proof}
\end{lemma}

\subsubsection{Solving (P1-2)} \label{sec:solving_P2_2}
Consider the problem (P1-2).
\begin{maxi}
  {}{f_2(\vec{p})}{}{\text{(P1-2)}:}
  \addConstraint{\vec{p} \in \Delta_{n},}{\ p_1F_1 \leq  3p_2F_2, \ p_1F_1 \geq p_2F_2}
  \addConstraint{p_2F_2}{ \geq p_kF_k \text{ } \forall 3 \leq k \leq n },
\end{maxi}
where the function $f_2$ is defined as
\begin{align}\label{eqn:f_2_def}
    f_2(\vec{p}) = \sum_{k =1}^n p_kF_k-\frac{p_1F_1}{2} -\frac{p_2F_2}{2}
\end{align}
Let us define 
\begin{align}\label{eqn:u_def_1}
    u = \arg\max_{2 \leq i \leq n} U_i
\end{align}
and
\begin{align}\label{eqn:v_def}
    v = \arg\max_{2 \leq i \leq n} V_i
\end{align}
where the sequences $(U_i; 1\leq i \leq n)$, and  $(V_i; 2\leq i \leq n)$ are defined in \eqref{eqn:def_r_i}, and \eqref{eqn:def_S_i}, respectively, and $\arg\max$ returns the least index in the case of ties. In this case, to define $u$, we only consider the indices of the $(U_i; 1\leq i \leq n)$ sequence starting from 2 in contrast to the definition of $u$ in the solution to (P1-1). The solution of (P1-2) can be described under two cases.

\noindent
\textbf{Case 1:} $V_v>U_u$: The solution to (P1-2) in this case is $\vec{\hat{p}}^*$ where
\begin{align}\label{eqn:s_geq_r}
\hat{p}^*_k = \begin{cases}
   \frac{\frac{1}{F_k}}{\sum_{j=1}^v \frac{1}{F_j}} &\text{ if } 1\leq k \leq v\\
   0 & \text{ otherwise,}
\end{cases}
\end{align}
with optimal objective value $V_v$.

\noindent
\textbf{Case 2:} $V_v\leq U_u$: The solution to (P1-2) in this case is $\vec{\bar{p}}^*$ where
\begin{align}\label{eqn:r_geq_s}
\bar{p}^*_k = \begin{cases}
   \frac{\frac{3}{F_1}}{\frac{3}{F_1}+\sum_{j=2}^u \frac{1}{F_j}} &\text{ if }  k = 1\\
   \frac{\frac{1}{F_k}}{\frac{3}{F_1}+\sum_{j=2}^u \frac{1}{F_j}} &\text{ if } 2\leq k \leq u\\
   0 & \text{ otherwise.}
\end{cases}
\end{align}
with optimal objective value $U_u$. 

Similar to the the solution of (P1-1), for each case, we construct a Lagrange multiplier vector such that the conditions of the Lemma~\ref{lemma:lagrange_lemma} are satisfied. For the Lagrange multiplier vector $\vec{\mu} \in \mathbb{R}^n$ we associate $\mu_1$ with the constraint $p_1F_1 \leq 3p_2F_2$, $\mu_2$ with the constraint $p_1F_1 \geq p_2F_2$, and $\mu_k$ for $k \in \{3,\dots,n\}$ with the constraint $p_2F_2 \geq p_kF_k$.
Now, we analyze the two cases separately.

\noindent
\textbf{Case 1}: $V_v>U_u$: Recall that we are required to prove $\vec{\hat{p}}^*$ where
\begin{align}\label{eqn:s_geq_r_1}
\hat{p}^*_k = \begin{cases}
   \frac{\frac{1}{F_k}}{\sum_{j=1}^v \frac{1}{F_j}} &\text{ if } 1\leq k \leq v\\
   0 & \text{ otherwise,}
\end{cases}
\end{align}
is the solution to (P1-2). Define the vector $\vec{\hat{\mu}}^* \in \mathbb{R}^n$ as
\begin{align}\label{eqn:def_mu}
\hat{\mu}^*_k = \begin{cases}
   \frac{1}{F_1} \frac{v-1}{\sum_{j=1}^{v}\frac{1}{F_j}} - \frac{1}{2} & \text{ if } k = 2,\\
    1 - \frac{1}{F_k}\frac{v-1}{\sum_{j=1}^{v}\frac{1}{F_j}} & \text{ if } 3 \leq k \leq v,\\
    0 & \text{ otherwise, }
\end{cases}
\end{align}
where $v$ is defined in \eqref{eqn:v_def}. Similar to the solution of (P1-1), we focus on establishing the four steps:
\begin{enumerate}
    \item $\hat{\mu}_k^* \geq 0$ for all $k \in [1:n]$.
    \item $\vec{\hat{p}}^{*}$ is feasible for (P1-2). In particular, we have that $\vec{\hat{p}}^{*} \in \Delta_{n}$ and $3\hat{p}^{*}_{2}F_{2} \geq \hat{p}^{*}_1F_1 \geq \hat{p}^{*}_2F_2$,  and $\hat{p}^{*}_2F_2 \geq \hat{p}^{*}_kF_k$ for $k \in \{3,\dots,n\}$.
    \item $\vec{\hat{p}}^{*}$ solves the unconstrained problem with Lagrange multiplier vector $\vec{\hat{\mu}}^{*}$.
    \item For $k \in \{1,\dots,n\}$, $\hat{\mu}^*_k>0$ implies the corresponding constraint of (P1-2) is met with equality.
\end{enumerate}
Similar to the solution of (P1-1), step 2 can be checked by direct substitution from \eqref{eqn:s_geq_r_1}. Also, for step 4, notice that  from the definition of $\vec{\hat{\mu}}^{*}$ in \eqref{eqn:def_mu}, $\hat{\mu}^*_k>0$  implies that $k \in \{2,\dots,v\}$. By substitution from the definition of $\vec{\hat{p}}^*$ in \eqref{eqn:s_geq_r_1}, it follows that $\tilde{p}^{*}_1F_1 = \tilde{p}^{*}_{2}F_{2}$ and $\tilde{p}^{*}_{2}F_{2} = \tilde{p}^{*}_{k}F_{k}$  for $k \in \{3,\dots,v\}$. We establish step 1 along with two other results that will be useful for step 3 in Lemma~\ref{lemma:a0b0_3}. Then we establish step 3 in Lemma~\ref{lemma:uncons_solve_lemma_1}.
\begin{lemma}\label{lemma:a0b0_3}
Consider the $\vec{\hat{\mu}}^*$ defined in \eqref{eqn:def_mu}. We have that
\begin{enumerate}
    \item[(a)] $\hat{\mu}^*_k \geq 0$ for all $k$ such that $1 \leq k \leq n$.
    \item[(b)] $F_1\left(\frac{1}{2} + \hat{\mu}^*_2\right) = \frac{v-1}{\sum_{j=1}^{v}\frac{1}{F_j}}$,  $F_2\left(\frac{1}{2} -\hat{\mu}^*_2 + \sum_{i=3}^v \hat{\mu}^*_i\right) = \frac{v-1}{\sum_{j=1}^{v}\frac{1}{F_j}}$ and  $F_k(1-\hat{\mu}^*_k) = \frac{v-1}{\sum_{j=1}^{v}\frac{1}{F_j}} \text{ for } 3 \leq k \leq v.$
    \item[(c)] $F_k  \leq  \frac{v-1}{\sum_{j=1}^{v}\frac{1}{F_j}}$ for $v+1 \leq k \leq n$
\end{enumerate}
\begin{proof}
 Notice that since $u = \arg\max_{2 \leq i \leq n} U_i$, and $v = \arg\max_{2 \leq i \leq n} V_i$, we have that $U_u \geq U_j$ for all $j \in [2:n]$, and $V_v \geq V_j$ for all $j \in [2:n]$. Since from the case description, we have that $V_v>U_u$, we should have that,
 \begin{align}\label{eqn:v_eq}
     V_v \geq V_j \text{ for all } j \in [2:n] \text{ and } V_v > U_j \text{ for all } j \in [2:n]
 \end{align}
\begin{enumerate}
\item[(a)] Notice that the result trivially follows for $k \not\in\{2,\dots,v\}$ since $\hat{\mu}^*_k = 0$ for such $k$. Hence, we focus on $k \in\{2,\dots,v\}$. We first prove that $\hat{\mu}^*_2 \geq 0$. Notice that from \eqref{eqn:v_eq}, we have that $V_v > U_v$. After substituting from \eqref{eqn:def_r_i} and \eqref{eqn:def_S_i}, $V_v > U_v$ translates to
\begin{align}
    \frac{v-1}{\sum_{j=1}^v \frac{1}{F_j}} \geq \frac{v}{\frac{1}{F_1}+\sum_{j=2}^v \frac{1}{F_j}}.
\end{align}
Simplifying the above, we get,
\begin{align}
   \frac{v-1}{\sum_{j=1}^v\frac{1}{F_j}} \geq \frac{F_1}{2}
\end{align}
as desired.

To obtain the result for $3 \leq k \leq v$, we can assume that $v \geq 3$. Notice that we are required to prove,
\begin{align}
    F_k \geq \frac{v-1}{\sum_{j=1}^{v}\frac{1}{F_j}}.
\end{align}
It is enough to prove the above for $k = v$, since $F_k \geq F_{k+1}$ for $k \geq 3$. From \eqref{eqn:v_eq} we have that $V_v \geq V_{v-1}$ (recall that $v \geq 3$). Substituting from \eqref{eqn:def_S_i}, $V_v \geq V_{v-1}$ translates to
\begin{align}
    \frac{v-1}{\sum_{j=1}^v \frac{1}{F_j}} \geq \frac{v-2}{\sum_{j=1}^{v-1} \frac{1}{F_j}}.
\end{align}
Simplifying the above, we get,
\begin{align}
  \frac{v-1}{\sum_{j=1}^v\frac{1}{F_j}} \leq F_v
\end{align}
as desired.

\item[(b)] Substituting from the definition of $\hat{\mu}^*_k$ from \eqref{eqn:def_mu} and simplifying yields the result. 

\item[(c)] If $v = n$, there is nothing to prove. Hence, we can assume $v<n$. Notice that it is enough to prove the result for $k = v+1$, since $F_k \geq F_{k+1}$ for $k \geq 3$. From \eqref{eqn:v_eq} we have that $V_v \geq V_{v+1}$ (recall that $v<n$). Substituting from \eqref{eqn:def_S_i}, $V_v \geq V_{v+1}$ translates to
\begin{align}
    \frac{v-1}{\sum_{j=1}^v \frac{1}{F_j}} \geq \frac{v}{\sum_{j=1}^{v+1} \frac{1}{F_j}}.
\end{align}
Simplifying the above, we get,
\begin{align}
   F_{v+1} \leq \frac{v-1}{\sum_{j=1}^v\frac{1}{F_j}}
\end{align}
as desired.   
\end{enumerate}
\end{proof}
\end{lemma}
\begin{lemma}\label{lemma:uncons_solve_lemma_1}
The vector $\vec{\hat{p}}^*$ defined in \eqref{eqn:s_geq_r_1} solves unconstrained problem with Lagrange multiplier vector $\vec{\hat{\mu}}^*$ defined in \eqref{eqn:def_mu}. In particular, $\vec{\hat{p}}^{*}$ solves
\begin{maxi}
  {}{f_2(\vec{p}) +  \hat{\mu}^*_1 (3p_2F_2-p_1F_1)+\hat{\mu}^*_2(p_1F_1-p_2F_2)+ \sum_{k=3}^n\hat{\mu}^*_k(p_2F_2-p_kF_k)}{}{}
  \addConstraint{\vec{p}}{ \in \Delta_{n}},
\end{maxi}
where the function $f_2$ is defined in \eqref{eqn:f_2_def}.
\begin{proof}
Noticing from the definition of $\vec{\hat{\mu}}^*$ in \eqref{eqn:def_mu} that $\hat{\mu}_k^* = 0$ for $k>v$, and using the definition of  function $f_2$ in \eqref{eqn:f_2_def}, the objective of the unconstrained problem simplifies as
\begin{align}
    &f_2(\vec{p}) +  \hat{\mu}^*_1 (3p_2F_2-p_1F_1)+\hat{\mu}^*_2(p_1F_1-p_2F_2)+ \sum_{k=3}^n\hat{\mu}^*_k(p_2F_2-p_kF_k) \nonumber\\&= p_1F_1\left(\frac{1}{2} + \hat{\mu}^*_2\right) +p_2F_2\left(\frac{1}{2} - \hat{\mu}^*_2+\sum_{i=3}^v\hat{\mu}^*_i\right) + \sum_{k=3}^v p_kF_k(1-\hat{\mu}^*_{k}) + \sum_{k=v+1}^np_kF_k \nonumber\\& = \sum_{i=1}^v p_iC + \sum_{k=v+1}^np_kF_k,
\end{align}
where 
\begin{align}
     C = \frac{v-1}{\sum_{i=1}^v\frac{1}{F_i}},
\end{align}
and the last inequality follows from Lemma~\ref{lemma:a0b0_3}-(b). From Lemma~\ref{lemma:a0b0_3}-(c), we have that $C \geq F_k$ for all $k \in \{v+1,\dots,n\}$. Hence, the optimal solution to the above defined unconstrained problem is any $\vec{p} \in \Delta_{n}$ such that $p_k = 0$ for all $k \in \{v+1,\dots,n\}$. In particular, $\vec{\hat{p}}^*$ given in \eqref{eqn:s_geq_r_1} is a solution to the unconstrained problem. 
\end{proof}
\end{lemma}

\noindent
\textbf{Case 2 } $U_u \geq V_v$: Recall that, we are required to prove $\vec{\bar{p}}^*$ given by
\begin{align}\label{eqn:r_geq_s_1}
\bar{p}^*_k = \begin{cases}
   \frac{\frac{3}{F_1}}{\frac{3}{F_1}+\sum_{j=2}^u \frac{1}{F_j}} &\text{ if }  k = 1\\
   \frac{\frac{1}{F_k}}{\frac{3}{F_1}+\sum_{j=2}^u \frac{1}{F_j}} &\text{ if } 2\leq k \leq u\\
   0 & \text{ otherwise.}
\end{cases}
\end{align}
is the solution to (P1-2). Consider the vector $\vec{\bar{\mu}}^* \in \mathbb{R}^n$ given by
\begin{align}\label{eqn:mu_2}
\bar{\mu}^*_k = \begin{cases}
  \frac{1}{2}- \frac{1}{F_1} \frac{u}{\frac{3}{F_1}+\sum_{j=2}^{u}\frac{1}{F_j}} & \text{ if } k = 1,\\
    1 - \frac{1}{F_k}\frac{u}{\frac{3}{F_1}+\sum_{j=2}^{u}\frac{1}{F_j}} & \text{ if } 3 \leq k \leq u,\\
    0 & \text{ otherwise. }
\end{cases}
\end{align}
where $u$ is defined in \eqref{eqn:u_def_1}. Similar to case 1, we focus on establishing the four steps:
\begin{enumerate}
    \item $\bar{\mu}_k^* \geq 0$ for all $k \in [1:n]$.
    \item $\vec{\bar{p}}^{*}$ is feasible for (P1-2). In particular, we have that $\vec{\bar{p}}^{*} \in \Delta_{n}$ and $3\bar{p}^{*}_{2}F_{2} \geq \bar{p}^{*}_1F_1 \geq \bar{p}^{*}_2F_2$,  and $\bar{p}^{*}_2F_2 \geq \bar{p}^{*}_kF_k$ for $k \in \{3,\dots,n\}$.
    \item $\vec{\bar{p}}^{*}$ solves the unconstrained problem with Lagrange multiplier vector $\vec{\bar{\mu}}^{*}$.
    \item For $k \in \{1,\dots,n\}$, $\bar{\mu}^*_k>0$ implies the corresponding constraint of (P1-2) is met with equality.
\end{enumerate}
Similar to case 1, step 2 can be checked by direct substitution from \eqref{eqn:r_geq_s_1}. For step 4, notice that from the definition of $\vec{\bar{\mu}}^{*}$ in \eqref{eqn:mu_2}, $\bar{\mu}^*_k>0$  implies that either $k = 1$ or $k \in \{3,\dots,u\}$. By substitution from the definition of $\vec{\bar{p}}^*$ in \eqref{eqn:r_geq_s_1}, it follows that $\bar{p}^{*}_1F_1 = 3\bar{p}^{*}_{2}F_{2}$ and $\bar{p}^{*}_{2}F_{2} = \bar{p}^{*}_{k}F_{k}$  for $k \in \{3,\dots,u\}$. Similar to case 1, we establish step 1 along with two other results that will be useful for step 3 in Lemma~\ref{lemma:lag_2}, after which we establish step 3 in Lemma~\ref{lemma:uncons_solve_lemma_2}.
\begin{lemma}\label{lemma:lag_2}
For the $\vec{\bar{\mu}}^*$ defined in \eqref{eqn:mu_2}, we have that
\begin{enumerate}
    \item[(a)]  $\bar{\mu}^*_k \geq 0$ for all $k$ such that $1 \leq k \leq n$.
    \item[(b)] We have $F_1\left(\frac{1}{2} - \bar{\mu}^*_1\right) = \frac{u}{\frac{3}{F_1}+\sum_{j=2}^{v}\frac{1}{F_j}}$,  $F_2\left(\frac{1}{2} + 3\bar{\mu}^*_1 + \sum_{i=3}^u \bar{\mu}^*_i\right) = \frac{u}{\frac{3}{F_1}+\sum_{j=2}^{v}\frac{1}{F_j}}$, and  $F_k(1-\bar{\mu}^*_k) = \frac{u}{\frac{3}{F_1}+\sum_{j=2}^{v}\frac{1}{F_j}}$ for  $3 \leq k \leq u$.
    \item[(c)] $F_k  \leq  \frac{u}{\frac{3}{F_1}+\sum_{j=2}^{u}\frac{1}{F_j}}$ for $u+1 \leq k \leq n$
\end{enumerate}
\begin{proof}
 Notice that since $u = \arg\max_{2 \leq i \leq n} U_i$, and $v = \arg\max_{2 \leq i \leq n} V_i$, we have that $U_u \geq U_j$ for all $j \in [2:n]$, and $V_v \geq V_j$ for all $j \in [2:n]$. Since from the case description, we have that $U_u \geq V_v$, we should have that,
 \begin{align}\label{eqn:v_1_eq}
     U_u \geq U_j \text{ for all } j \in [2:n] \text{ and } U_u \geq V_j \text{ for all } j \in [2:n]
 \end{align}
\begin{enumerate}
    \item[(a)] Notice that this condition is trivially satisfied for $k \in \{2\}\cup\{u+1,\dots,n\}$ since $\bar{\mu}^*_k = 0$ for such $k$. Hence, we focus on $k \not\in \{2\}\cup\{u+1,\dots,n\}$. First, we prove that $\bar{\mu}^*_1 \geq 0$. Notice that from \eqref{eqn:v_1_eq}, we have that $U_u \geq V_u$. Substituting from \eqref{eqn:def_r_i} and \eqref{eqn:def_S_i}, $U_u \geq V_u$ translates to
    \begin{align}
        \frac{u}{\frac{3}{F_1}+ \sum_{k=2}^u \frac{1}{F_k}} \geq  \frac{u-1}{\sum_{k=1}^u \frac{1}{F_k}}.
    \end{align}
    Simplifying the above, we have that
    \begin{align}
        \frac{u}{\frac{3}{F_1}+ \sum_{j=2}^u \frac{1}{F_j}} \leq \frac{F_1}{2}
    \end{align}
    as desired.
  
    To obtain the result for $3 \leq k \leq u$, notice that we can assume $u\geq 3$. Notice that we are required to prove
    \begin{align}
        F_k \geq \frac{u}{\frac{3}{F_1}+ \sum_{j=2}^u \frac{1}{F_j}}.
    \end{align}
    Since $F_k \geq F_{k+1}$ for $k \geq 3$, it is enough to prove the above for $k=u$. Notice that from \eqref{eqn:v_1_eq}, we have that $U_u \geq U_{u-1}$ (recall that $u \geq 3$). Substituting from \eqref{eqn:def_r_i}, $U_u \geq U_{u-1}$ translates to
    \begin{align}
        \frac{u}{\frac{3}{F_1}+ \sum_{k=2}^u \frac{1}{F_k}} \geq  \frac{u-1}{\frac{3}{F_1}+\sum_{k=2}^{u-1} \frac{1}{F_k}}.
    \end{align}
    Simplifying the above, we have that
    \begin{align}
         F_u \geq \frac{u}{\frac{3}{F_1}+ \sum_{k=2}^u \frac{1}{F_k}}
    \end{align}
    as desired.

    \item[(b)] Substituting from the definition of $\bar{\mu}^*_k$ simplifying yields the result. 

    \item[(c)] If $u = n$, there is nothing to prove. Hence, we can assume $u<n$. Notice that it is enough to prove the result for $k = u+1$, since $F_k \geq F_{k+1}$ for $k \geq 3$. From \eqref{eqn:v_1_eq} we have that $U_u \geq U_{u+1}$ (recall that $u<n$). Substituting from \eqref{eqn:def_r_i},  $U_u \geq U_{u+1}$ translates to
    \begin{align}
        \frac{u}{\frac{3}{F_1}+ \sum_{k=2}^u \frac{1}{F_k}} \geq  \frac{u+1}{\frac{3}{F_1}+\sum_{k=2}^{u+1} \frac{1}{F_k}}.
    \end{align}
    Simplifying the above, we have that
    \begin{align}
         F_{u+1} \leq \frac{u}{\frac{3}{F_1}+ \sum_{k=2}^u \frac{1}{F_k}}
    \end{align}
    as desired. 
\end{enumerate}
\end{proof}
\end{lemma}
\begin{lemma}\label{lemma:uncons_solve_lemma_2}
The vector $\vec{\bar{p}}^*$ defined in \eqref{eqn:r_geq_s_1} solves unconstrained problem with Lagrange multiplier vector $\vec{\bar{\mu}}^*$ defined in \eqref{eqn:mu_2}. In particular, $\vec{\bar{p}}^{*}$ solves
\begin{maxi}
  {}{f_2(\vec{p}) +  \bar{\mu}^*_1 (3p_2F_2-p_1F_1)+\bar{\mu}^*_2(p_1F_1-p_2F_2)+ \sum_{k=3}^n\bar{\mu}^*_k(p_2F_2-p_kF_k)}{}{}
  \label{prob:dual_2}
  \addConstraint{\vec{p}}{ \in \Delta_{n}},
\end{maxi}
where the function $f_2$ is defined in \eqref{eqn:f_2_def}.
\begin{proof}
Noticing from the definition of $\vec{\bar{\mu}}^*$ in \eqref{eqn:mu_2} that $\bar{\mu}_k^* = 0$ for $k>u$, and using the definition of function $f_2$ in \eqref{eqn:f_2_def}, the objective of the above unconstrained problem simplifies as
\begin{align}
    &f_2(\vec{p}) +  \bar{\mu}^*_1 (3p_2F_2-p_1F_1)+\bar{\mu}^*_2(p_1F_1-p_2F_2)+ \sum_{k=3}^n\bar{\mu}^*_k(p_2F_2-p_kF_k) \nonumber\\&= p_1F_1\left(\frac{1}{2} - \bar{\mu}^*_1\right) +p_2F_2\left(\frac{1}{2} +3\bar{\mu}^*_1+\sum_{i=3}^u\bar{\mu}^*_i\right) + \sum_{k=3}^u p_kF_k(1-\bar{\mu}^*_{k}) + \sum_{k=u+1}^np_kF_k, \nonumber\\& = \sum_{i=1}^u p_iC + \sum_{k=u+1}^np_kF_k,
\end{align}
where 
\begin{align}
    C = \frac{u}{\frac{3}{F_1}+\sum_{i=2}^u\frac{1}{F_i}},
\end{align}
and the last inequality follows from Lemma~\ref{lemma:lag_2}-(b). From Lemma~\ref{lemma:lag_2}-(c), we have that $C \geq F_k$ for all $k \in \{u+1,\dots,n\}$. Hence, the optimal solution to the above defined unconstrained problem is any $\vec{p} \in \Delta_{n}$ such that $p_k = 0$ for all $k \in \{u+1,\dots,n\}$. In particular, $\vec{\bar{p}}^*$ given in \eqref{eqn:r_geq_s} is a solution to the unconstrained problem. 
\end{proof}
\end{lemma}

\subsubsection{Finding $\vec{p}^*$}
Finally, we are ready to combine the solutions of (P1-1) and (P1-2) to find $\vec{p}^* \in \arg\max_{\vec{p} \in \Delta_{n}} f^{\text{worst}}(\vec{p})$. Notice that since we solved (P1-1) and (P1-2), we have solved all of the $n^2$ problems $\text{(P1-}i\text{)}$, and $\text{(P1-}(i,j)\text{)}$ for $i,j \in [1:n]$ such that $i \neq j$ defined in \eqref{prob:p1_i} and \eqref{prob:p1_i_j}, respectively. Hence, we can find $\vec{p}^*$ by solving all the above problems and finding the one that gives the highest optimal objective. But, it turns out that it is, in fact, enough to solve $\text{(P1-}1\text{)}$, and $\text{(P1-}(1,2)\text{)}$. To prove this, consider arbitrary $(i,j)$ such that $1 \leq i,j \leq n$ such that $i \neq j$. Define, $\vec{D} \in \mathbb{R}^n$ to be the vector obtained by permuting the entries of $\vec{E}$ such that $D_1 = E_i, D_2 = E_j$, and $D_k \geq D_{k+1}$ for $k \in [3:n-1]$. Notice that due to the solution of (P1-2), the optimal value of (P1-($i,j$)) is given by
\begin{align}
    \gamma^* = \max\left\{ \frac{a-1}{\sum_{k=1}^{a}\frac{1}{D_k}}, \frac{b}{\frac{3}{D_1}+\sum_{k=2}^{b}\frac{1}{D_k}} \Bigg{|} 2 \leq a,b \leq n\right\},
\end{align}
Notice that, 
\begin{align}\label{eqn:gamma_st}
    \max\left\{ \frac{a-1}{\sum_{k=1}^{a}\frac{1}{E_k}}, \frac{b}{\frac{3}{E_1}+\sum_{k=2}^{b}\frac{1}{E_k}} \Bigg{|} a,b \in [2:n]\right\} \geq \gamma^*,
\end{align}
where the inequality follows since $\sum_{k=1}^{a}\frac{1}{E_k} \leq \sum_{k=1}^{a}\frac{1}{D_k}$, and $\frac{3}{E_1}+\sum_{k=2}^{b}\frac{1}{E_k} \leq \frac{3}{D_1}+\sum_{k=2}^{b}\frac{1}{D_k}$ for all $a,b \in [2:n]$. This follows since, $E_k \geq E_{k+1}$ for all $k \in [1:n-1]$. But notice that the left-hand side of \eqref{eqn:gamma_st} is the optimal value of (P1-($1,2$)). Hence, the optimal value of (P1-($1,2$)) is at least as that of (P1-($i,j$)). Hence, it is enough to solve (P1-($1,2$)). With similar reasoning, we can establish that solving (P1-1) suffices. Considering the solutions (P1-($1,2$)) and (P1-1), we have the result.

\subsection{$m=2$, arbitrary $r$}\label{subsec:solving_a0b0_non_singleton} 
The general two-player case can be reduced to a linear program. Again $f^{\text{worst}}$ can be found explicitly in this case. It can be easily seen that 
\begin{align}
    f^{\text{worst}}(\vec{p}) &= \sum_{k =1}^n p_kE_k  -\frac{1}{2}\left(\sum_{j=1}^r\text{max}_{(j)}\{ p_kE_k; 1 \leq k \leq n \}\right),
\end{align}
where $\text{max}_{(j)}$ returns the $j$-th largest element in a set. Consider the following $n \choose r$ linear programs, each indexed by a size $r$ ordered subset  of $[1:n]$ containing distinct elements, where the problem (P-$a_1,a_2,..,a_r$) with $a_k \in [1:n]$ for each $k \in [1:r]$ and $a_k < a_{k+1}$ for $k \in [1:r-1]$, is given by,
\begin{maxi}
  {\vec{p},\gamma}{\sum_{j =1}^n p_jE_j  -\frac{1}{2}\left(\sum_{j =1}^r p_{a_j}E_{a_j}\right)}{}{\text{(P-}a_1,a_2,..,a_r\text{):}}
 \addConstraint{\vec{p}}{\in \Delta_{n,r} }  
 \addConstraint{p_{a_j}E_{a_j}}{\geq \gamma \  \forall 1\leq j\leq r}
 \addConstraint{\gamma}{\geq p_{t}E_{t} \ \forall t \in [1:n]\setminus \{a_1,a_2,..,a_r\}}
\end{maxi}
Notice that the solution of (P2) is the solution of the problem out of the above ${n \choose r}$ problems with the maximum objective value. Hence, solving (P2) amounts to solving $n \choose r$ linear programs. In the below lemma, we prove that it is, in fact, enough to solve $\text{(P-}1,2,..,r\text{)}$
\begin{lemma}\label{lemma:a0b0_1}
The optimal objective value of $\text{(P-}1,2,..,r\text{)}$ is at least the optimal objective value of $\text{(P-}a_1,a_2,..,a_r\text{)}$, where $a_k \in [1:n]$ for each $k \in [1:r]$ and $a_k < a_{k+1}$ for each $k \in [1:r-1]$. 
\begin{proof}
    See Appendix~\ref{app:a0b0_1} 
\end{proof}
\end{lemma}
For $1 \leq a,b \leq n$, define,
\begin{align}
S_{a,b} = \begin{cases}
    \sum_{i=a}^b \frac{1}{E_i} & \text{ if } b \geq a\\
    0 & \text{ otherwise }
\end{cases} ,
\end{align}
Now, we focus on constructing the solution for the problem subjected to two assumptions.
\begin{enumerate}
\item [\textbf{A1}] $E_1,E_2,\dots,E_n$ are distinct real numbers
\item [\textbf{A2}] for all $a \in [1:r]$, and $b \in [r+1,n]$, $E_bS_{a,b}$ is not an integer.
\end{enumerate}
The solution for this case is defined in terms of three functions $h:[0:r-1]\times[r+1:n] \to \mathbb{R}$, $e:[0:r-1]\times[r+1:n] \to \mathbb{R}$, and $g: [r+1:n]\times [r+1:n] \to \mathbb{R}$. Before introducing the three functions, we begin with a few definitions.

\noindent
\textbf{Good triplets and bad triplets: }We call a triplet $(a,b,c) \in [0:r-1] \times [r+1:n] \times [r+1:n]$ a \textit{good-triplet} if $r-a+b-c < E_bS_{a+1,b}$. If the reverse inequality is true, we call $(a,b,c)$ a \textit{bad-triplet} index.

The following lemma introduces certain properties regarding triplets.
\begin{lemma}\label{lemma:heg_def}
Consider the following scenarios.
\begin{enumerate}
\item If $(a,b,c)$ is a \textit{good-triplet} then,
\begin{enumerate}
\item If $b>r+1$, then $(a,b-1,c)$ is a \textit{good-triplet}
\item If $c<n$, then $(a,b,c+1)$ is a \textit{good-triplet}
\item If $a<r-1$, then $(a+1,b,c)$ is a \textit{good-triplet}
\item If $a>0$, and $c<n$, then $(a-1,b,c+1)$ is a \textit{good-triplet}
\end{enumerate}
\item If $(a,b,c)$ is a \textit{bad-triplet} then,
\begin{enumerate}
\item If $b<n$, then $(a,b+1,c)$ is a \textit{bad-triplet}
\item If $c>r+1$, then $(a,b,c-1)$ is a \textit{bad-triplet}
\item If $a>0 $, then $(a-1,b,c)$ is a \textit{bad-triplet}
\item If $a<r-1$, and $c>0$, then $(a+1,b,c-1)$ is a \textit{bad-triplet}
\end{enumerate}
\end{enumerate}
\begin{proof}
    See Appendix~\ref{app:heg_def}
\end{proof}
\end{lemma}
\noindent
\textbf{Function $h$}: From Lemma~\ref{lemma:heg_def}-1-a, 2-a, we have that, for fixed $(a,c) \in [0:r-1]\times [r+1,n]$, either $(a,b,c)$ are \text{good-triplets} for all $b \in [r+1,n]$, $(a,b,c)$ are \text{bad-triplets} for all $b \in [r+1,n]$, or there exists a unique $b \in [r+1,n-1]$ such that $(a,b,c)$ is a \textit{good-triplet} and $(a,b+1,c)$ is a \textit{bad-triplet}. Define $h(a,c) = n$ in the first case, $h(a,c) = r$ in the second case, and $h(a,c) = b$ where $b$ is the unique index in the third case.

\noindent
\textbf{Function $e$}: Similarly, from Lemma~\ref{lemma:heg_def}-1-b, 2-b, we have that, for fixed $(a,b) \in [0:r-1]\times [r+1,n]$, either $(a,b,c)$ are \text{good-triplets} for all $c \in [r+1,n]$, $(a,b,c)$ are \text{bad-triplets} for all $c \in [r+1,n]$, or there exists a unique $c \in [r+2,n]$ such that $(a,b,c)$ is a \textit{good-triplet} and $(a,b,c-1)$ is a \textit{bad-triplet}. Define $e(a,b) =  r+1$ in the first case, $e(a,b) = n+1$ in the second case, and $e(a,b) = c$ where $c$ is the unique index in the third case.

\noindent
\textbf{Function $g$}: Similarly, from Lemma~\ref{lemma:heg_def}-1-c, 2-c, we have that, for fixed $(b,c) \in [r+1,n]\times [r+1,n]$, either $(a,b,c)$ are \text{good-triplets} for all $a \in [0,r-1]$, $(a,b,c)$ are \text{bad-triplets} for all $a \in [0,r-1]$, or there exists a unique $a \in [1,r-1]$ such that $(a,b,c)$ is a \textit{good-triplet} and $(a-1,b,c)$ is a \textit{bad-triplet}. Define $g(b,c) =  0$ in the first case, $g(b,c) = r$ in the second case, and $g(b,c) = a$ where $a$ is the unique index in the third case.
    
Now, we construct the explicit solution using the functions defined above.

\begin{theorem}\label{lemma:a0b0explicit_1}
Assume that we are given the two assumptions \textbf{A1}, and \textbf{A2} are true. Define the three sets $\mathcal{X}_1,\mathcal{X}_2,\mathcal{X}_3$, as
\begin{align}
&\mathcal{X}_1 = \{(a,c) \in [0:r-1] \times [r+1:n] | r<h(a,c)\}
\nonumber\\
&\mathcal{X}_2 = \{(a,b) \in [0:r-1] \times [r+1:n] | b< e(a,b) \leq n\} 
\nonumber\\
&\mathcal{X}_3 = \{(b,c) \in [r+1:n] \times [r+1:n] | b \leq c,  0< g(b,c) \leq r-1 \},
\end{align}
and define the vectors $\vec{p}^{1,a,c}$ for $(a,c) \in \mathcal{X}_1$, $\vec{p}^{2,a,b}$ for $(b,c) \in \mathcal{X}_2$ and $\vec{p}^{3,b,c}$ for $(b,c) \in \mathcal{X}_3$, where,
\begin{enumerate}
\item for $(a,c) \in \mathcal{X}_1$
\begin{align}\label{eqn:p1ac}
p^{1,a,c}_k = \begin{cases}
    1 & \text{ if } 1\leq k \leq a\\
    \frac{r-a+b-c}{E_kS_{a+1,b}}  & \text{ if } a+1\leq k \leq b\\
    1 & \text{ if } b+1\leq k \leq c\\
    0 & \text{ otherwise},
\end{cases}
\end{align}
where $b = \min\{h(a,c),c\}$,
\item  for $(a,b) \in  \mathcal{X}_2$,
\begin{align}\label{eqn:p2ab}
p^{2,a,b}_k = \begin{cases}
    1 & \text{ if } 1\leq k \leq a\\
    \frac{E_b}{E_k}  & \text{ if } a+1\leq k \leq b\\
    1 & \text{ if } b+1\leq k \leq c-1\\
    (r-a)+b-c - E_bS_{a+1,b-1} & \text{ if } k = c\\
    0 & \text{ otherwise},
\end{cases}
\end{align}
where $c = e(a,b)$
\item for $(b,c) \in \mathcal{X}_3$,
\begin{align}\label{eqn:p3bc}
p^{3,b,c}_k = \begin{cases}
    1 & \text{ if } 1\leq k \leq a-1\\
    r-a+b-c - E_bS_{a+1,b-1} & \text{ if } k = a\\
    \frac{E_b}{E_k}  & \text{ if } a+1\leq k \leq b\\
    1 & \text{ if } b+1\leq k \leq c\\
    0 & \text{ otherwise},
\end{cases}
\end{align}
where $a = g(b,c)$.
\end{enumerate}
\begin{enumerate}
\item We have that, 
\begin{enumerate}
    \item $\vec{p}^{1,a,c}$ for all $(a,c) \in \mathcal{X}_1$ are all valid vectors belonging to $\Delta_{n,r}$.
    \item $\vec{p}^{2,a,b}$ for all $(a,b) \in \mathcal{X}_2$ are all valid vectors belonging to $\Delta_{n,r}$.
    \item $\vec{p}^{3,b,c}$ for all $(b,c) \in \mathcal{X}_3$ are all valid vectors belonging to $\Delta_{n,r}$.
\end{enumerate}
where $\vec{p}^{1,a,c},\vec{p}^{2,a,b}$ and $\vec{p}^{3,b,c}$ are defined in \eqref{eqn:p1ac}, \eqref{eqn:p2ab}, and \eqref{eqn:p3bc}, respectively.
\item We have that,
\begin{enumerate}
\item for $(a,c) \in \mathcal{X}_1$, $(\vec{p}^{1,a,c},\gamma)$ is feasible for \text{(P-}1,2,..,r\text{)}, where $\gamma = \frac{\delta}{S_{a+1,b}}$, and $b = \text{min}\{h(a,c),c\}$.
\item for $(a,b) \in \mathcal{X}_2$, $(\vec{p}^{2,a,b},\gamma)$ is feasible for \text{(P-}1,2,..,r\text{)}, where $\gamma = E_b$
\item for $(b,c) \in \mathcal{X}_3$, $(\vec{p}^{3,b,c},\gamma)$ is feasible for \text{(P-}1,2,..,r\text{)}, where $\gamma = E_b$
\item the pair $(\vec{p}^0,\gamma)$, where
\begin{align}\label{eqn:def_vecp0}
p^0_k = \begin{cases}
    1 & \text{ if } 1\leq k \leq r\\
    0 & \text{ otherwise}
\end{cases}
\end{align}
and $\gamma = E_{r+1}$ is feasible for \text{(P-}1,2,..,r\text{)}.
\end{enumerate}

\item The solution to $\text{(P-}1,2,..,r\text{)}$ is the one that produces the maximum objective value out of the elements in the following set
\begin{align}
\mathcal{A} = &\{\vec{p}^0\}\cup\{\vec{p}^{1,a,c}: (a,c) \in \mathcal{X}_1\} \cup \{\vec{p}^{2,a,b}: (a,b) \in \mathcal{X}_2 \}\cup \{\vec{p}^{3,b,c}: (b,c) \in \mathcal{X}_3\},
\end{align}
along with the $\gamma$ values defined in part-2.
\end{enumerate}
\begin{proof}
    See Appendix~\ref{app:a0b0explicit_1}
\end{proof}
\end{theorem}
Fig.~\ref{fig:200} denotes the optimal probabilities found for $n,r = 10,3$, along with $\vec{E}$ given by,
\begin{align}
E_j = \begin{cases}
    7 & \text{ if } j = 1\\
    6.7 & \text{ if } j = 2\\
    1 + \frac{60-3j}{19} & \text{ otherwise } 
\end{cases}
\end{align}
\begin{figure*}
\centering
\input{opt_sol_m2}
\end{figure*} 
In Fig.~\ref{fig:200}, it can be seen that player $A_1$ will always choose a subset of resources with the highest mean rewards, and the probabilities of choosing the remaining resources follow a similar pattern to the $m=3,r=1$ case described in Section~\ref{subsec:solving_a0b0}. The intuition behind this is also very similar to the three-player singleton case.
\section{Simulation Results}
\begin{figure}[ht!]
\centering
\begin{minipage}{0.33\linewidth}
\resizebox {\textwidth} {!} {
\includegraphics[]{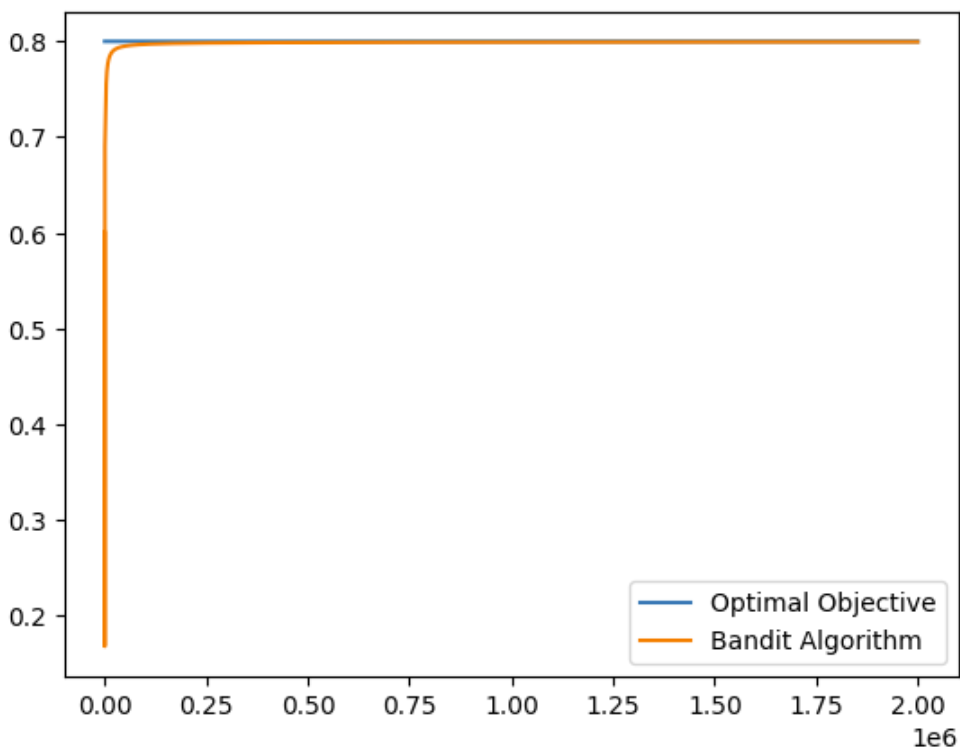}
}
\end{minipage}\hfill
\begin{minipage}{0.33\linewidth}
\resizebox {\textwidth} {!} {
\includegraphics[]{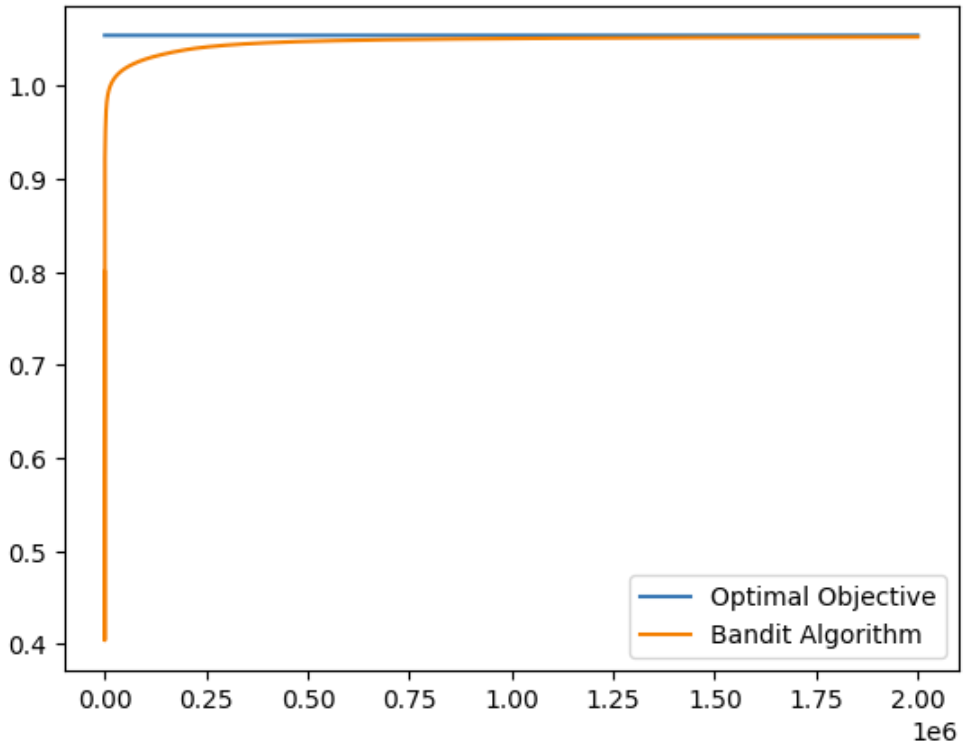}
}
\end{minipage}
\hfill
\begin{minipage}{0.33\linewidth}
\resizebox {\textwidth} {!} {
\includegraphics[]{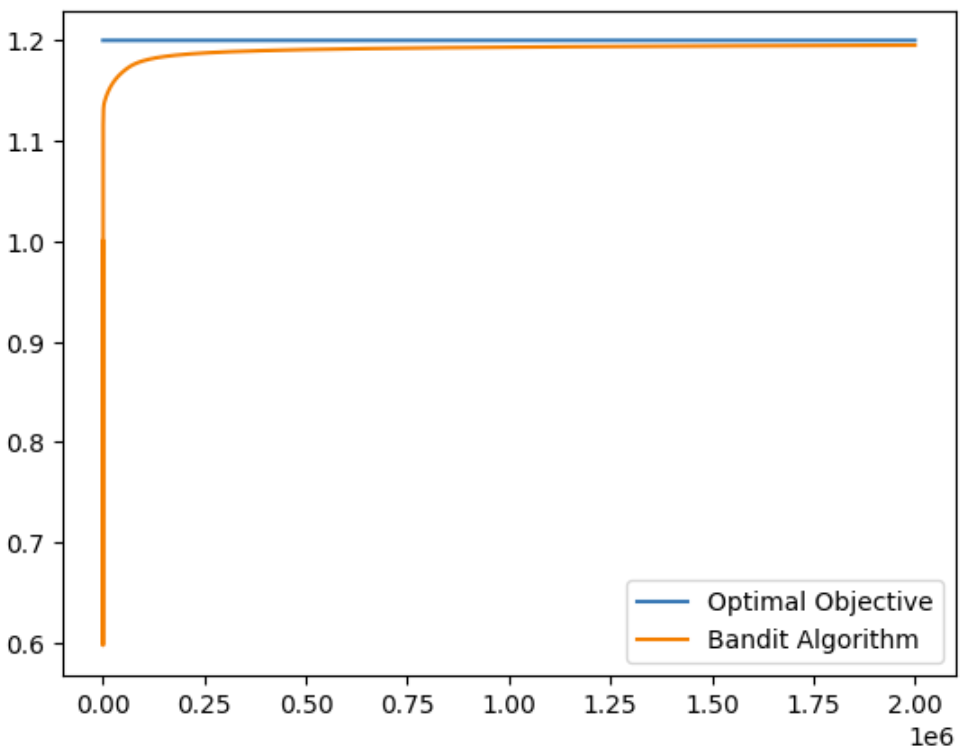}
}
\end{minipage}
\begin{minipage}{0.33\linewidth}
\resizebox {\textwidth} {!} {
\includegraphics[]{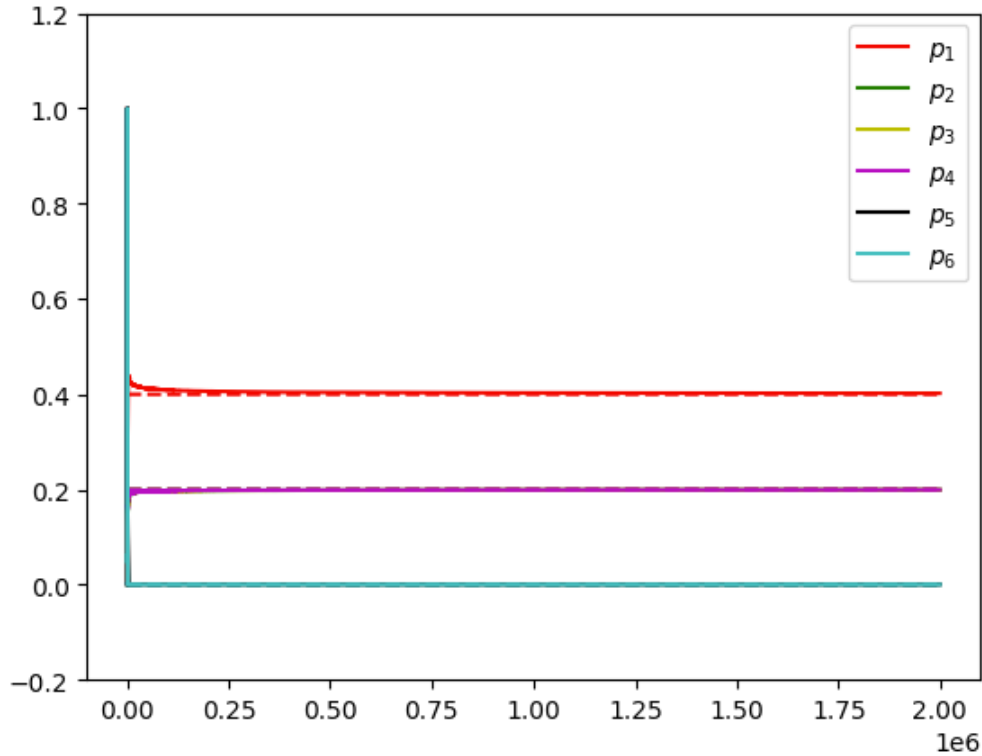}
}
\end{minipage}\hfill
\begin{minipage}{0.33\linewidth}
\resizebox {\textwidth} {!} {
\includegraphics[]{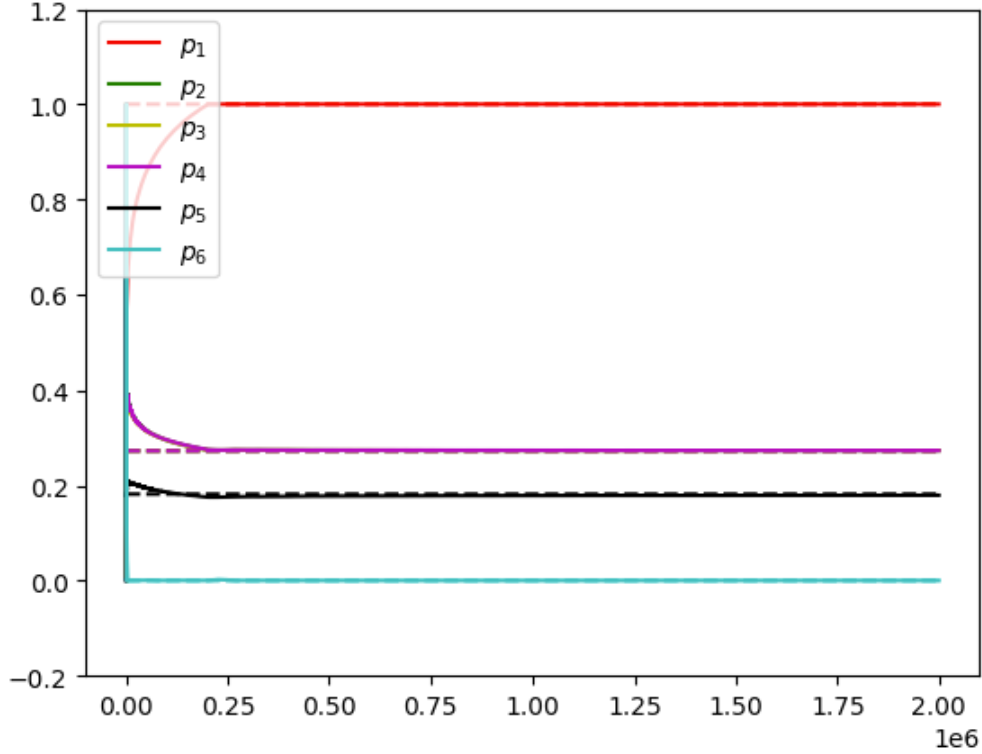}
}
\end{minipage} \hfill
\begin{minipage}{0.33\linewidth}
\resizebox {\textwidth} {!} {
\includegraphics[]{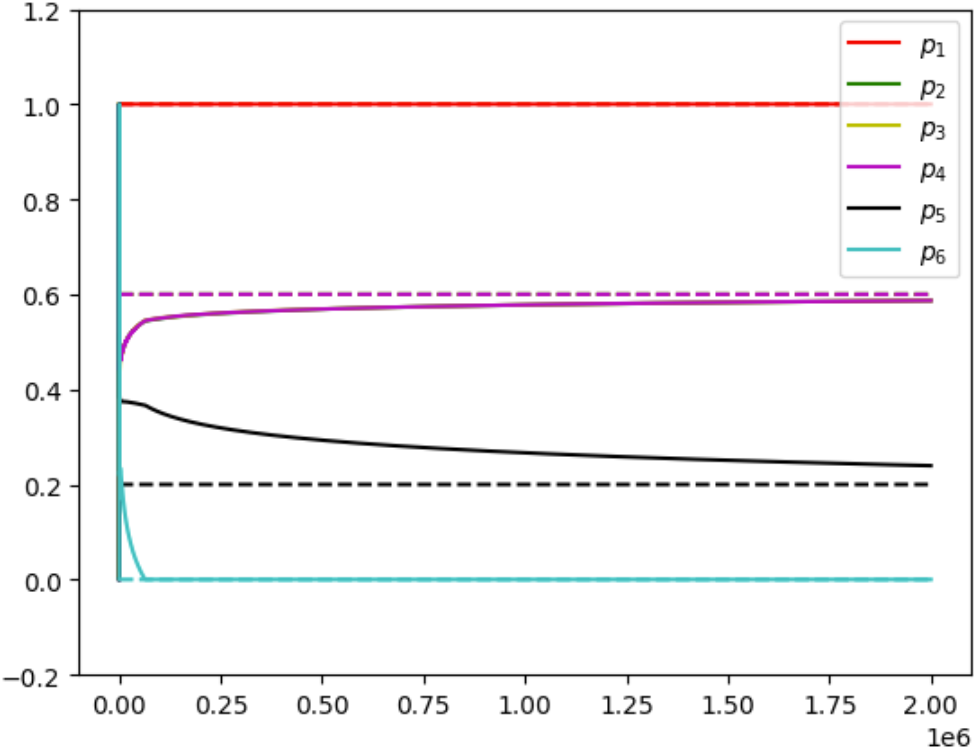}
}
\end{minipage}
\caption{Scenario $\vec{E} =[3,1,1,1,0.5,0.1]$. \textbf{Top:} $\frac{1}{t}\sum_{\tau=1}^tf^{\text{worst}}(\vec{p}(\tau))$ and $f^{\text{worst},*}$ vs $t$, \textbf{Bottom: }Components of $\vec{p}(t)$ and components of $\vec{p}^*$ vs $t$ for , \textbf{Left:} $r = 1$, \textbf{Middle:} $r = 2$, \textbf{Right:} $r = 3$}\label{fig:e_1}
\end{figure}

\begin{figure}[ht!]
\centering
\begin{minipage}{0.33\linewidth}
\resizebox {\textwidth} {!} {
\includegraphics[]{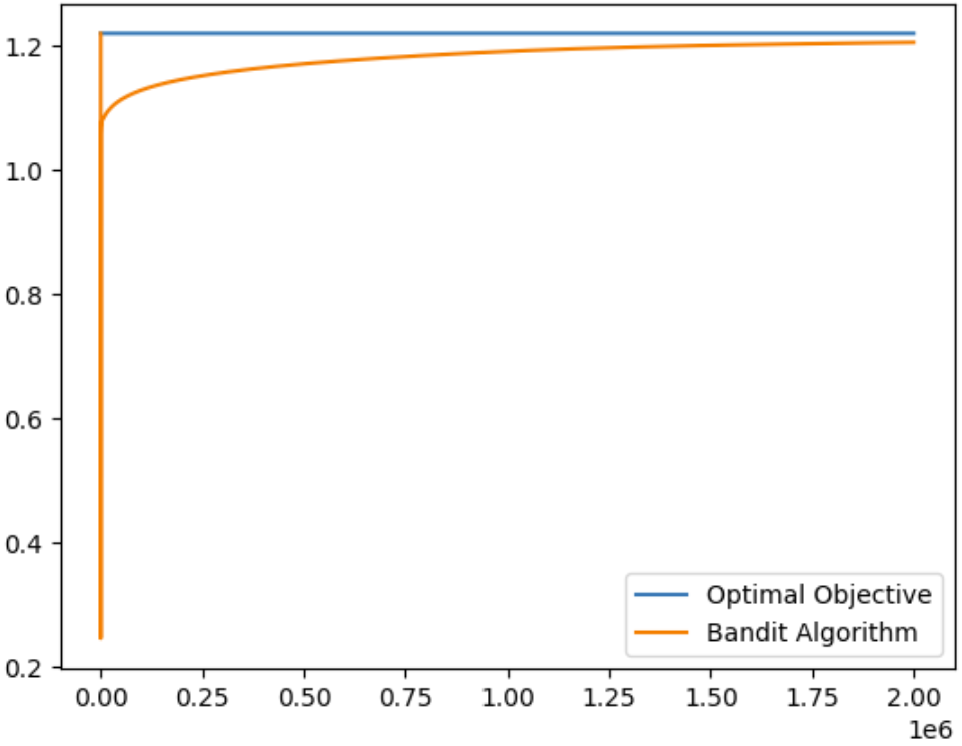}
}
\end{minipage}\hfill
\begin{minipage}{0.33\linewidth}
\resizebox {\textwidth} {!} {
\includegraphics[]{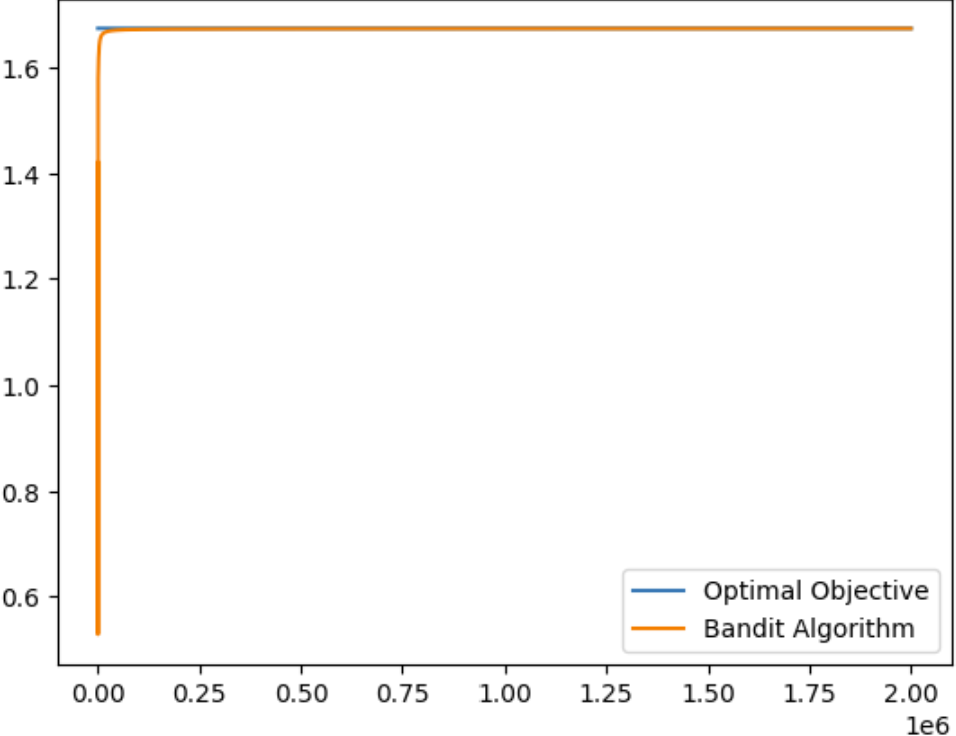}
}
\end{minipage}\hfill
\begin{minipage}{0.33\linewidth}
\resizebox {\textwidth} {!} {
\includegraphics[]{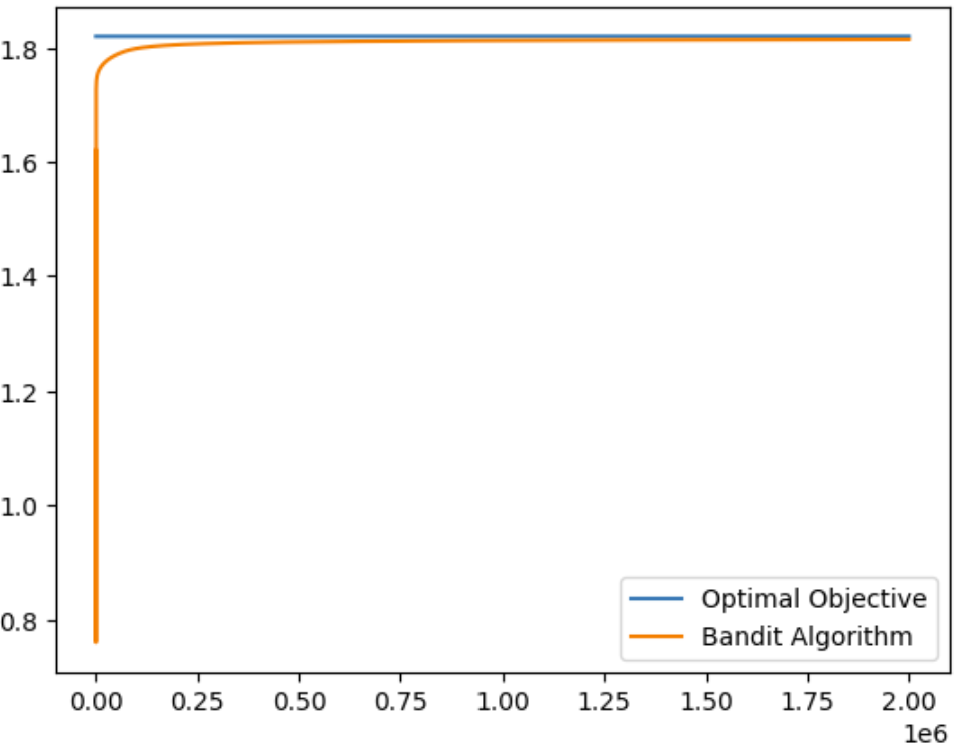}
}
\end{minipage}
\begin{minipage}{0.33\linewidth}
\resizebox {\textwidth} {!} {
\includegraphics[]{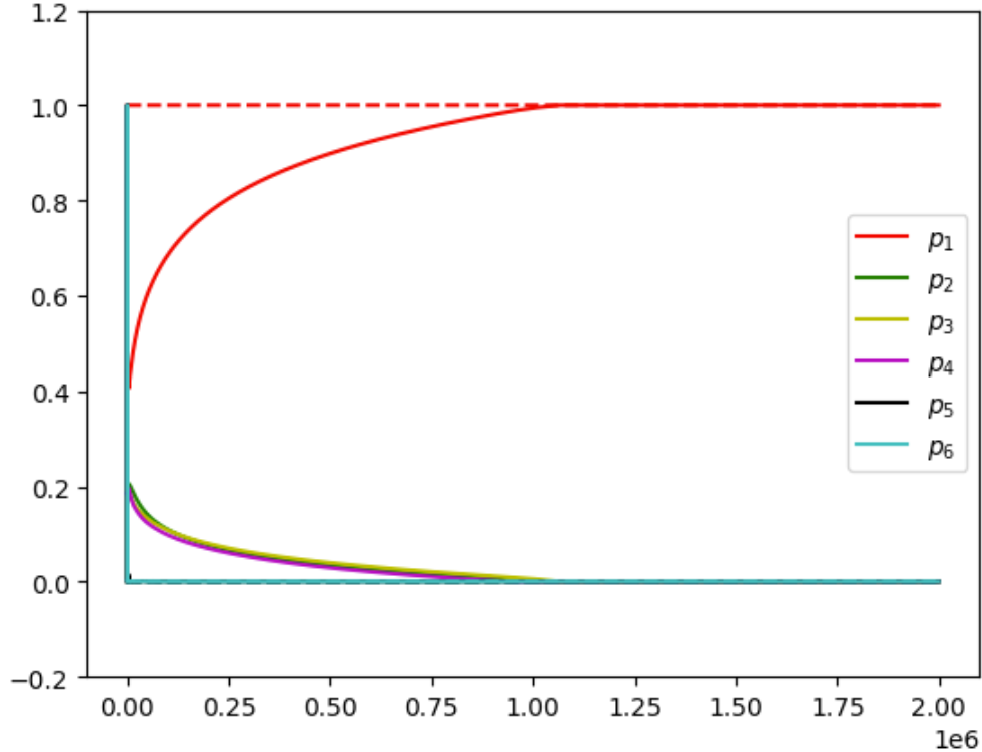}
}
\end{minipage}\hfill
\begin{minipage}{0.33\linewidth}
\resizebox {\textwidth} {!} {
\includegraphics[]{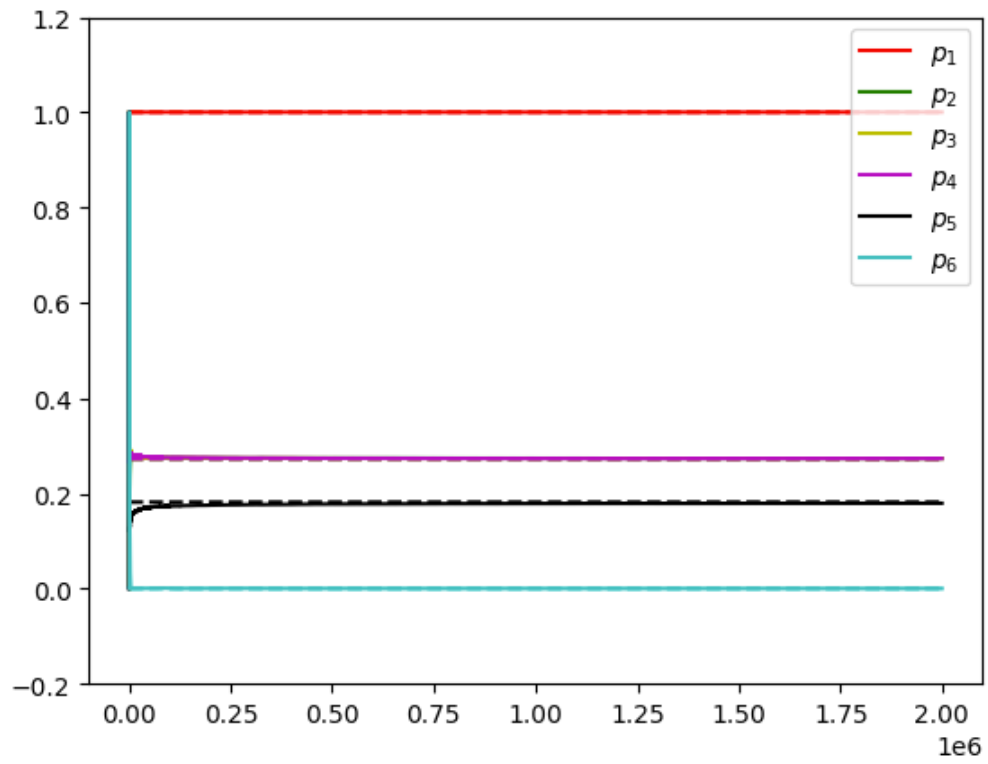}
}
\end{minipage} \hfill
\begin{minipage}{0.33\linewidth}
\resizebox {\textwidth} {!} {
\includegraphics[]{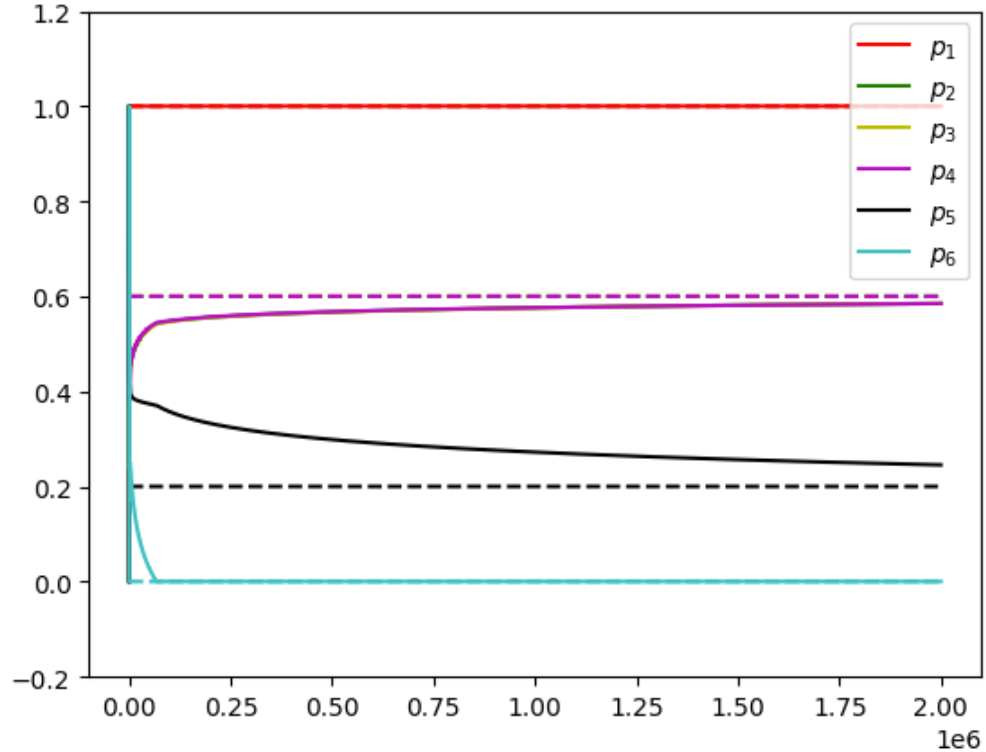}
}
\end{minipage}
\caption{Scenario $\vec{E} = [6.1,1,1,1,0.5,0.1]$.  \textbf{Top:} $\frac{1}{t}\sum_{\tau=1}^tf^{\text{worst}}(\vec{p}(\tau))$ and $f^{\text{worst},*}$ vs $t$, \textbf{Bottom: }Components of $\vec{p}(t)$ and components of $\vec{p}^*$ vs $t$, \textbf{Left:} $r = 1$, \textbf{Middle:} $r = 2$, \textbf{Right:} $r = 3$}\label{fig:e_2}
\end{figure}

In this section we present our simulation results. In Figure~\ref{fig:e_1}, we simulate the performance of our algorithm for $n = 6, m =5, \vec{E} = [3,1,1,1,0.5,0.1]$ for $r \in \{1,2,3\}$. For each value of $r$, we run Algorithm~\ref{algo:0} for $2 \times 10^6$ iterations. We first plot $\frac{1}{t}\sum_{\tau=1}^tf^{\text{worst}}(\vec{p}(\tau))$ vs $t$, after which we plot the entries of $\vec{p}(t)$ vs $t$, where $t$ is the iteration number. In the plots, we also plot the optimal objective value $f^{\text{worst},*}$, and the optimal $\vec{p}^*$ for reference. In Figure~\ref{fig:e_2}, we repeat the above with parameters $n = 6, m =5, \vec{E} = [6.1,1,1,1,0.5,0.1]$ for $r \in \{1,2,3\}$.

Notice that in both cases, we use $\vec{E}$ sorted in decreasing order. Comparing the case $r = 1$ for the two values of $\vec{E}$ it can be seen that when $\vec{E} = [6.1,1,1,1,0.5,0.1]$, player 1 chooses resource 1 with probability 1 while when $\vec{E} = [3,1,1,1,0.5,0.1]$, player 1 chooses several resources with nonzero probability. This is because when $\vec{E} = [6.1,1,1,1,0.5,0.1]$, the mean reward of the first resource is higher than five times the mean reward of the second resource. Hence, even if all the other players choose resource 1, player 1 will not benefit by choosing a different resource. From Figure~\ref{fig:e_1}-Bottom-Left and Figure~\ref{fig:e_2}-Bottom-Left, it can be seen that the online algorithm learns this behavior. However, when $r>1$, player 1 chooses resource 1 with probability 1 for both values of $\vec{E}$. In all cases, it can be seen that the worst-case expected utility of the online algorithm converges to the optimal value.

Another interesting observation is the slower convergence of the algorithm for $r = 1$ with $\vec{E} = [6.1,1,1,1,0.5,0.1]$. This may be due to the fact that this is the only case where the optimal solution $\vec{p}^*$ is an extreme point of $\Delta_{n,r}$ and it chooses all resources except resource 1 with zero probability. In particular, using $\vec{p}(t)$ close to $\vec{p}^*$ in the initial phases of the algorithm reduces exploration required to learn the $E_k$ values.

\section{Conclusions}
In this paper, we considered the problem of worst-case expected utility maximization for the first player of multi-player resource-sharing games with fair reward allocation under two settings. In the first setting, we provided an algorithmic solution to a one-slot game, where we also provided explicit solutions for two special cases. For the second setting, we considered an online scenario, for which we provided an upper confidence bound algorithm that achieves a worst-case regret of $\mathcal{O}(\sqrt{T\log{(T)}})$. The simulations and the explicit solutions depict interesting variations of the probability of choosing a resource when the mean of the considered resource is changed while holding the mean reward of other resources fixed. 
\newpage
\appendices
\section{Finding $\vec{x}^* \in \arg\min_{\vec{x} \in \mathcal{J}} f(\vec{p},\vec{x})$}\label{app:solving_selection}
Given $\vec{F} \in \mathbb{R}_+^n$, and $\vec{p} \in \Delta_{n,r}$, we focus on finding $\vec{x}^* \in \arg\min_{\vec{x} \in \mathcal{J}}\sum_{k=1}^n \frac{p_kF_k }{1+x_k}$. This is an optimization over a nonconvex discrete set $\vec{x} \in \mathcal{J}$. However, it has a classical separable structure that is well studied in the literature and can be solved exactly using either a greedy $\mathcal{O}(n+mr\log(n))$ incremental algorithm or an improved $\mathcal{O}(n\log(mr))$ algorithm. For completeness, we summarize an $\mathcal{O}(nmr)$ algorithm in Algorithm~\ref{algo:0}. For improved algorithms, refer to the work of~\cite{Ibaraki1988}.  
\begin{algorithm}[t]
\label{algo:0}
\SetAlgoLined
\DontPrintSemicolon
Initialize $\vec{x} = [0,0,\dots,0] \in \mathbb{N}^n$.\;
\For{each iteration $k \in [1:(m-1)r]$}{
Increase $x_{i}$ by $1$ where $i \in \arg\min_{\substack{k \in [1:n]\\x_k<m-1}} \left\{\frac{p_kF_k}{1+x_k}-\frac{p_kF_k}{2+x_k}\right\}$.
}
Output  $\vec{x}$.\\
\caption{Algorithm for Appendix~\ref{app:solving_selection}}
\end{algorithm}
\section{Madow's Sampling Technique}\label{app:madows}
In this section, we present the Madow's sampling technique (Algorithm~\ref{alg:madows}). The algorithm takes as an input a vector $\vec{p} \in \Delta_{n,r}$ and outputs a set $\mathcal{A} \subset [1:n]$ such that $|\mathcal{A}| = r$, and $\mathbb{E}\{\mathbbm{1}_{k \in \mathcal{A}}\} = p_k$ for all $k \in [1:n]$. See~\cite{pmlr-v151-mukhopadhyay22a} for the proof of the correctness of the algorithm.
\begin{algorithm}
\SetAlgoLined
\DontPrintSemicolon
Define $\Pi_0 = 0, \text{ and } \Pi_k = \Pi_{k-1}+p_k \ \forall k \in [1:n]$.\;
Sample $U \sim \text{Uniform}(0,1)$.\;
Define the set $\mathcal{S}_0 = \varnothing$, where $\varnothing$ denotes the empty set.\;
\For{each $k \in \{0,1,\dots,r-1\}$}{
Find the unique $i \in [1:n]$ such that $\Pi_{i-1} \leq U+k < \Pi_i$.\;
Define $\mathcal{S}_{k+1} = \mathcal{S}_{k} \cup \{i\}$.\;
}
Output $\mathcal{A} = \mathcal{S}_r$.\;
\caption{Madow's sampling technique}\label{alg:madows}
\end{algorithm}

\section{Proof of Lemma~\ref{lemma:a0b0_1}}\label{app:a0b0_1}
Let $\mathcal{A} =(a_1,a_2,..,a_r)$ be a subset of $[1:n]$ containing distinct elements such that $a_k < a_{k+1}$ for each $k \in [1:r-1]$. Consider the problem $\text{(P-}\mathcal{A}\text{)}$. Let us $\mathcal{B} = [1:n]\setminus \mathcal{A}$. Denote $\mathcal{A}_{\text{bad}} = \mathcal{A}\setminus[1:r]$ as the set of \textit{bad-1} indices and the set, $\mathcal{B}_{\text{bad}} = \mathcal{B} \cap [1:r]$ as the set of \textit{bad-2} indices. Notice that for any given problem, there are an equal number of \textit{bad-1} and \textit{bad-2} indices. We intend to prove that there is an optimal solution with no \textit{bad-1} (or \textit{bad-2}) indices. For this, we establish that for any problem with $k>0$ \textit{bad-1} elements, there exists another problem with $k-1$ \textit{bad-1} indices with an objective value at least as the objective value of the problem with $k$ \textit{bad-1} indices. Assume $\text{(P-}\mathcal{A}\text{)}$ has $k$ \textit{bad-1} indices. Let $(\vec{p},\gamma)$ be the optimal solution of $\text{(P-}a_1,a_2,..,a_r\text{)}$. We consider two cases.

\noindent
\textbf{Case 1:} There is no pair $(a^{'},b^{'})$ such that $a^{'}$ is a \textit{bad-1} index and $b^{'}$ is a \textit{bad-2} such that $E_{a^{'}} <E_{b^{'}}$.

Notice that any \textit{bad-1} index is greater than any \textit{bad-2} index. Hence, for pair $(i,j)$ such that $i$ is a \textit{bad-1} index, and $j$ is a \textit{bad-2} index, we have that $E_j \geq E_i$ (Since $\vec{E}$ is assumed to be decreasing). Hence, the above condition would mean that $E_j = E_i$ for all $i,j$ such that $i$ is \textit{bad-1}, and $j$ is \textit{bad-2}. Hence $(\vec{p},\gamma)$ will be feasible for $\text{(P-}(\mathcal{A}\setminus\{i\} \cup \{j\}\text{))}$ as well. Moreover, $(\vec{p},\gamma)$ will give the same objective value for $\text{(P-}(\mathcal{A}\setminus\{i\} \cup \{j\}\text{))}$ as $\text{(P-}\mathcal{A})$, and $\text{(P-}(\mathcal{A}\setminus\{i\} \cup \{j\}\text{))}$ will have $k-1$ \textit{bad-1} indices.

\noindent
\textbf{Case 2:} There exists a pair $(a^{'},b^{'})$ such that $a^{'}$ is a \textit{bad-1} index and $b^{'}$ is a \textit{bad-2} such that $E_{a^{'}} <E_{b^{'}}$.

We begin with the following two lemmas.
\begin{lemma}\label{lemma:atleatonebad}
There exists $t \in \mathcal{A}$, and $s \in \mathcal{B}$ such that $E_tp_t = E_sp_s =  \gamma$.
\begin{proof}
We only prove the existence of $t \in \mathcal{A}$ such that $E_tp_t = \gamma$. The other part can be solved by repeating the same argument. Assume the contrary. Let $\gamma^* = (\min\{p_jE_j;j \in \mathcal{A}\}+\gamma)/2$. We have that $\gamma^* >\gamma$, and $p_jE_j > \gamma^*$ for all $j \in \mathcal{A}$. Notice that $p_{a^{'}}>0$ and $p_{b^{'}}<1$ (The first inequality follows since $p_{a^{'}}E_{a^{'}} > p_{b^{'}}E_{b^{'}}$, and the second inequality follows since $p_{a^{'}}E_{a^{'}} > p_{b^{'}}E_{b^{'}}$, and  $E_{b^{'}} > E_{a^{'}}$). Hence, there exists $\delta>0$, small enough such that,
\begin{align}
 &E_{a^{'}} (p_{a^{'}} -\delta) \geq \gamma^*\\
 &E_{b^{'}} (p_{b^{'}} +\delta) \leq \gamma^*\\
 &(p_{a^{'}} -\delta) \geq 0\\
 &(p_{b^{'}} +\delta) \leq 1.
\end{align}
Hence $(\vec{\tilde{p}},\gamma^*)$, where $\vec{\tilde{p}}$ is given by,
\begin{align}
\tilde{p}_k = \begin{cases}
 p_k & \text{ if } k \in [1:n]\setminus\{a^{'},b^{'}\}\\
 p_{a^{'}} -\delta & \text{ if } k =a^{'}\\
 p_{b^{'}} +\delta & \text{ if } k =b^{'}
\end{cases}
\end{align}
is feasible for $\text{(P-}a_1,a_2,..,a_r\text{)}$, and also achieves a higher optimal objective value since $E_{b^{'}} > E_{a^{'}}$. This is a contradiction.
\end{proof}
\end{lemma}
\begin{lemma}\label{lemma:bad12equgam}
For $\text{(P-}a_1,a_2,..,a_r\text{)}$, there exists an optimal solution with at least one \textit{bad-1} element $a$ such that, $E_a p_a = \gamma$, and at least one bad \textit{bad-2} element $b$ such that $E_b p_b = \gamma$. 
\begin{proof}
Notice that for all $k \in \mathcal{A} \setminus\mathcal{A}_{\text{bad}}$, and $j \in \mathcal{A}_{\text{bad}}$, we have that $E_k \geq E_j$. 

\noindent
Notice that the entries of $\vec{p}$ can be rearranged without affecting the objective and feasibility for $\text{(P-}a_1,a_2,..,a_r\text{)}$ such that the following two conditions are satisfied. 

\begin{enumerate}
    \item [\textbf{C1}] For $k \in \mathcal{A} \setminus \mathcal{A}_{\text{bad}}$, and $j \in \mathcal{A}_{\text{bad}}$, if we have $E_k = E_j$, then $p_k \geq p_j$.
    \item [\textbf{C2}] For $k \in \mathcal{B} \setminus\mathcal{B}_{\text{bad}}$, and $j \in \mathcal{B}_{\text{bad}}$ if we have $E_k = E_j$, then $p_j \geq p_k$.
\end{enumerate}

Now we establish that any optimal $\vec{p}$ reordered such that both \textbf{C1} and \textbf{C2} are met satisfy the conditions of the lemma. We show only the \textit{bad-1} case. The \textit{bad-2} case can be solved using the same argument.

Assume the contrary. Hence, all \textit{bad-1} elements $j$ satisfy $E_j p_j > \gamma$.  Consider $t \in \mathcal{A}$ such that $E_tp_t = \gamma$ (Such a $t$ always exists from Lemma~\ref{lemma:atleatonebad}). Notice that $t$ cannot be \textit{bad-1}. Hence $E_t \geq E_j$ for all \textit{bad-1} indices $j$. In this case, we have the following claim.

\noindent
\textbf{Claim}: There exists a \textit{bad-1} index $i$ such that $E_i < E_t$.
\begin{proof}
    If no such \textit{bad-1} index $i$ exists, then we should have $E_t = E_j$ for all \textit{bad-1} indices $j$. From \textbf{C1}, this would imply that $p_t \geq p_j$ for all \textit{bad-1} indices $j$. Hence, we should have $E_tp_t \geq E_jp_j>\gamma$, which contradicts $E_tp_t = \gamma$.
\end{proof}
 
Consider the $i$ described in the Claim. Since $E_tp_t = \gamma < E_ip_i$, and $E_t > E_i$, we have that, $p_t<p_i \leq 1$. Also we have that $p_i > 0$ since $E_ip_i>\gamma$. Hence, it is possible to find $\delta>0$, small enough such that,
 \begin{align}
     &E_i (p_i -\delta) \geq \gamma\\
     &(p_i -\delta) \geq 0\\
     &(p_t +\delta) \leq 1.
 \end{align}
 Since $E_t > E_i$ it is easy to see that, $(\vec{\tilde{p}},\gamma)$ given by,
 \begin{align}
     \tilde{p}_k = \begin{cases}
         p_k & \text{ if } k \in [1:n]\setminus\{i,t\}\\
         p_i -\delta & \text{ if } k =i\\
         p_t +\delta & \text{ if } k =t
     \end{cases}
 \end{align}
 is a better solution to $\text{(P-}a_1,a_2,..,a_r\text{)}$. This is a contradiction.
 \end{proof}
 \end{lemma}
 Now, let $a$, $b$ be the indices such that $a$ is \textit{bad-1} and $E_ap_a = \gamma$, and $b$ is \textit{bad-2} and $E_bp_b = \gamma$, which are guaranteed to exists due to Lemma~\ref{lemma:bad12equgam}. Consider the problem,
 $\text{(P-}(\mathcal{A}\setminus\{a\}) \cup\{b\}\text{)}$, which has $k-1$ \textit{bad-1} elements. Notice that since $E_ap_a = E_bp_b = \gamma$, we have that $(\vec{p},\gamma)$ is feasible for $\text{(P-}(\mathcal{A}\setminus\{a\}) \cup\{b\}\text{)}$. Also, the objective values of $\text{(P-}(\mathcal{A}\setminus\{a\}) \cup\{b\}\text{)}$ and $\text{(P-}a_1,a_2,..,a_r\text{)}$ evaluated at $(\vec{p},\gamma)$ are equal. Hence, we are done.
\section{Proof of Lemma~\ref{lemma:heg_def}}\label{app:heg_def}
Let $\delta = r-a+b-c$.
\begin{enumerate}
\item Recall that $(a,b,c)$ being a \textit{good-triplet} is equivalent to,
\begin{align}\label{eqn:pref_1}
    \delta < E_bS_{a+1,b}
\end{align}
\begin{enumerate}
\item Notice that,
\begin{align}
    r-a+b-1-c = \delta - 1 <_{(a)} E_bS_{a+1,b}-1 = E_{b}S_{a+1,b-1} \leq E_{b-1}S_{a+1,b-1},
\end{align}
where (a) follows from \eqref{eqn:pref_1}, and the last inequality follows since $E_{b-1} \geq E_b$.
\item Notice that,
\begin{align}
    r-a+b-(c+1) = \delta - 1 <_{(a)} E_bS_{a+1,b}-1 < E_{b}S_{a+1,b}
\end{align}
where (a) follows from \eqref{eqn:pref_1}
\item Notice that,
\begin{align}
    r-(a+1)+b-c = \delta - 1 <_{(a)} E_bS_{a+1,b}-1 = E_{b}S_{a+1,b-1}< E_{b}S_{a+2,b}.
\end{align}
where (a) follows from \eqref{eqn:pref_1} and the last inequality follows since $S_{a+2,b} \geq S_{a+1,b-1}$, which follows since $\vec{E}$ is non-increasing in it's components.
\item Notice that,
\begin{align}
    r-(a-1)+b-(c+1) = \delta  <_{(a)} E_bS_{a+1,b} < E_{b}S_{a,b},
\end{align}
where (a) follows from \eqref{eqn:pref_1}.
\end{enumerate}
\item All the claims in this part follow from the contra-positives of the corresponding claims in part 1. 
\end{enumerate}
\section{Proof of Theorem~\ref{lemma:a0b0explicit_1}}\label{app:a0b0explicit_1}
\noindent
1) 

\noindent
a) Recall that, $\mathcal{X}_1 = \{(a,c) \in [0:r-1] \times [r+1:n] | r<h(a,c)\}$, and $\vec{p}^{1,a,c}$ for $(a,c) \in \mathcal{X}_1$ is defined as,
\begin{align}\label{eqn:p1ac_1}
p^{1,a,c}_k = \begin{cases}
    1 & \text{ if } 1\leq k \leq a\\
    \frac{r-a+b-c}{E_kS_{a+1,b}}  & \text{ if } a+1\leq k \leq b\\
    1 & \text{ if } b+1\leq k \leq c\\
    0 & \text{ otherwise},
\end{cases}
\end{align}
where $b = \min\{h(a,c),c\}$. We first prove that $\vec{p}^{1,a,c}$, is a valid vector in $\Delta_{n,r}$.

Since $(a,c) \in \mathcal{X}_1$, we have that $h(a,c) >r$ and $c>r$, which implies that,
\begin{align}\label{eqn:b_geq_r}
    0 \leq a < r<b \leq  c \leq n. 
\end{align} 
Notice that since $h(a,c) >r$, from the definition of $h$, we have that $(a,h(a,c),c)$ is a \textit{good-triplet}. Since $b \leq h(a,c)$, combining with Lemma~\ref{lemma:heg_def}-1-a, we have that, 
\begin{align}\label{eqn:good_c1}
    (a,b,c)  \text{ is a \textit{good-triplet}}.
\end{align}
This means that,
\begin{align}\label{eqn:10101011}
     r-a+b-c < E_bS_{a+1,b}.
\end{align}
Also, notice that if $b < c$, then we should have $b = h(a,c)$ and $b<n$, which implies from the definition of $h$ that $(a,b+1,c)$ is a \textit{bad-triplet}. Hence,
\begin{align}\label{eqn:bad_c1}
    \text{if } b<c, (a,b+1,c) \text{ is a \textit{bad-triplet}}.
\end{align}
Hence, if $b<c$ we have that,
\begin{align}\label{eqn:11101011}
    E_{b+1}S_{a+1,b} < r-a+b-c,
\end{align}
where the inequality is strict due to assumption \textbf{A2}. Since $(a,c) \in \mathcal{X}_1$, we have that $a+1 \leq r < b \leq c$, which implies that \eqref{eqn:p1ac_1} is a valid definition. Now we check the conditions for $\vec{p}^{1,a,c} \in \Delta_{n,r}$. The sum constraint can be checked by direct substitution. The constraint, $0 \leq p^{1,a,c}_k \leq 1$ follows trivially for $k \not\in [a+1,b]$. For $k \in [a+1,b]$, the constraint $0 \leq p^{1,a,c}_k$ holds if and only if $r-a+b-c>0$. Notice that this holds whenever $b<c$ due to \eqref{eqn:11101011}. If $b = c$, the above reduces to $r-a>0$, which holds since $a<r$ by the definition of $\mathcal{X}_1$. Hence, we have,
\begin{align}\label{eqn:del_geq_0}
    \delta \geq 0
\end{align}

Now, to establish that $p^{1,a,c}_k \leq 1$, we have 
\begin{align}
    p^{1,a,c}_k = \frac{\delta}{E_kS_{a+1,b}} \leq \frac{\delta}{E_bS_{a+1,b}} \leq 1,
\end{align}
where the last inequality follows due to \eqref{eqn:10101011}. 

\noindent
b) Recall that, $\mathcal{X}_2 = \{(a,b) \in [0:r-1] \times [r+1:n] | b< e(a,b) \leq n\}$, and $\vec{p}^{2,a,b}$ for $(a,b) \in \mathcal{X}_2$ is defined as,
\begin{align}\label{eqn:p2ab_1}
p^{2,a,b}_k = \begin{cases}
    1 & \text{ if } 1\leq k \leq a\\
    \frac{E_b}{E_k}  & \text{ if } a+1\leq k \leq b\\
    1 & \text{ if } b+1\leq k \leq c-1\\
    (r-a)+b-c - E_bS_{a+1,b-1} & \text{ if } k = c\\
    0 & \text{ otherwise},
\end{cases}
\end{align}
where $c = e(a,b)$. Now, we prove that $\vec{p}^{1,a,b}$, is a valid vector in $\Delta_{n,r}$. Since $(a,b) \in \mathcal{X}_2$, we have that,
\begin{align}\label{eqn:c_ge_b}
    n \geq e(a,b) = c >b > r > a \geq 0
\end{align} 
Notice that since the definition of function $e$, and the fact that $e(a,b) > b \geq r+1$, we have that, 
\begin{align}\label{eqn:good_b1}
    (a,b,c)  \text{ is a \textit{good-triplet}},
\end{align}
and 
\begin{align}\label{eqn:bad_b1}
  (a,b,c-1) \text{ is a \textit{bad-triplet}}.
\end{align}
This means that,
\begin{align}\label{eqn:100001}
    E_bS_{a+1,b}-1 < r-a+b-c < E_bS_{a+1,b},
\end{align}
where the first inequality is strict due to assumption \textbf{A2}. Notice that, $a+1 \leq r < b < c \leq n$. Hence, $\vec{p}^{2,a,b}$ defined in \eqref{eqn:p2ab_1} is a valid definition. Now we check the conditions for $\vec{p}^{2,a,b} \in \Delta_{n,r}$ The sum constraint can be checked using direct substitution. Since for $k \in [a+1,b]$ we have,
\begin{align}
    p^{2,a,b}_k = \frac{E_b}{E_k} \leq \frac{E_b}{E_b} = 1,
\end{align}
the constraint, $0 \leq p^{2,a,b}_k \leq 1$ follows trivially for $k \neq c$.

For $k = c$, notice that,
\begin{align}
    p^{2,a,b}_c = \delta - E_bS_{a+1,b-1} =  \delta +1 - E_bS_{a+1,b}\in [0,1],
\end{align}
where last inequality follows from \eqref{eqn:100001}.

\noindent
c) Recall that, $\mathcal{X}_3 = \{(b,c) \in [r+1:n] \times [r+1:n] | b \leq c,  0< g(b,c) \leq r-1 \}$, and $\vec{p}^{3,b,c}$ for $(b,c) \in \mathcal{X}_3$ is defined as,
\begin{align}\label{eqn:p3bc_1}
p^{3,b,c}_k = \begin{cases}
    1 & \text{ if } 1\leq k \leq a-1\\
    r-a+b-c - E_bS_{a+1,b-1} & \text{ if } k = a\\
    \frac{E_b}{E_k}  & \text{ if } a+1\leq k \leq b\\
    1 & \text{ if } b+1\leq k \leq c\\
    0 & \text{ otherwise},
\end{cases}
\end{align}
where $a = g(b,c)$. Since $(b,c) \in \mathcal{X}_3$, we have that,
\begin{align}\label{eqn:a_ge_zer}
    0<a <r< b \leq c \leq n.
\end{align} 
Notice that since the definition of $g$, and the fact that $g(b,c) > 0$, we have that, 
\begin{align}\label{eqn:good_a1}
    (a,b,c)  \text{ is a \textit{good-triplet}},
\end{align}
and 
\begin{align}\label{eqn:bad_a1}
  (a-1,b,c) \text{ is a \textit{bad-triplet}}.
\end{align}
This means that,
\begin{align}\label{eqn:1000011}
    E_bS_{a,b}-1 < r-a+b-c < E_bS_{a+1,b},
\end{align}
where the first inequality is strict due to assumption \textbf{A2}.

Notice that the definition of $\vec{p}^{3,b,c}$ in \eqref{eqn:p3bc_1} is a valid since $0< a < r < b \leq c \leq n$. Now we check the conditions for $\vec{p}^{3,b,c} \in \Delta_{n,r}$. The sum constraint can be checked using direct substitution. Due to the same argument as case 2, in this case, the constraint, $0 \leq p^{3,b,c}_k \leq 1$ follows trivially for $k \neq a$. For $k = a$, notice that,
\begin{align}
    p^{3,b,c}_a = \delta - E_bS_{a+1,b-1} =  \delta +1 - E_bS_{a+1,b} \leq 1,
\end{align}
where the last inequality follows from~\eqref{eqn:1000011}.

\noindent
2) 

\begin{enumerate}
\item[a)]  For $k \in [1:a]$ we have that, $p^{1,a,c}_kE_k = E_k \geq p_{a+1}E_{a+1} = \gamma$. For $k \in [a+1:b]$ we have that, $p^{1,a,c}_kE_k = \gamma$. Finally, for $k \in [b+1:c]$, we can assume that $b < c$, in which case we have that $p^{1,a,c}_kE_k = E_k \leq E_{b+1} \leq \gamma$, where the last inequality follows from \eqref{eqn:10101011}.
\item[b)] For $k \in [1:a]$ we have that, $p^{2,a,b}_kE_k = E_k \geq E_b = \gamma$. For $k \in [a+1:b-1]$ we have that, $p^{2,a,b}_kE_k = E_b =\gamma$. For $k \in [b+1:n]$, we have that $p^{2,a,b}_kE_k  \leq  E_k \leq E_b = \gamma$. For $k = b$, we have that, $p^{2,a,b}_bE_b  \leq  E_b = \gamma$.
\item[c)] For $k \in [1:a-1]$ we have that, $p^{3,b,c}_kE_k = E_k \geq E_b = \gamma$. For $k \in [a+1:b]$ we have that, $p^{3,b,c}_kE_k = E_b =\gamma$. For $k \in [b+1:n]$, we have that $p^{3,b,c}_kE_k  \leq  E_k \leq E_b = \gamma$. For $k = a$, we have that,
 \begin{align}
    p^{3,b,c}_bE_a -E_b &= E_a(r-a+b-c-E_bS_{a+1,b-1})-E_b \nonumber\\&= E_a(r-a+b-c - E_bS_{a,b-1}) \geq 0,
\end{align}
where the last inequality follows due to \eqref{eqn:1000011}.
\item[d)] This follows trivially, by substitution, due to the non-increasing property of $\vec{E}$.
\end{enumerate}

\noindent
3) Define the three sets,
\begin{align}
 &\mathcal{A}_1 = \{\vec{p}^{1,a,c}: (a,c) \in \mathcal{X}_1\},\ 
 \mathcal{A}_2 = \{\vec{p}^{2,a,b}: (a,b) \in  \mathcal{X}_2\},\ 
 \mathcal{A}_3 =  \{\vec{p}^{3,b,c}: (b,c) \in \mathcal{X}_3\},\nonumber\\& 
 \mathcal{A} =\{\vec{p}^0\} \cup \mathcal{A}_1 \cup \mathcal{A}_2 \cup \mathcal{A}_3
\end{align}
Let us denote by $z(\vec{q})$ the objective value of $\text{(P-}1,2,..,r\text{)}$ for $\vec{q} \in \Delta_{n,r}$. We solve the problem under four cases. The four cases can be summarized as,
\begin{enumerate}
\item[\textbf{C1}] Best vector in $\mathcal{A}$ comes from $\mathcal{A}_1$
\item[\textbf{C2}] Best vector in $\mathcal{A}$ comes from $\mathcal{A}_2$
\item[\textbf{C3}] Best vector in $\mathcal{A}$ comes from $\mathcal{A}_3$
\item[\textbf{C4}] Best vector in $\mathcal{A}$ is $\vec{p}^0$, where $\vec{p}^0$ is defined in \eqref{eqn:def_vecp0}.
\end{enumerate}
In each of the above cases, we focus on constructing a Lagrange multiplier vector $\vec{\mu} \in \mathbb{R}^n$ that will establish the best vector is optimal from Lagrange Multiplier Lemma (Lemma~\ref{lemma:lagrange_lemma}).

\noindent
\textbf{Case 1: }Best vector in $\mathcal{A}$ comes from $\mathcal{A}_1$

Let $\vec{p}^{1,a,c}$ denote the best vector where $(a,c) \in \mathcal{X}_1$ (See the definition in \eqref{eqn:p1ac_1}). Define, $b = \min\{h(a,c),c\}$.  
\begin{align}\label{eqn:gamma_def}
    \theta = \frac{r-a}{2}+b-r
\end{align}
and $\delta = r-a+b-c$. Hence, we have,
\begin{align}\label{eqn:z1ac}
    z(\vec{p}^{1,a,c}) = \sum_{i=1}^a \frac{E_i}{2} + \frac{\theta \delta}{S_{a+1,b}} + \sum_{i=b+1}^cE_i.
\end{align}
We introduce the following lemma, which will be useful in handling this case.
\begin{lemma}\label{lemma:comes_from_A1}
We have that,
\begin{enumerate}
\item $\frac{\theta}{S_{a+1,b}} \leq E_c$
\item $\frac{E_{a+1}}{2} \leq  \frac{\theta}{S_{a+1,b}}$
\item If $a>0$, we have, $\frac{E_{a}}{2} \geq \frac{\theta}{S_{a+1,b}}$
\item If $c<n$, we have, $E_{c+1} \leq \frac{\theta}{S_{a+1,b}}$,
\end{enumerate}
where $\theta$ is defined in \eqref{eqn:gamma_def}.
\begin{proof}

\noindent
1) We prove this part in several cases. The cases make sense since $c \geq b \geq r+1$ from \eqref{eqn:b_geq_r},

\noindent
\textbf{Case 1} $c = r+1$: Combining $b\leq c$ and \eqref{eqn:b_geq_r}, we should have $b = r+1$. We are required to prove that $E_{r+1}S_{a+1,r+1} \geq \frac{r-a}{2}+1$. Notice that in this case, \eqref{eqn:10101011} simplifies to $r-a< E_{r+1}S_{a+1,r+1}$. Hence, we are done if $r-a \geq 2$. Hence, the only case to check is $a = r-1$. In this case, the required statement simplifies to  $E_r \leq 2E_{r+1}$, which follows from $z(\vec{p}^0) \leq z(\vec{p}^{1,a,c})$, where $\vec{p}^0$ is defined in \eqref{eqn:def_vecp0}.

\noindent
\textbf{Case 2 } $c > r+1$, $b = c$, and $(a,c-1,c-1)$ is a \textit{good-triplet}: From \eqref{eqn:b_geq_r} and $c-1\geq r+1$, we have that $(a,c-1)$ belongs to the domain of function $h$. Since $(a,c-1,c-1)$ is a \textit{good-triplet}, from the definition of function $h$, we have that $h(a,c-1) \geq c-1$. Since, $c-1 \geq r+1$, we have that $(a,c-1) \in \mathcal{X}_1$, and $\min\{h(a,c-1),c-1\} = c-1$. Hence,
\begin{align}
    z(\vec{p}^{1,a,c-1}) = \sum_{i=1}^a \frac{E_i}{2} + \frac{(\theta-1)\delta}{S_{a+1,c-1}}
\end{align}
Simplifying $z(\vec{p}^{1,a,c-1}) \leq z(\vec{p}^{1,a,c})$ we have the result.

\noindent
\textbf{Case 3 } $c > r+1$, $b = c$, and $(a,c-1,c-1)$ is a \textit{bad-triplet}: From \eqref{eqn:b_geq_r} and $c-1\geq r+1$, we have that $(a,c-1)$ belongs to the domain of function $e$. Combining \eqref{eqn:good_c1} with Lemma~\ref{lemma:heg_def}-1-a, we have that, $(a,c-1,c)$ is a \text{good-triplet}. Combining this with the case description, we have that $e(a,c-1) = c$. Since $c-1 < c \leq n$, where the last inequality follows from \eqref{eqn:b_geq_r}, we have that $(a,c-1) \in \mathcal{X}_2$. Notice that,
\begin{align}
    z(\vec{p}^{2,a,c-1}) = \sum_{i=1}^{a}\frac{E_i}{2} + E_{c-1}\left(\theta - 1\right)+E_c(\delta -E_{c-1}S_{a+1,c-1})
\end{align}
Substituting for  $z(\vec{p}^{2,a,c-1}) \leq z(\vec{p}^{1,a,c})$, we get,
\begin{align}
    (E_cS_{a+1,c} - \theta)(E_{c-1}S_{a+1,c}-\delta)>0.
\end{align}
Since $E_{c-1}S_{a+1,c} \geq E_cS_{a+1,c} > \delta$, where the last inequality follows from \eqref{eqn:10101011}, we are done.

 \noindent
\textbf{Case 4 } $c > r+1$, $b < c$, and $(a,b,c-1)$ is a \textit{good-triplet}: From \eqref{eqn:b_geq_r} and $c-1\geq r+1$, we have that $(a,c-1)$ belongs to the domain of function $h$. Since $b<c \leq n$, from \eqref{eqn:bad_c1}, we have that $(a,b+1,c)$ is a \textit{bad-triplet}. Combining with $c>r+1$, from Lemma~\ref{lemma:heg_def}-2-b, we have that $(a,b+1,c-1)$ is a \textit{bad-triplet}. Since $(a,b,c-1)$ is a \textit{good-triplet}, we have that $h(a,c-1) = b$. Since $c-1 \geq r+1$, we have that $(a,c-1) \in \mathcal{X}_1$. Also, $\min\{h(a,c-1),c-1\} = c-1$, since $b \leq c-1$. Hence, 
\begin{align}
    z(\vec{p}^{1,a,c-1}) = \sum_{i=1}^a \frac{E_i}{2} + \frac{\theta(\delta+1)}{S_{a+1,b}} + \sum_{i=b+1}^{c-1}
\end{align}
Substituting to $z(\vec{p}^{1,a,c-1}) \leq z(\vec{p}^{1,a,c})$ and simplifying, we get the desired result.

\noindent
\textbf{Case 5 }$c > r+1$, and $b <c$, $(a,b,c-1)$ is a \textit{bad-triplet}: From \eqref{eqn:b_geq_r}, we have that $(a,b)$ belongs to the domain of $e$. Combining $(a,b,c-1)$ is a \textit{bad-triplet} with \eqref{eqn:good_c1}, we have that $e(a,b) = c$. Since $b<c \leq n$, where the last inequality follows from \eqref{eqn:b_geq_r}, we have that $(a,b) \in \mathcal{X}_2$. Hence,
\begin{align}
    z(\vec{p}^{2,a,b}) = \sum_{i=1}^a \frac{E_i}{2} + E_b\theta + \sum_{i=b+1}^{c-1}E_i+E_c(\delta-E_bS_{a+1,b-1}).
\end{align}
Substituting for $z(\vec{p}^{2,a,b}) \leq z(\vec{p}^{1,a,c})$ and simplifying yields,
\begin{align}
    (E_cS_{a+1,b} - \theta)(E_bS_{a+1,b}-\delta) > 0
\end{align}
Combining with \eqref{eqn:10101011}, we have the desired result.

\noindent
2) We consider four cases. The cases make sense since $a \leq r-1$, $b \leq c$ from \eqref{eqn:b_geq_r}.

\noindent
\textbf{Case 1 } $a = r-1$: Notice that from \eqref{eqn:del_geq_0}, in this case we should have $c-b \in \{0,1\}$. Also, since $z(\vec{p}^0) \leq z(\vec{p}^{1,a,c})$, we have,
\begin{align}
    \frac{E_r}{2} &\leq \frac{\theta \delta}{S_{r,b}} + \sum_{i=b+1}^cE_i  =  \frac{\theta (1+b-c)}{S_{r,b}}  + \sum_{i=b+1}^cE_i \leq \frac{\theta}{S_{r,b}}+ (c-b)\left(E_{b+1} - \frac{\theta}{S_{r,b}}\right) 
\end{align}
Now, notice that if $c - b = 0$, we have the desired result. If $c-b = 1$, we have,
\begin{align}\label{eqn:er2_ebp1}
    \frac{E_r}{2} \leq E_{b+1}.
\end{align}
Hence,
\begin{align}
    \frac{E_r}{2}S_{r,b} = \frac{1}{2}+ \frac{E_r}{2}\sum_{i=r+1}^b\frac{1}{E_i} \leq_{(a)} \frac{1}{2}+ \frac{E_r}{2}\left(\frac{b-r}{E_{b+1}}\right) \leq_{(b)}\frac{1}{2}+b-r = \theta.
\end{align}
where (a) follows since, $E_i \geq E_{b+1}$ for $i \in [r+1:b]$, and (b) follows from \eqref{eqn:er2_ebp1}.

\noindent
\textbf{Case 2} $a < r-1$ and $b=c$: 

\noindent
\textbf{Case 3 } $a<r-1$, $c>b$, $(a+1,b+1,c)$ is a \textit{bad-triplet}: We handle the above two cases together. In both the above cases, due to $a < r-1$, and \eqref{eqn:b_geq_r}, we have that $(a+1,c)$ belongs to the domain of $h$. We prove that in both of the above cases, $(a+1,c) \in \mathcal{X}_1$ and $\min\{h(a+1,c),c\} = b$. First, notice that from \eqref{eqn:good_c1}, $a<r-1$, and Lemma~\ref{lemma:heg_def}-1-c, we have that $(a+1,b,c)$ is a \textit{good-triplet}.
\begin{itemize}
    \item If $b=c$, since $(a+1,b,c)$ is a \textit{good-triplet}, we have $h(a+1,c) \geq b = c>r$, where last inequality follows from \eqref{eqn:b_geq_r}. Hence, $(a+1,c) \in \mathcal{X}_1$, and $\min\{h(a+1,c) ,c\} = c = b$.
    \item If $c>b$ and $(a+1,b+1,c)$ is a \textit{bad-triplet}, we have that $h(a+1,c) = b > r$, where the last inequality follows from \eqref{eqn:b_geq_r}. Hence $(a+1,c) \in \mathcal{X}_1$, and $\min\{h(a+1,c),c\} = b$.
\end{itemize} 
Hence,
\begin{align}
    z(\vec{p}^{1,a+1,c}) = \sum_{i=1}^{a+1} \frac{E_i}{2} + \frac{\left(\theta-\frac{1}{2}\right) (\delta-1)}{S_{a+2,b}} + \sum_{i=b+1}^{c}E_i.
\end{align}
Substituting and simplifying $z(\vec{p}^{1,a+1,c}) \leq z(\vec{p}^{1,a,c})$ we get,
\begin{align}
    \left(\frac{E_{a+1}}{2}S_{a+1,b} - \theta \right)\left(E_{a+1}S_{a+1,b} - \delta\right) < 0
\end{align}
Notice that $E_{a+1}S_{a+1,b} \geq E_bS_{a+1,b} > \delta$, where last inequality follows from \eqref{eqn:10101011}.

\noindent
\textbf{Case 4: } $a<r-1$, $b<c$, and $(a+1,b+1,c)$ is a \textit{good-triplet}: Since $c>b$, we have that from \eqref{eqn:bad_c1} that $(a,b+1,c)$ is a \textit{bad-triplet}. Due to $b+1\leq c$, and \eqref{eqn:b_geq_r}, we have that $(b+1,c)$ belongs to the domain of $g$. Combining the above with the case description, we have $g(b+1,c) = a+1$. Since $b+1 \leq c$ and $a+1 <r$, we have that $(b+1,c) \in \mathcal{X}_3$. Hence,
\begin{align}
    z(\vec{p}^{3,b+1,c}) = \sum_{i=1}^{a} \frac{E_i}{2} + \frac{E_{a+1}}{2}\left(\delta - E_{b+1}S_{a+2,b} \right)+ E_{b+1}\left(\theta +\frac{1}{2}\right) + \sum_{i=b+2}^{c}E_i.
\end{align}
Substituting and simplifying $z(\vec{p}^{3,b+1,c}) \leq z(\vec{p}^{1,a,c})$ we get,
\begin{align}
    \left(\frac{E_{a+1}}{2}S_{a+1,b} - \theta \right)\left(E_{b+1}S_{a+1,b} - \delta\right) > 0
\end{align}
Combining with \eqref{eqn:11101011}, we are done.

\noindent
3) We consider two cases. The cases make sense since $a-1\geq 0$ from the statement description.

\noindent
\textbf{Case 1} $(a-1,b,c)$ is a \textit{good-triplet}: Notice that since $a>0$ from the statement and \eqref{eqn:b_geq_r}, we have that $(a-1,c)$ belongs to the domain of $h$. Since, $(a-1,b,c)$ is a \textit{good-triplet}, we have that $h(a-1,c) \geq b>r$, where the last inequality follows from \eqref{eqn:b_geq_r}. Hence, $(a-1,c) \in \mathcal{X}_1$.

If $b = c$, we have that $\min\{h(a-1,c),c\} = c = b$. If $c>b$, we have from \eqref{eqn:bad_c1} that $(a,b+1,c)$ is a \textit{bad-triplet}, which when combined with $a>0$ and Lemma~\ref{lemma:heg_def}-2-c, gives $(a-1,b+1,c)$ is a \textit{bad-triplet}. Combining with the case description, we have that $h(a-1,c) = b$. Hence, $\min\{h(a-1,c),c\} = b$. Hence in either case we have that, $(a-1,c) \in \mathcal{X}_1$ and $\min\{h(a-1,c),c\} = b$. Hence,
\begin{align}
    z(\vec{p}^{1,a-1,c}) = \sum_{i=1}^{a-1} \frac{E_i}{2} + \frac{\left(\theta+\frac{1}{2}\right) (\delta+1)}{S_{a,b}} + \sum_{i=b+1}^{c}E_i.
\end{align}
Substituting and simplifying $z(\vec{p}^{1,a-1,c}) \leq z(\vec{p}^{1,a,c})$ we get,
\begin{align}
    \left(\frac{E_a}{2}S_{a+1,b} - \theta \right)\left(E_aS_{a+1,b} - \delta\right) \geq 0
\end{align}
Combining \eqref{eqn:10101011}, and $E_a \geq E_b$, we have $E_aS_{a+1,b} \geq E_bS_{a+1,b} > \delta$ which gives the result.

\noindent
\textbf{Case 2} $(a-1,b,c)$ is a \textit{bad-triplet}: Notice that from \eqref{eqn:b_geq_r}, we have that $(b,c)$ belongs to the domain of $g$. Combining the case description with \eqref{eqn:good_c1}, we have $g(b,c) = a$. Combining \eqref{eqn:b_geq_r}, and $0 < a$, we have that $(b,c) \in \mathcal{X}_3$. Hence,
\begin{align}
z(\vec{p}^{3,b,c}) = \sum_{i=1}^{a-1} \frac{E_i}{2} + \frac{E_a}{2}\left(\delta-E_bS_{a+1,b-1}\right) + E_b\theta + \sum_{i=b+1}^{c}E_i
\end{align}
Using $z(\vec{p}^{3,b,c}) \leq z(\vec{p}^{1,a,c})$, yields the inequality,
\begin{align}
    \left(\frac{E_a}{2}S_{a+1,b} - \theta \right)\left(E_bS_{a+1,b} - \delta\right) > 0
\end{align}
Using \eqref{eqn:10101011}, we have the desired result.

\noindent
4) We consider the following two cases. The cases make sense since $b+1 \leq c+1 \leq n$, where the first inequality follows from \eqref{eqn:b_geq_r}, and the second follows from the statement description. 

\noindent
\textbf{Case 1} $(a,b+1,c+1)$ is a \textit{bad-triplet}: Combining $c<n$ from the statement description with \eqref{eqn:b_geq_r}, we have that $(a,c+1)$ belongs to the domain of $h$. Notice that from \eqref{eqn:good_c1}, and Lemma~\ref{lemma:heg_def}-1-b, we have that $(a,b,c+1)$ is a \textit{good-triplet}. Hence, we have $h(a,c+1) = b>r$, where the last inequality follows from \eqref{eqn:b_geq_r}. Hence, $(a,c+1) \in \mathcal{X}_1$, and $\min\{h(a,c+1),c+1\} = b$. Hence,
\begin{align}
    z(\vec{p}^{1,a,c+1}) = \sum_{i=1}^{a} \frac{E_i}{2} + \frac{\theta (\delta-1)}{S_{a+1,b}} + \sum_{i=b+1}^{c+1}E_i.
\end{align}
Substituting and simplifying $z(\vec{p}^{1,a,c+1}) \leq z(\vec{p}^{1,a,c})$ we get the desired result.
         
\noindent
\textbf{Case 2} $b = c$,  $(a,b+1,c+1)$ is a \textit{good-triplet}: Combining $c<n$ from the statement description with \eqref{eqn:b_geq_r}, we have that $(a,c+1)$ belongs to the domain of $h$. Since, $(a,b+1,c+1)$ is a \textit{good-triplet}, we should have $h(a,c+1) \geq b+1=c+1>r$, where the last inequality follows from \eqref{eqn:b_geq_r}. Hence, $(a,c+1) \in \mathcal{X}_1$, and $\min\{h(a,c+1),c+1\} = c+1$. Hence,
\begin{align}
    z(\vec{p}^{1,a,c+1}) = \sum_{i=1}^{a} \frac{E_i}{2} + \frac{(\theta+1) \delta}{S_{a+1,c+1}}.
\end{align}
Substituting and simplifying $z(\vec{p}^{1,a,c+1}) \leq z(\vec{p}^{1,a,c})$ we get the desired result.
         
\noindent
\textbf{Case 3} $b< c$, $(a,b+1,c+1)$ is a \textit{good-triplet}: Since $b<c$, from \eqref{eqn:bad_c1}, we have that $(a,b+1,c)$ is a \textit{bad-triplet}. From \eqref{eqn:b_geq_r}, we have $b+1 \leq c \leq n$ and $a \geq 0$, which implies that $(a,b+1)$ belongs to the domain of $e$. Combining with $(a,b+1,c+1)$ is a \textit{good-triplet}, we should have $e(a,b+1) = c+1$. Since $n \geq c+1 > b+1$ where the first inequality follows from \eqref{eqn:b_geq_r}, we have that $(a,b+1) \in \mathcal{X}_2$. Hence,
 \begin{align}
    z(\vec{p}^{2,a,b+1}) = \sum_{i=1}^{a} \frac{E_i}{2} + E_{c+1}\left(\delta-E_{b+1}S_{a+1,b}\right) + \theta E_{b+1} + \sum_{i=b+1}^{c}E_i
\end{align}
Using $z(\vec{p}^{2,a,b+1}) \leq z(\vec{p}^{1,a,c})$, yields the inequality,
\begin{align}
    \left(E_{c+1}S_{a+1,b} - \theta \right)\left(E_{b+1}S_{a+1,b} - \delta\right) > 0
\end{align}
Combining with \eqref{eqn:11101011}, we have the desired result.
\end{proof}
\end{lemma}
Now, we construct a Lagrange multiplier that satisfies the conditions of Lemma~\ref{lemma:lagrange_lemma}. Consider $\vec{\mu} \in \mathbb{R}^n$, given by,
\begin{align}
\mu_k = \begin{cases}
    \frac{C}{E_k} - \frac{1}{2} & \text{ if } a+1 \leq k \leq r\\
    1-\frac{C}{E_k} & \text{ if } r+1 \leq k \leq b\\
    0 & \text{ otherwise }
\end{cases},
\end{align}
where,
\begin{align}
    C = \frac{\theta}{S_{a+1,b}}.
\end{align}
The above $\vec{\mu}$ satisfies $\vec{\mu} \geq 0$. If $k \in [a+1,r]$, we have that,
\begin{align}
    \mu_k =  \frac{C}{E_k} - \frac{1}{2} \geq \frac{C}{E_a} - \frac{1}{2} \geq 0,
\end{align}
where the last inequality follows due to Lemma~\ref{lemma:comes_from_A1}-2.  If $k \in [r+1,b]$, we have that,
\begin{align}
    \mu_k = 1- \frac{C}{E_k} \geq 1-\frac{C}{E_c} \geq 0,
\end{align}
where the last inequality follows due to Lemma~\ref{lemma:comes_from_A1}-1.

Using the above $\mu$ as a Lagrange multiplier for problem \text{(P-}1,2,..,r\text{)}, we have the problem,
\begin{maxi}
  {\vec{p},\gamma}{\sum_{j =1}^a p_j\frac{E_j}{2} + \sum_{j =a+1}^b C p_j +\sum_{j =b+1}^n E_j}{}{}
     \addConstraint{\vec{p}}{\in \Delta_{n,r}, \lambda \in \mathbb{R}} 
\end{maxi}
Notice that due to Lemma~\ref{lemma:comes_from_A1}, we have that, $E_j/2 \geq C$ for $j \in [1:a]$, $E_j\geq C$, for $j \in [b+1:c]$, and $E_j\leq C$ for $j \in [c+1,n]$. Hence, an optimal solution $\vec{p}$ for the above problem is $\vec{p} = \vec{p}^{1,a,c}$ with arbitrary $\gamma$. Let, $\gamma = \frac{\delta}{S_{a+1,b}}$. Notice that from Lemma~\ref{lemma:a0b0explicit_1}, part-2-a, we have that $(\vec{p}^{1,a,c}, \gamma)$ is feasible for  \text{(P-}1,2,..,r\text{)}. Also, notice that from the definition of $\vec{\mu}$, we have $\mu_k > 0$ implies  $p^{1,a,c}_kE_k = \gamma$. Hence, from Lemma~\ref{lemma:lagrange_lemma}, we have that $(\vec{p}^{1,a,c}, \gamma)$ solves \text{(P-}1,2,..,r\text{)}, as desired.

\noindent
\textbf{Case 2: }Best vector in $\mathcal{A}$ comes from $\mathcal{A}_2$

Let $\vec{p}^{2,a,b}$ denote the best vector where $(a,b) \in \mathcal{X}_2$ and let $c = e(a,b)$. Define 
\begin{align}\label{eqn:gamma_def_1}
    \theta = \frac{r-a}{2}+b-r
\end{align}
and $\delta = r-a+b-c$. We have the following claim.

\noindent
\textbf{Claim: } We should have $a < r-1$. 
\begin{proof}
Assume the contrary. Hence, from \eqref{eqn:c_ge_b} we have $a = r-1$. Hence, we have $p^{2,a,b}_k = 1$ for all $1 \leq k \leq r-1$. Additionally from \eqref{eqn:b_geq_r}, notice that $b\geq r+1$. Also from the definition of $\vec{p}^{2,a,b}$, we have $p^{2,a,b}_{b} = 1$, and  $p^{2,a,b}_{r} = E_{r+1}/E_r > 0$, which implies that, $\sum_{k=1}^{n}p^{2,a,b}_k > r$. This is a contradiction. 
\end{proof}
Combining the claim, and \eqref{eqn:c_ge_b}, we should have in this case, that 
\begin{align}\label{eqn:ap1cm1}
(a+1,b,c-1) \in [0:r-1] \times [r+1:n]\times [r+1:n]
\end{align}
Now we prove the following lemma.
\begin{lemma}\label{lemma:comes_from_A2}
We have that,
\begin{enumerate}
    \item $\frac{E_c}{E_{a+1}} \geq \frac{1}{2}$
    \item If $a>0$, then $\frac{E_c}{E_a} \leq \frac{1}{2}$
    \item $E_{c}S_{a+1,b-1} +1\geq \theta$
    \item $E_b\left(\theta - E_cS_{a+1,b-1}\right) \geq E_c$,
\end{enumerate}
where $\theta$ is defined in \eqref{eqn:gamma_def_1}.
\begin{proof}

\noindent
1) We complete the proof using two cases. Notice that the following two cases make sense due to \eqref{eqn:ap1cm1}.

\noindent
\textbf{Case 1} $(a+1,b,c-1)$ is a \textit{bad-triplet}: From \eqref{eqn:ap1cm1}, we have $(a+1,b)$ belongs to the domain of $e$. Combining \eqref{eqn:good_b1} and $a+1 \leq r-1$ with Lemma~\ref{lemma:heg_def}-1-c, we have that $(a+1,b,c)$ is a \textit{good-triplet}. Hence, $e(a+1,b) = c$. Since, $b< c \leq n$ from \eqref{eqn:c_ge_b}, we have that, $(a+1,b) \in \mathcal{X}_2$. Hence, 
\begin{align}
z(\vec{p}^{2,a+1,b}) = \sum_{i=1}^{a+1} \frac{E_i}{2} +E_b \left(\theta-\frac{1}{2}\right)+\sum_{i=b+1}^{c}E_i + E_c\left(\delta-1 - E_bS_{a+2,b}\right).
\end{align}
Using $z(\vec{p}^{2,a,b}) \geq z(\vec{p}^{2,a+1,b})$, yields the inequality,
\begin{align}
   (2E_c-E_{a+1})(E_{a+1}-E_b) \geq  0,
\end{align}
which yields the result since $E_{a+1} > E_b$ (the inequality is strict due to assumption \textbf{A1}).

\noindent
\textbf{Case 2} $(a+1,b,c-1)$ is a \textit{good-triplet}: Notice that due to \eqref{eqn:ap1cm1}, we have that $(b,c-1)$ belongs to the domain of $g$. Combining \eqref{eqn:bad_b1} with the case description, we have that $g(b,c-1)=a+1$. Combining the claim, \eqref{eqn:c_ge_b}, and $a+1>0$, we have that $(b,c-1) \in \mathcal{X}_3$. Hence,
\begin{align}
z(\vec{p}^{3,b,c-1}) = \sum_{i=1}^{a} \frac{E_i}{2} +E_b\left(\theta-\frac{1}{2}\right)+\sum_{i=b+1}^{c-1}E_i + \frac{E_{a+1}}{2}\left(\delta - E_bS_{a+2,b-1}\right).
\end{align}
Using $z(\vec{p}^{2,a,b}) \geq z(\vec{p}^{3,b,c+1})$, yields the inequality,
\begin{align}
\left(E_c - \frac{E_{a+1}}{2}\right)\left(\delta+1-E_bS_{a+1,b}\right) > 0,
\end{align}
which establishes the result combined with \eqref{eqn:100001}.

\noindent
2) We consider two cases. The two cases make sense since $a>0$ by the statement description.  

\noindent
\textbf{Case 1 }$(a-1,b,c)$ is a \textit{good-triplet}:  Combining $a>0$ from the statement description and \eqref{eqn:c_ge_b}, we have that, $(a-1,b)$ belongs to the domain of $e$. Combining $a>0$, \eqref{eqn:bad_b1}, and Lemma~\ref{lemma:heg_def}-2-c, $(a-1,b,c-1)$ is a \textit{bad-triplet}. Combining the above with the case description, we have that $e(a-1,b) = c$. Since $b< c \leq n$ from \eqref{eqn:c_ge_b}, we have that $(a-1,b) \in \mathcal{X}_2$. Hence,
\begin{align}
    z(\vec{p}^{2,a-1,b}) = \sum_{i=1}^{a-1} \frac{E_i}{2} +E_b \left(\theta+\frac{1}{2}\right)+\sum_{i=b+1}^{c}E_i + E_c\left(\delta+1 - E_bS_{a+1,b}\right).
\end{align}
Using $z(\vec{p}^{2,a,b}) \geq z(\vec{p}^{2,a-1,b})$, yields the inequality,
\begin{align}\label{eqn:e_ageqe_b}
   (E_a-2E_c)(E_a-E_b) \geq 0.
\end{align}
which establishes the result since $E_a > E_b$ (the inequality is strict by assumption \textbf{A1}).

\noindent
\textbf{Case 2} $(a-1,b,c)$ is a \textit{bad-triplet}: From \eqref{eqn:ap1cm1}, we have that $(b,c)$ belongs to the domain of $g$. Combining the case description with \eqref{eqn:good_b1}, we have that $g(b,c)=a$.Combining \eqref{eqn:c_ge_b} and $a>0$, we have that $(b,c) \in \mathcal{X}_3$. Hence,
\begin{align}
    z(\vec{p}^{3,b,c}) = \sum_{i=1}^{a-1} \frac{E_i}{2} +E_b\theta+\sum_{i=b+1}^{c}E_i + \frac{E_a}{2}\left(\delta - E_bS_{a+1,b-1}\right).
\end{align}
Using $z(\vec{p}^{2,a,b}) \geq z(\vec{p}^{3,b,c})$, yields the inequality,
\begin{align}
   \left(E_c-\frac{E_a}{2}\right)\left(\delta-E_bS_{a+1,b}\right) > 0,
\end{align}
which establishes the result from \eqref{eqn:100001}.

\noindent
3) We consider three cases. The cases make sense since, $b \geq r+1$ from \eqref{eqn:c_ge_b}, and if $b>r+1$, $(a,b-1,c-1) \in [0:r-1] \times [r+1:n] \times [r+1:n]$ from \eqref{eqn:c_ge_b}.

\noindent
\textbf{Case 1} $b = r+1$: This case reduces to, $\sum_{i=a+1}^r \frac{E_c}{E_i} \geq \frac{r-a}{2}$, which is true due to 1.

\textbf{Case 2} $b>r+1$, $(a,b-1,c-1)$ is a \textit{bad-triplet}: Combining $b -1 \geq r+1$ and \eqref{eqn:c_ge_b}, we have that $(a,b-1)$ belongs to the domain of $e$. From $b-1 \geq r+1$, \eqref{eqn:good_b1}, and Lemma~\ref{lemma:heg_def}-1-a, we have that $(a,b-1,c)$ is a \textit{good-triplet}. Hence, we have $e(a,b-1) = c$. Combining with \eqref{eqn:c_ge_b}, we have $(a,b-1) \in \mathcal{X}_2$. Hence,
\begin{align}
z(\vec{p}^{2,a,b-1}) = \sum_{i=1}^{a} \frac{E_i}{2} +E_{b-1} \left(\theta-1\right)+\sum_{i=b}^{c}E_i + E_c\left(\delta-1 - E_{b-1}S_{a+1,b-1}\right).
\end{align}
Using $z(\vec{p}^{2,a,b}) \geq z(\vec{p}^{2,a,b-1})$, yields the inequality,
\begin{align}
   (E_b-E_{b-1})(\theta-1-E_cS_{a+1,b-1}) > 0,
\end{align}
yields the result since $E_{b-1} > E_b$ (the inequality is strict due to assumption \textbf{A1}).

\noindent
\textbf{Case 3} $b>r+1$, $(a,b-1,c-1)$ is a \textit{good-triplet}: Combining $c-1 \geq b-1geq r+1$, with \eqref{eqn:c_ge_b}, we have that $(a,c-1)$ belongs to the domain of $h$. Combining \eqref{eqn:bad_b1}, and the case description, we have that $h(a,c-1)=b-1$. Notice that $b-1 \geq r+1$. Hence, $(a,c-1) \in \mathcal{X}_1$, and $\min\{h(a,c-1),c-1\} = b-1$. Hence,
\begin{align}
    z(\vec{p}^{1,a,c-1}) = \sum_{i=1}^{a} \frac{E_i}{2} +\frac{\left(\theta-1\right)\delta}{S_{a+1,b-1}} +\sum_{i=b}^{c-1}E_i
\end{align}
Using $z(\vec{p}^{2,a,b}) \geq z(\vec{p}^{1,a,c-1})$, yields the inequality,
\begin{align}
   \left(\theta-1-E_cS_{a+1,b-1}\right)\left(\delta-E_bS_{a+1,b-1}\right) \leq 0,
\end{align}
which establishes the result due to \eqref{eqn:100001}.

\noindent
4) Notice that the above reduces to,
\begin{align}
    \theta \geq E_cS_{a+1,b}.
\end{align}
We consider three cases. The cases make sense since \eqref{eqn:c_ge_b} tells us $b+1 \leq c$ and hence $(a,b+1,c) \in [0:r-1] \times [r+1:n] \times [r+1:n]$, and $b \leq c$ from \eqref{eqn:c_ge_b}.

\noindent
\textbf{Case 1} $(a,b+1,c)$ is a \textit{bad-triplet}: Due to \eqref{eqn:c_ge_b}, we have that $(a,b)$ belongs to the domain of $h$. Combining the case description with \eqref{eqn:good_b1}, we have that $h(a,c) = b$. Since, $b \geq r+1$ from \eqref{eqn:c_ge_b}, we have that $(a,c) \in \mathcal{X}_1$, and $\min\{h(a,c),c\} = b$. Hence,
\begin{align}
    z(\vec{p}^{1,a,c}) = \sum_{i=1}^{a} \frac{E_i}{2} +\frac{\theta\delta}{S_{a+1,b}} +\sum_{i=b+1}^{c}E_i
\end{align}
Using $z(\vec{p}^{2,a,b}) \geq z(\vec{p}^{1,a,c})$, yields the inequality,
\begin{align}
   \left(E_cS_{a+1,b}-\theta\right)\left(E_bS_{a+1,b}-\delta\right) < 0,
\end{align}
which establishes the result due to \eqref{eqn:100001}.

\noindent
\textbf{Case 2} $b+1 = c$, and $(a,b+1,c)$ is a \textit{good-triplet}: Due to \eqref{eqn:c_ge_b}, we have that $(a,b)$ belongs to the domain pf $h$. Since, $(a,b+1,c)$ is a \textit{good-triplet}, we have $h(a,c) \geq b+1 = c$. Since, $b+1 > r+1$ from \eqref{eqn:c_ge_b}, we have $(a,c) \in \mathcal{X}_1$, and $\min\{h(a,c),c\} = c = b+1$. Hence,
 \begin{align}
    z(\vec{p}^{1,a,b+1}) = \sum_{i=1}^{a} \frac{E_i}{2} +\frac{(\theta+1)(\delta+1)}{S_{a+1,b+1}}
\end{align}
Using $z(\vec{p}^{2,a,b}) \geq z(\vec{p}^{1,a,b+1})$, yields the inequality,
\begin{align}
   (E_bS_{a+1,b+1}-\delta-1)(E_{b+1}S_{a+1,b}-\theta) \leq 0,
\end{align}
which yields the result since $E_bS_{a+1,b+1} \geq E_bS_{a+1,b}+1 > \delta+1$ ($E_b > E_{b+1}$ and \eqref{eqn:100001}).

\noindent
\textbf{Case 3} $b+1 < c$, and $(a,b+1,c)$ is a \textit{good-triplet}: Due to \eqref{eqn:c_ge_b}, we have that, $(a,b+1)$ belongs to the domain of $e$. Combining \eqref{eqn:bad_b1}, $c-1 \geq r+1$ from \eqref{eqn:ap1cm1}, and Lemma~\ref{lemma:heg_def}-2-a, we have that, $(a,b+1,c-1)$ is a \textit{bad-triplet}. Combining with the case description, we have that $e(a,b+1) = c$. Since, $b+1<c \leq n$, we have that, $(a,b+1) \in \mathcal{X}_2$. Hence,
 \begin{align}
    z(\vec{p}^{2,a,b+1}) = \sum_{i=1}^{a} \frac{E_i}{2} +E_{b+1} \left(\theta+1\right)+\sum_{i=b+2}^{c}E_i + E_c\left(\delta+1 - E_{b+1}S_{a+1,b+1}\right).
\end{align}
Using $z(\vec{p}^{2,a,b}) \geq z(\vec{p}^{2,a,b+1})$, yields the inequality,
\begin{align}
   (E_b-E_{b+1})\left(\theta-E_cS_{a+1,b}\right) > 0,
\end{align}
yields the result since $E_{b} > E_{b+1}$ (the inequality is strict due to assumption \textbf{A1}).   
\end{proof}
\end{lemma}

Now, we construct a Lagrange multiplier, similar to case 1. Consider $\vec{\mu} \in \mathbb{R}^n$, given by,
\begin{align}
\mu_k = \begin{cases}
    \frac{E_c}{E_k} - \frac{1}{2} & \text{ if } a+1 \leq k \leq r\\
    1-\frac{E_c}{E_k} & \text{ if } r+1 \leq k \leq b-1\\
    E_cS_{a+1,b-1}+1-\theta & \text{ if } k = b\\
    0 & \text{ otherwise }
\end{cases},
\end{align}

The above $\vec{\mu}$ satisfies $\vec{\mu} \geq 0$. If $k \in [a+1,r]$, we have that,
\begin{align}
    \mu_k =  \frac{E_c}{E_k} - \frac{1}{2} \geq \frac{E_c}{E_{a+1}} - \frac{1}{2} \geq 0,
\end{align}
where the last inequality follows due to Lemma~\ref{lemma:comes_from_A2}-1.  If $k \in [r+1,b-1]$, we have that,
\begin{align}
    \mu_k = 1- \frac{E_c}{E_k} \geq 1-\frac{E_c}{E_c} \geq 0,
\end{align}
If $k = b$, $\mu_k \geq 0$, is Lemma~\ref{lemma:comes_from_A2}-3.

Using the above $\mu$ as a Lagrange multiplier for problem \text{(P-}1,2,..,r\text{)}, we have the problem,
\begin{maxi}
  {\vec{p},\gamma}{\sum_{j =1}^a p_j\frac{E_j}{2} + \sum_{j =a+1}^{b-1}p_j E_c  +p_bE_b(\theta - E_cS_{a+1,b-1})+\sum_{j =b+1}^n p_jE_j}{}{}
     \addConstraint{\vec{p}}{\in \Delta_{n,r}, \lambda\in \mathbb{R} } 
\end{maxi}
Notice that due to Lemma~\ref{lemma:comes_from_A2}, we have that $E_j/2 \geq E_c$ for $j \in [1:a]$, $E_j\geq E_c$ for $j \in [b+1:c]$, $E_j\leq E_c$ for $j \in [c+1:n]$, and $(\theta - E_cS_{a+1,b-1})\geq E_c$. Hence, an optimal solution $\vec{p}$ for the above problem is $\vec{p} = \vec{p}^{2,a,b}$ with arbitrary $\gamma$. Let, $\gamma = E_b$. Notice that from Lemma~\ref{lemma:a0b0explicit_1}, part-2-b, we have that $(\vec{p}^{1,a,b}, \gamma)$ is feasible for \text{(P-}1,2,..,r\text{)}. Also, notice that from the definition of $\vec{\mu}$, we have $\mu_k > 0$ implies $p^{2,a,b}_kE_k = E_b$. Hence, from Lemma~\ref{lemma:lagrange_lemma}, we have that $(\vec{p}^{2,a,b}, \gamma)$ solves \text{(P-}1,2,..,r\text{)}, as desired.

\noindent
\textbf{Case 3: }Best vector in $\mathcal{A}$ comes from $\mathcal{A}_3$

Let $\vec{p}^{3,b,c}$ denote the best vector where $(b,c) \in \mathcal{X}_3$, and let $a = g(b,c)>0$. Define 
\begin{align}\label{eqn:gamma_def_2}
    \theta = \frac{r-a}{2}+b-r
\end{align}
and $\delta = r-a+b-c$.  Now we prove the following lemma.
\begin{lemma}\label{lemma:comes_from_A3}
We have that,
\begin{enumerate}
    \item If $c<n$, then $\frac{E_a}{2} \geq E_{c+1}$
    \item $\frac{E_{a}}{2}S_{a+1,b-1} +1\geq \theta$
    \item $E_b\left(\theta - \frac{E_a}{2}S_{a+1,b-1}\right) \geq \frac{E_a}{2}$
    \item $2E_{c}\geq E_a$
\end{enumerate}
\begin{proof}

\noindent
1) From \eqref{eqn:a_ge_zer} we have that $(a-1,b)$ belongs to the domain of $e$. Combining $a>0$ from \eqref{eqn:a_ge_zer}, $c+1 \leq n$ from the statement description, Lemma~\ref{lemma:heg_def}-1-d, and \eqref{eqn:good_a1}, we have that, $(a-1,b,c+1)$ is a \textit{good-triplet}. Combining this with \eqref{eqn:bad_a1}, we have that $e(a-1,b) = c+1$. Notice that, $b < c+1 \leq n$ from \eqref{eqn:a_ge_zer}. Hence, $(a-1,b) \in \mathcal{X}_2$. Hence, 
\begin{align}
z(\vec{p}^{2,a-1,b}) = \sum_{i=1}^{a-1} \frac{E_i}{2} +E_b\left(\theta+\frac{1}{2}\right)+\sum_{i=b+1}^{c}E_i + E_{c+1}\left(\delta + 1- E_bS_{a+1,b}\right).
\end{align}
Using $z(\vec{p}^{3,b,c}) \geq z(\vec{p}^{2,a-1,b})$, yields,
\begin{align}
\left(\frac{E_a}{2}-E_{c+1}\right)\left(\delta+1-E_bS_{a,b}\right) \geq 0,
\end{align}
which establishes the desired inequality from \eqref{eqn:1000011}.

\noindent
2) We consider three cases. The cases make sense since $b \geq r+1$, and if $b>r+1$, we have that $(a-1,b-1,c) \in [0:r+1] \times [r+1:n] \times [r+1:n]$ from \eqref{eqn:a_ge_zer}.

\noindent
\textbf{Case 1} $b = r+1$: This case reduces to $E_aS_{a+1,r} \geq r-a$, which follows since $E_a \geq E_i\ \forall i \in [a+1:r]$.

\noindent
\textbf{Case 2} $b>r+1$ and $(a-1,b-1,c)$ is a \textit{bad-triplet}: Due to $b-1\geq r+1$, and \eqref{eqn:a_ge_zer}, we have that $(b-1,c)$ belongs to the domain of $g$. From \eqref{eqn:good_a1}, $b-1 \geq r+1$, and Lemma~\ref{lemma:heg_def}-1-a, we have that, $(a,b-1,c)$ is a \textit{good-triplet}. Combining with the case description, we have that $g(b-1,c) = a$. Notice that $b-1<c \leq n$, and $0 < a \leq r-1$ from \eqref{eqn:a_ge_zer}. Hence, $(b-1,c) \in \mathcal{X}_3$. Hence,
\begin{align}
    z(\vec{p}^{3,b-1,c}) = \sum_{i=1}^{a} \frac{E_i}{2} +E_{b-1} \left(\theta-1\right)+\sum_{i=b}^{c}E_i + \frac{E_a}{2}\left(\delta-1 - E_{b-1}S_{a+1,b-1}\right).
\end{align}
Using $z(\vec{p}^{3,b,c}) \geq z(\vec{p}^{3,b-1,c})$, yields the inequality,
\begin{align}
   (E_b-E_{b-1})\left(\theta-1-\frac{E_a}{2}S_{a+1,b-1}\right) > 0,
\end{align}
yields the result since $E_{b-1} > E_b$ (the inequality is strict due to assumption \textbf{A1}).

\noindent
\textbf{Case 3} $b>r+1$, $(a-1,b-1,c)$ is a \textit{good-triplet}: Due to \eqref{eqn:a_ge_zer}, we have that $(a-1,c)$ belongs to the domain of $h$. Combining the case description with \eqref{eqn:bad_a1}, we have that $h(a-1,c)=b-1$. Since $b-1 \geq r+1$, from the case description we have, $(a-1,c) \in \mathcal{X}_1$, and $\min\{h(a-1,c),c\} = b-1$. Hence,
\begin{align}
    z(\vec{p}^{1,a-1,c}) = \sum_{i=1}^{a-1} \frac{E_i}{2} +\frac{\left(\theta-\frac{1}{2}\right)\delta}{S_{a,b-1}} +\sum_{i=b}^{c}E_i
\end{align}
Using $z(\vec{p}^{3,b,c}) \geq z(\vec{p}^{1,a-1,c})$, yields the inequality,
\begin{align}
   \left(\frac{E_a}{2}S_{a+1,b-1}-\theta+1\right)\left(E_bS_{a,b-1}-\delta\right) \leq 0,
\end{align}
which establishes the result from \eqref{eqn:1000011}.

\noindent
3) Notice that the above reduces to,
\begin{align}
    \theta \geq \frac{E_a}{2}S_{a+1,b}.
\end{align}
We consider three cases. The cases make sense since, $b \leq c$ by \eqref{eqn:a_ge_zer}, and if $c>b$, we have to have that $(a,b+1,c) \in [0:r+1] \times [r+1:n] \times [r+1:n]$ from \eqref{eqn:a_ge_zer}.

\noindent
\textbf{Case 1} $b=c$: From \eqref{eqn:good_a1}, we have that, $h(a,c) \geq b = c$. Since $c \geq r+1$, we have that, $(a,c) \in \mathcal{X}_1$. Moreover, $\min\{h(a,c) ,c\} = c= b$. Hence, 
\begin{align}
    z(\vec{p}^{1,a,b}) = \sum_{i=1}^{a} \frac{E_i}{2} +\frac{\theta\delta}{S_{a+1,b}} 
\end{align}
Using $z(\vec{p}^{3,b,c}) \geq z(\vec{p}^{1,a,b})$, yields the inequality,
\begin{align}
   \left(\frac{E_a}{2}S_{a+1,b}-\theta\right)\left(E_bS_{a+1,b}-\delta\right) < 0,
\end{align}
which establishes the result from \eqref{eqn:1000011}.

\noindent
\textbf{Case 2} $b<c$, and $(a,b+1,c)$ is a \textit{good-triplet}:  Since $b+1 \leq c \leq n$, where the last inequality follows from \eqref{eqn:a_ge_zer}, we have that $(b+1,c)$ belongs to the domain of $g$. Combining \eqref{eqn:bad_a1}, $b+1\leq c \leq n$, with Lemma~\ref{lemma:heg_def}-2-a, we have that $(a-1,b+1,c)$ is a \textit{bad-triplet}. Combining with the case description, we have that $g(b+1,c) = a$. Notice that $b+1 \leq c$, and $0 < a \leq r-1$ from \eqref{eqn:a_ge_zer}. Hence, $(b+1,c) \in \mathcal{X}_3$. Hence,
\begin{align}
z(\vec{p}^{3,b+1,c}) = \sum_{i=1}^{a} \frac{E_i}{2} +E_{b+1} \left(\theta+1\right)+\sum_{i=b+2}^{c}E_i + \frac{E_a}{2}\left(\delta+1 - E_{b+1}S_{a+1,b+1}\right).
\end{align}
Using $z(\vec{p}^{3,b,c}) \geq z(\vec{p}^{3,b+1,c})$, yields the inequality,
\begin{align}
   (E_b-E_{b+1})\left(\theta-\frac{E_a}{2}S_{a+1,b}\right) > 0,
\end{align}
yields the result since $E_{b} > E_{b+1}$ (the inequality is strict due to assumption \textbf{A1}).

\noindent
\textbf{Case 3} $b<c$, and $(a,b+1,c)$ is a \textit{good-triplet}: From \eqref{eqn:a_ge_zer}, we have that $(a,c)$ belongs to the domain of $h$. Combining the case description with \eqref{eqn:good_a1}, we have that $h(a,c)=b$. Notice that $b \geq r+1$ from \eqref{eqn:a_ge_zer}. Hence, $(a,c) \in \mathcal{X}_1$, and $\min\{h(a,c),c \} = b$. Hence,
\begin{align}
    z(\vec{p}^{1,a,c}) = \sum_{i=1}^{a} \frac{E_i}{2} +\frac{\theta\delta}{S_{a+1,b}} +\sum_{i=b+1}^{c}E_i
\end{align}
Using $z(\vec{p}^{3,b,c}) \geq z(\vec{p}^{1,a,c})$, yields the inequality,
\begin{align}
   \left(\frac{E_a}{2}S_{a+1,b}-\theta\right)\left(E_bS_{a+1,b}-\delta\right) < 0,
\end{align}
which establishes the result from \eqref{eqn:1000011}.

\noindent
4) We consider two cases. The cases make sense since $b \leq c$ from \eqref{eqn:a_ge_zer}.

\noindent
\textbf{Case 1} $b = c$: Notice that from part 2 of the lemma,
\begin{align}
    \theta - \frac{E_a}{2}S_{a+1,b-1} \leq 1.
\end{align}
Substituting this in part 3, we have the result.

\noindent
\textbf{Case 2} $b < c$: From \eqref{eqn:a_ge_zer}, we have that, $(a,b)$ belongs to the domain of $e$. Combining \eqref{eqn:bad_a1}, $c-1 \geq b \geq r+1$, from the case description, with Lemma~\ref{lemma:heg_def}-2-d, we have that, $(a,b,c-1)$ is a \textit{bad-triplet}. Combining with \eqref{eqn:good_a1}, we have that  $e(a,b) = c$. Notice that $b < c \leq n$, where the last inequality follows from \eqref{eqn:a_ge_zer}. Hence, $(a,b) \in \mathcal{X}_2$. Hence, 
\begin{align}
    z(\vec{p}^{2,a,b}) = \sum_{i=1}^{a} \frac{E_i}{2} +E_b\theta+\sum_{i=b+1}^{c}E_i + E_{c}\left(\delta- E_bS_{a+1,b}\right).
\end{align}
Using $z(\vec{p}^{3,b,c}) \geq z(\vec{p}^{2,a,b})$, yields,
\begin{align}
    \left(\frac{E_a}{2}-E_{c}\right)\left(\delta-E_bS_{a+1,b}\right) \geq 0,
\end{align}
which establishes the desired inequality due to \eqref{eqn:1000011}
\end{proof}
\end{lemma}

Now, we construct a Lagrange multiplier, similar to case 1. Consider $\vec{\mu} \in \mathbb{R}^n$, given by,
\begin{align}
\mu_k = \begin{cases}
    \frac{E_a}{2E_k} - \frac{1}{2} & \text{ if } a+1 \leq k \leq r\\
    1-\frac{E_2}{2E_k} & \text{ if } r+1 \leq k \leq b-1\\
    \frac{E_a}{2}S_{a+1,b-1}+1-\theta & \text{ if } k = b\\
    0 & \text{ otherwise }
\end{cases},
\end{align}

The above $\vec{\mu}$ satisfies $\vec{\mu} \geq 0$. If $k \in [a+1,r]$, we have that,
\begin{align}
    \mu_k =  \frac{E_a}{2E_k} - \frac{1}{2} \geq \frac{E_a}{2E_{a}} - \frac{1}{2} = 0,
\end{align}
If $k \in [r+1,b-1]$, we have that,
\begin{align}
    \mu_k = 1- \frac{E_a}{2E_k} \geq 1-\frac{E_a}{2E_c} \geq 0,
\end{align}
where the last inequality follows due to Lemma~\ref{lemma:comes_from_A3}-4

If $k = b$, $\mu_k \geq 0$, is Lemma~\ref{lemma:comes_from_A3}-2.

Using the above $\mu$ as a Lagrange multiplier for problem \text{(P-}1,2,..,r\text{)}, we have the problem,
\begin{maxi}
  {\vec{p},\gamma}{\sum_{j =1}^a p_j\frac{E_j}{2} + \sum_{j =a+1}^{b-1} p_j\frac{E_a}{2}  +p_bE_b(\theta - \frac{E_a}{2}S_{a+1,b-1})+\sum_{j =b+1}^n p_jE_j}{}{}
     \addConstraint{\vec{p}}{\in \Delta_{n,r} , \lambda \in \mathbb{R}} 
\end{maxi}

Notice that due to Lemma~\ref{lemma:comes_from_A3}, we have that, $E_j/2 \geq E_a/2$ for $j \in [1:a]$, $E_j\leq E_{c+1} \leq E_a/2$ for $j \in [c+1:n]$, $E_j\geq E_{c} \geq \frac{E_a}{2}$ for $j \in [b+1:c]$, and $E_b\left(\theta - \frac{E_a}{2}S_{a+1,b-1}\right)\geq \frac{E_a}{2}$. Hence, an optimal solution $\vec{p}$ for the above problem is $\vec{p} = \vec{p}^{3,b,c}$ with arbitrary $\gamma$. Let, $\gamma = E_b$. Notice that from Lemma~\ref{lemma:a0b0explicit_1}, part-2-c, we have that $(\vec{p}^{3,b,c}, \gamma)$ is feasible for \text{(P-}1,2,..,r\text{)}. Also, notice that from the definition of $\vec{\mu}$, we have $\mu_k > 0$ implies  $p^{3,b,c}_kE_k = E_b$. Hence, from Lemma~\ref{lemma:lagrange_lemma}, we have that $(\vec{p}^{3,b,c}, \gamma)$ solves \text{(P-}1,2,..,r\text{)}, as desired.

\noindent
\textbf{Case 4: } Best vector in $\mathcal{A}$ is $\vec{p}^0$.
\begin{lemma}\label{lemma:comes_from_A4}
We have that,
\begin{align}
    \frac{E_r}{2} \geq E_{r+1}.
\end{align}
\begin{proof}
    Notice that,
    \begin{align}
        r-(r-1) = 1 <E_{r+1}S_{r,r+1} = 1 + \frac{E_{r+1}}{E_r}.
    \end{align}
    Hence, $(r-1,r+1,r+1)$ is a \textit{good-triplet}. Hence, $h(r-1,r+1) \geq r+1$. Clearly, $(r-1,r+1) \in \mathcal{X}_1$. Also, $\min\{h(r-1,r+1),r+1\} = r+1$.
    Hence,
     \begin{align}
        z(\vec{p}^{1,r-1,r+1}) = \sum_{i=1}^{r-1} \frac{E_i}{2} + \frac{3}{2S_{r,r+1}}
    \end{align}
    Using $z(\vec{p}^0) > z(\vec{p}^{1,r-1,r+1})$, yields the desired inequality.
\end{proof}
\end{lemma}

In this case we can use $\vec{\mu} = \vec{0}$ as a Lagrange multiplier vector for \text{(P-}1,2,..,r\text{)}, which gives the problem,
\begin{maxi}
  {\vec{p},\gamma}{\sum_{j =1}^r p_j\frac{E_j}{2} +\sum_{j =r+1}^n p_jE_j}{}{}
     \addConstraint{\vec{p}}{\in \Delta_{n,r},\lambda \in \mathbb{R} } 
\end{maxi}

Due to Lemma~\ref{lemma:comes_from_A4}, we have that $\vec{p}=\vec{p}^0$ is an optimal solution for the above problem with arbitrary $\gamma$. Let, $\gamma = E_{r+1}$. From Lemma~\ref{lemma:a0b0explicit_1}, part-2-d we have that $(\vec{p}^0, \gamma)$ is feasible for 
 \text{(P-}1,2,..,r\text{)}. Also, notice that $\mu_k = 0$ for all $k \in [1:n]$. Hence from Lemma~\ref{lemma:lagrange_lemma}, we have that $(\vec{p}^0, \gamma)$ solves \text{(P-}1,2,..,r\text{)}, as desired.
\section{Algorithm to project onto $\Delta_{n,r}$}\label{app:projecting_to_gem_simp}
\SetKwRepeat{Do}{do}{while}
\begin{algorithm}
\label{algo:333}
\SetAlgoLined
\DontPrintSemicolon
Define for all $1 \leq a \leq b \leq n$,
\vspace{-4mm}
\begin{align}\label{eqn:mu_ab}
    &\mu_{a,b} = \frac{\sum_{j=a}^by_j - (r-a+1)}{b-a+1}, \mathcal{A}_{a,b} = \mathbbm{1}\{y_b \geq \mu_{a,b} \geq y_a - 1\} \nonumber\\
    &\mathcal{B}_{a,b} = \mathbbm{1}\{(b = n) \text{ or }[(b < n) \text{ and } (y_{b+1} < \mu_{a,b})]\}\nonumber\\
    &\mathcal{C}_{a,b} = \mathbbm{1}\{(a = 1) \text{ or }[(a > 1)\text{ and }(y_{a-1} -1 > \mu_{a,b})]\}\nonumber\\
    & g(a,b) =  \min\{ c : c \geq b, \mathcal{B}_{a,c} = 1\}, h(a,b) =  \max\{ c : c \leq a, \mathcal{C}_{c,b} = 1\}.
\end{align}\;
\vspace{-7mm}
Initialize $(a_1,b_1) = (r,r)$.\;
\For{each $t \in \{1,2,\dots\}$}{
Set $(a_{t+1},b_{t+1}) = (h(a_{t},g(a_{t},b_{t})),g(a_{t},b_{t}))$.\;
\If{$(a_{t+1},b_{t+1}) = (a_{t},b_{t})$}{
  Output $\vec{x} \in \mathbb{R}^n$, where $x_i = \Pi_{[0,1]}(y_i-\mu_{a_t,b_t})$.\;
}
}
\caption{Projecting $\vec{y}$ sorted in the nonincreasing order onto $\Delta_{n,r}$}\label{alg:proj_I_r}
\end{algorithm}

\noindent
\textbf{Analysis of Algorithm~\ref{alg:proj_I_r}:} Fix $\vec{y} \in \mathbb{R}$. Notice that the problem of projection of $\vec{y} \in \mathbb{R}^n$ onto $\Delta_{n,r}$ is,  
\begin{mini}
	  {\vec{z}}{\frac{1}{2}\lVert\vec{z} - \vec{y} \rVert^2}{}{}
         \addConstraint{\vec{y}}{\in \Delta_{n,r}}
\end{mini}
We assume, without loss of generality, that $\vec{y}$ is sorted in non-increasing order (Notice that if $\vec{y}$ is not sorted, we could sort $\vec{y}$, perform the projection, and rearrange the elements according to the original order. This works since the set $\Delta_{n,r}$ is closed under the permutation of entries of its element vectors). 

Now consider $L(\vec{z}, \mu)$ for $\mu \in \mathbb{R}$ given by $L(\vec{z}, \mu) = \frac{1}{2}\lVert\vec{z} - \vec{y} \rVert^2 + \mu\left(\sum_{j=1}^n z_j - r\right)$, and the problem,
\begin{mini}
	  {\vec{z}}{ L(\vec{z}, \mu)}{}{\text{(P6-}\mu\text{)}}
        \label{prob:p6_mu}
         \addConstraint{\vec{z}}{\in [0,1]^n}
\end{mini}
for a fixed $\mu \in \mathbb{R}$. Let us assume the existence of a $\mu^* \in \mathbb{R}$ such that the solution $\vec{z}^*$ of $\text{(P6-}\mu^*\text{)}$ defined in \eqref{prob:p6_mu} satisfies, $\sum_{j=1}^n z^*_j = r$. Notice that $\vec{z}^*$ is optimal for the original problem since for any $\vec{z} \in \Delta_{n,r}$,
\begin{align}
     \frac{1}{2}\lVert\vec{z} - \vec{y} \rVert^2 = L(\vec{z}, \mu^*) \geq L(\vec{z}^*, \mu^*) = \frac{1}{2}\lVert\vec{z}^* - \vec{y} \rVert^2.
\end{align}
Hence, we focus on finding such a $\mu^*$ and the corresponding $\vec{z}^*$. First, we focus on solving $\text{(P6-}\mu\text{)}$ defined in \eqref{prob:p6_mu} for a fixed $\vec{\mu} \in \mathbb{R}$. Notice that $\text{(P6-}\mu\text{)}$ is a separable quadratic program in the entries of $\vec{z}$. Hence, the optimal $z_j$ can be obtained by projecting the unconstrained optimal value for each entry of $\vec{z}$ onto $[0,1]$. Hence, the solution is $z_j = \Pi_{[0,1]}(y_j - \mu)$ for all $j \in [1:n]$, where $\Pi_{[0,1]}$ denotes the projection operator onto $[0,1]$.

Now we need to find $\mu^*$ such that the optimal solution $\vec{z}^*$ of $\text{(P6-}\mu^*\text{)}$ defined in \eqref{prob:p6_mu} satisfies $\vec{z}^* \in \Delta_{n,r}$. Hence, we require $\mu^* \in \mathbb{R}$ such that
\begin{align}\label{eqn:proj_eq}
    \sum_{j=1}^n \Pi_{[0,1]}(y_j - \mu^*) = r.
\end{align}
For $\mu \in \mathbb{R}$, define the set $\mathcal{K}_{\mu} = \{i ; 1\leq i \leq n, \mu +1 \geq y_i \geq \mu\}$. Notice that for each $\mu \in \mathbb{R}$, $\mathcal{K}_{\mu}$ is either the empty set or a set of the form $[a:b]$ where $1 \leq a\leq b\leq n$.

We have two possibilities if $\mathcal{K}_{\mu^*}$ is the empty set. The first is $\mu^* > y_j$ for all $j \in [1:n]$ in which case we have $\sum_{j=1}^n \Pi_{[0,1]}(y_j-\mu^*) = 0$ which does not agree with \eqref{eqn:proj_eq}. The second is $\mu^* < y_j-1$ for all $j \in [1:n]$ in which case we have  $\sum_{j=1}^n \Pi_{[0,1]}(y_j-\mu^*) = n$. This is only possible when $n = r$, in which case the only solution to the problem is the trivial solution of player 1 choosing all the resources. 

Hence, we will focus on the case of non-empty $K_{\mu^*}$. Let $\mathcal{K}_{\mu^*} = [a^*:b^*]$ where $1 \leq a^*\leq b^*\leq n$. This is equivalent to  $\mu^*$ satisfying the conditions,
\begin{align}\label{eqn:conds_mu}
    &y_{b^*} \geq \mu^* \geq y_{a^*}-1\nonumber\\
    & (b^* = n) \text{ or } [(b^*<n) \text{ and }(y_{b^*+1} < \mu^*)] \nonumber\\
    & (a^* = 1) \text{ or } [(a^* > 1) \text{ and }(y_{a^*-1}-1 > \mu^*)]
\end{align}
Define for each $a,b \in [1:n]$ the real number $\mu_{a,b}$ as
\begin{align}\label{eqn:def_mu_ab}
    \mu_{a,b} = \frac{\sum_{j=a}^b y_j - (r - a+1)}{b-a+1}.
\end{align}
Now, notice that \eqref{eqn:proj_eq} translates to,
\begin{align}\label{eqn:mu_eq_muab}
    \mu^* = \mu_{a^*,b^*},
\end{align}
where $\mathcal{K}_{\mu^*} = [a^*:b^*]$. Combining \eqref{eqn:mu_eq_muab} and \eqref{eqn:conds_mu}, we have that if we can find $a^*,b^*$ ($1 \leq a^*\leq b^*\leq n$) such that
\begin{align}
    &y_{b^*} \geq \mu_{a^*,b^*} \geq y_{a^*}-1\nonumber\\
    & (b^* = n) \text{ or } [(b^*<n) \text{ and }(y_{b^*+1} < \mu_{a^*,b^*})] \nonumber\\
    & (a^* = 1) \text{ or } [(a^* > 1) \text{ and }(y_{a^*-1}-1 > \mu_{a^*,b^*})]
\end{align}
are all satisfied, then we are guaranteed that the solution $\vec{z}^*$ of (P6-$\mu_{a^*,b^*}$) defined in \eqref{prob:p6_mu} satisfies $\vec{z}^* \in \Delta_{n,r}$. For each $a,b \in [1:n]$, we will denote the three conditions,
\begin{align}
    &\mathcal{A}_{a,b} = \mathbbm{1}\{y_b \geq \mu_{a,b} \geq y_a-1\}\nonumber\\
    &\mathcal{B}_{a,b} = \mathbbm{1}\{(b = n) \text{ or } [(b<n) \text{ and }(y_{b+1} < \mu_{a,b})]\}\nonumber\\
    &\mathcal{C}_{a,b} = \mathbbm{1}\{(a = 1) \text{ or } [(a > 1) \text{ and }(y_{a-1}-1 > \mu_{a,b})]\}
\end{align}
Hence, our goal is to find $(a^*,b^*)$ such that $\mathcal{A}_{a^*,b^*} = 1$, $\mathcal{B}_{a^*,b^*} = 1$, and $\mathcal{C}_{a^*,b^*} = 1$.

An easy way to find $a^*,b^*$ is to go through all $a,b \in [1:n]$ and check whether the above three conditions are satisfied. This approach has to go through $n^2$ pairs $(a,b)$. We will provide an alternative approach that is efficient and goes through at most $n$ pairs $(a,b)$. With this approach, we can also establish the existence of $a^*,b^* \in [1:n]$ satisfying $\mathcal{A}_{a^*,b^*} = 1$, $\mathcal{B}_{a^*,b^*} = 1$, and $\mathcal{C}_{a^*,b^*} = 1$.

Given  $a,b \in [1:n]$, define $g(a,b)$ as the minimum integer in $[b:n]$ such that $\mathcal{B}_{a,g(a,b)} = 1$ (Notice that $\mathcal{B}_{a,n} = 1$, so such an integer always exists). Similarly, define $h(a,b)$ as the maximum integer in $[1:a]$ such that $\mathcal{C}_{h(a,b),b} = 1$ (Notice that $\mathcal{C}_{1,b} = 1$, so such an integer always exists).

We have the following claim.

\noindent
\textbf{Claim 1: } If $\mathcal{A}_{a,b} = 1$ then we have that $\mathcal{A}_{a,g(a,b)} = 1$ and  $\mathcal{A}_{h(a,b),b} = 1$
\begin{proof}
    We only prove that $\mathcal{A}_{a,g(a,b)}= 1$. The other part follows from a similar argument. First, notice that if $g(a,b) = b$, we are done. Hence, we will assume $g(a,b) > b$. We prove a stronger statement. We prove that $\mathcal{A}_{a,c} = 1$ for all $c \in [b:g(a,b)]$. We use induction for the proof. Notice that the base case $c = b$ is true. Now assume that $\mathcal{A}_{a,c} = 1$ for some $c \in [b:g(a,b)-1]$. We prove that  $\mathcal{A}_{a,c+1} = 1$. Since $c \in  [b:g(a,b)-1]$, from the definition of function $g$, we have that $\mathcal{B}_{a,c} = 0$. Also since $c \leq g(a,b)-1$, we have that $c< n$. Hence, using the definition of $\mathcal{B}_{a,c}$, we have that $y_{c+1} \geq \mu_{a,c}$. Hence,
    \begin{align}
        \mu_{a,c+1} = \frac{\mu_{a,c}(c-a+1) + y_{c+1}}{c-a+2} \leq \frac{y_{c+1}(c-a+1) + y_{c+1}}{c-a+2} = y_{c+1},
    \end{align}
    where for the first equation we have used the definition of $\mu_{a,c+1}$ from \eqref{eqn:def_mu_ab}. Also,
    \begin{align}
        \mu_{a,c+1} = \frac{\mu_{a,c}(c-a+1) + y_{c+1}}{c-a+2}  = \mu_{a,c} + \frac{y_{c+1} - \mu_{a,c} }{c-a+2} \geq_{(a)} \mu_{a,c} \geq_{(b)} y_a-1,
    \end{align}
where (a) follows since $y_{c+1} \geq \mu_{a,c}$ and (b) follows since $\mathcal{A}_{a,c}$ is true by assumption. From the above two inequalities, we have that $\mathcal{A}_{a,c+1} = 1$ as desired.
    
\end{proof}

Now consider the following sequence $\mathcal{S}$ of tuples $\mathcal{S} = \{ (a_1,b_1),(a_2,b_2),\dots\}$, where $(a_1,b_1) = (r,r)$, and $(a_i,b_i) = (h(a_{i-1},g(a_{i-1},b_{i-1})),g(a_{i-1},b_{i-1}))$ for each $i >1$. We have the following claim regarding $\mathcal{S}$.

\noindent
\textbf{Claim 2: } We have that $\mathcal{A}_{a_i,b_i} = 1$ and $\mathcal{C}_{a_i,b_i} = 1$ for all $i \in \{2,3,\dots\}$.
\begin{proof}
    The fact that $\mathcal{C}_{a_i,b_i} = 1$ for all  $i \in \{2,3,\dots\}$ follows from the definition of $a_i, b_i$ and the function $h$, since $(a_i,b_i) = (h(a_{i-1},g(a_{i-1},b_{i-1})),g(a_{i-1},b_{i-1}))$ for all $i > 1$. For the other part we use induction. It can be easily checked that $\mathcal{A}_{a_1,b_1} = \mathcal{A}_{r,r} = 1$. Assume $\mathcal{A}_{a_i,b_i} = 1$ for some $i \geq 1$. Hence, we have from claim 1 that $\mathcal{A}_{a_i,g(a_{i},b_{i})} = 1$. Applying claim 1 again we have that $\mathcal{A}_{h(a_i,g(a_{i},b_{i})),g(a_{i},b_{i})} = 1$ which completes the induction.
\end{proof}

Now notice that the sequence $\mathcal{S}$ satisfies,
\begin{align}\label{eqn:1111}
    a_{i+1} \leq a_i, b_{i+1} \geq b_i
\end{align}
for all $i \in \{1,2,\dots\}$. This is because $ b_{i+1}  = g(a_{i},b_{i}) \geq b_i$ by definition of function $g$ and $a_{i+1}  = h(a_i,g(a_{i},b_{i})) \leq a_i$ by definition of function $h$. Additionally, from the definition of sequence $\mathcal{S}$, it can be easily seen that if $(a_{i+1},b_{i+1}) = (a_i,b_i)$ for some $i \geq 1$, then we have $(a_j,b_j) = (a_i,b_i)$ for all $j \geq i$. Combining the above property with \eqref{eqn:1111}, we have that the sequence $\mathcal{S}$ is eventually constant. In particular, there exists $i \geq 1$ such that $(a_j,b_j) = (\bar{a},\bar{b})$ for all $j \geq i$. It is also not difficult to see that the minimum such $i$ satisfies $i \leq n$. To see this, notice that,
\begin{align}\label{eqn:7383}
   n -1 &\geq  b_i - a_i  = \sum_{j=1}^{i-1}[b_{j+1} - b_j + a_{j} - a_{j+1}] \geq (i-1),
\end{align}
where the last inequality follows since for each $j < i$, we should have $a_{j+1} \leq  a_j$ and $b_{j+1} \geq b_j$, and at least one of the two inequalities is strict (if not we will have $(a_{j+1}, b_{j+1}) =  (a_j,b_j)$ which will contradict the minimality of $i$).

From claim 2 we have that $\mathcal{A}_{\bar{a},\bar{b}} = 1$ and $\mathcal{C}_{\bar{a},\bar{b}} = 1$. We also prove that $\mathcal{B}_{\bar{a},\bar{b}} = 1$. To prove this, pick any $j > i$. We have that $(a_{j+1},b_{j+1}) = (h(a_{j},g(a_{j},b_{j})),g(a_{j},b_{j}))$, which reduces to $(\bar{a} ,\bar{b}) = (h(\bar{a},g(\bar{a},\bar{b})),g(\bar{a},\bar{b}))$. Hence, we have $\bar{b} = g(\bar{a},\bar{b})$. Notice that since from the definition of $g$, we have that $\mathcal{B}_{\bar{a},g(\bar{a},\bar{b})} = 1$ we have that $\mathcal{B}_{\bar{a},\bar{b}} = 1$ as desired. Hence, $(a^*,b^*)$ exists and is equal to $(\bar{a},\bar{b})$.

To find $(\bar{a},\bar{b})$ we enumerate the sequence $\mathcal{S}$. As established by \eqref{eqn:7383}, the sequence becomes constant before $n$ steps. Hence, this process is more efficient compared to the naive scheme which evaluates $\mu_{a,b}$ values for all $a,b \in [1:n]$.

\noindent
\textbf{Note: } In Algorithm~\ref{algo:333} although we have defined $\mu_{a,b}$, $g(a,b)$, $h(a,b)$, $\mathcal{A}_{a,b}$, $\mathcal{B}_{a,b}$, and $\mathcal{C}_{a,b}$ for all $a,b \in [1:n]$, we only require computing above for $(a,b)$ tuples in $\mathcal{S}$.
\bibliographystyle{IEEEtran}
\bibliography{main}
\end{document}

%% file: opt_sol_m3r1.tex
\centering
\begin{minipage}{0.48\textwidth}
\resizebox {\textwidth} {!} {
\begin{tikzpicture}[spy using outlines = {magnification = 1, rectangle, size=2cm,black, connect spies}]
        \begin{axis}[
            ylabel={$E_j$},
            xlabel={$j$},
            xmin= 1, xmax=10,
            ymin= 0, ymax=10,
            xtick={1,2,3,4,5,6,7,8,9,10},
            ytick={0,2,4,6,8},
            ymajorgrids=true,
            xmajorgrids=true,
            grid style=dashed,
            legend pos=north east,
            every axis plot/.append style={thin},
            label style={font=\small},
            tick label style={font=\tiny},
            legend style={nodes={scale=0.9, transform shape}}
        ]  
        \addplot[
            color=red,
            only marks,
            mark options={scale=1.5}
            ] coordinates {
            (1,9)(2,6.7)(3,5.5)(4,4.5)(5,1.263157894736842)(6,1.2105263157894737)(7,1.1578947368421053)(8,1.1052631578947367)(9,1.0526315789473684)(10,1.0)
            };
            \addlegendentry{Scenario 1};
        \addplot[
            color=blue,
            only marks,
            mark options={scale=1.5},
            mark = *, 
            ] coordinates {
            (1,9.1)(2,6.7)(3,5.5)(4,4.5)(5,1.263157894736842)(6,1.2105263157894737)(7,1.1578947368421053)(8,1.1052631578947367)(9,1.0526315789473684)(10,1.0)
            };
            \addlegendentry{Scenario 2};
        \end{axis}
    \end{tikzpicture}
}
\end{minipage}
\begin{minipage}{0.48\textwidth}
\resizebox {\textwidth} {!} {
\begin{tikzpicture}[spy using outlines = {magnification = 1, rectangle, size=2cm,black, connect spies}]
        \begin{axis}[
            ylabel={$p_j$},
            xlabel={$j$},
            xmin= 1, xmax=10,
            ymin=-0.1, ymax=1.1,
            xtick={1,2,3,4,5,6,7,8,9,10},
            ytick={0,0.25,0.5,0.75,1},
            ymajorgrids=true,
            xmajorgrids=true,
            grid style=dashed,
            legend pos=north east,
            every axis plot/.append style={thin},
            label style={font=\small},
            tick label style={font=\tiny},
            legend style={nodes={scale=0.9, transform shape}}
        ] 
        \addplot[
            color=red,
            only marks,
            mark options={scale=1.5}
            ] coordinates {
          (1,0.2512785543811797)(2,0.3375383566314354)(3,0.4111830889873849)(4,0)(5,0)(6,0)(7,0)(8,0)(9,0)(10,0)
            };
            \addlegendentry{Scenario 1};
        \addplot[
            color=blue,
            only marks,
            mark options={scale=1.5}
            ] coordinates {
         (1,0.4989393871011419)(2,0.2258879812248951)(3,0.27517263167396305)(4,0)(5,0)(6,0)(7,0)(8,0)(9,0)(10,0)
            };
            \addlegendentry{Scenario 2};
        \end{axis}
        
    \end{tikzpicture}
}
\end{minipage}
\caption{\textbf{Left: } The mean rewards of the resources, \textbf{Right: } Probabilities of choosing the resources}\label{fig:201}

%% file: opt_sol_m2.tex
\centering
\begin{minipage}{0.48\textwidth}
\resizebox {\textwidth} {!} {
\begin{tikzpicture}[spy using outlines = {magnification = 1, rectangle, size=2cm,black, connect spies}]
        \begin{axis}[
            ylabel={$E_j$},
            xlabel={$j$},
            xmin= 1, xmax=10,
            ymin= 0, ymax=8,
            xtick={1,2,3,4,5,6,7,8,9,10},
            ytick={0,2,4,6,8},
            ymajorgrids=true,
            xmajorgrids=true,
            grid style=dashed,
            legend pos=north west,
            every axis plot/.append style={thin},
            label style={font=\small},
            tick label style={font=\tiny},
            legend style={nodes={scale=0.9, transform shape}}
        ]  
        \addplot[
            color=blue,
            only marks,
            mark options={scale=1.5}
            ] coordinates {
            (1,7)(2,6.7)(3,3.6842105263157894)(4,3.526315789473684)(5,3.3684210526315788)(6,3.210526315789474)(7,3.0526315789473686)(8,2.894736842105263)(9,2.736842105263158)(10,2.5789473684210527)
            };
        \end{axis}
    \end{tikzpicture}
}
\end{minipage}
\begin{minipage}{0.48\textwidth}
\resizebox {\textwidth} {!} {
\begin{tikzpicture}[spy using outlines = {magnification = 1, rectangle, size=2cm,black, connect spies}]
        \begin{axis}[
            ylabel={$p_j$},
            xlabel={$j$},
            xmin= 1, xmax=10,
            ymin=-0.1, ymax=1.1,
            xtick={1,2,3,4,5,6,7,8,9,10},
            ytick={0,0.25,0.5,0.75,1},
            ymajorgrids=true,
            xmajorgrids=true,
            grid style=dashed,
            legend pos=north west,
            every axis plot/.append style={thin},
            label style={font=\small},
            tick label style={font=\tiny},
            legend style={nodes={scale=0.9, transform shape}}
        ]  
        \addplot[
            color=red,
            only marks,
            mark options={scale=1.5}
            ] coordinates {
            (1,1)(2,1)(3,0.18205108623615748)(4,0.1902026274109108)(5,0.19911837557079723)(6,0.20891108256608232)(7,0.2197168282160521)(8,0)(9,0)(10,0)    
            };
        \end{axis}
    \end{tikzpicture}
}
\end{minipage}
\caption{\textbf{Left: } The mean rewards of different resources. \textbf{Right: } Probabilities of choosing different resources for the considered $\vec{E}$.}\label{fig:200}